\documentclass[11pt,a4paper]{article}
\pdfoutput=1

\usepackage{jheppub-nosort,amsthm}

\usepackage{amsmath,amssymb,amsthm,amscd}
\usepackage{tikz}
\tikzset{node distance=2cm, auto}
\usepackage{epsfig}
\usepackage{psfrag,comment}
\usepackage{graphicx,caption,subcaption}
\newcommand{\CCD}{{\cal D}}

\newcommand{\CF}{{\cal F}}

\newcommand{\CN}{{\cal N}}

\newcommand{\CX}{{\cal X}}

\def\BN{{\mathbb N}}
\def\BZ{{\mathbb Z}}
\def\BR{{\mathbb R}}
\def\BC{{\mathbb C}}
\def\BP{{\mathbb P}}

\newcommand{\be}{\begin{equation}}
\newcommand{\ee}{\end{equation}}
\newcommand{\ba}{\begin{aligned}}
\newcommand{\ea}{\end{aligned}}
\newcommand{\bea}{\begin{eqnarray}}
\newcommand{\eea}{\end{eqnarray}}
\newcommand{\bean}{\begin{eqnarray*}}
\newcommand{\eean}{\end{eqnarray*}}

\newcommand{\p}{\partial}

\def\r{\right\rangle}

\def\1{\mathbf{1}}
\def\0{|\1\r}

\def\re{{\mathbb{R}}{\mathrm{e}}}

\newcommand{\rme}{{\rm e}}
\newcommand{\rmi}{{\rm i}}
\newcommand{\rmd}{{\rm d}}

\def\G{{\Gamma}}
\def\hol{{\textrm{hol}}}
\def\bfk{{\boldsymbol{k}}}

\def\bfa{{\boldsymbol\alpha}}
\def\bfs{{\boldsymbol\sigma}}
\def\bfe{{\boldsymbol{e}}}
\def\bf0{{\boldsymbol{0}}}
\def\bfm{{\boldsymbol{m}}}
\def\bfn{{\boldsymbol{n}}}
\def\th{{\theta}}
\def\tmu{{\widetilde{\mu}}}
\def\bfep{{\boldsymbol{\varepsilon}}}

\def\Xint#1{\mathchoice
   {\XXint\displaystyle\textstyle{#1}}%
   {\XXint\textstyle\scriptstyle{#1}}%
   {\XXint\scriptstyle\scriptscriptstyle{#1}}%
   {\XXint\scriptscriptstyle\scriptscriptstyle{#1}}%
   \!\int}
\def\XXint#1#2#3{{\setbox0=\hbox{$#1{#2#3}{\int}$}
     \vcenter{\hbox{$#2#3$}}\kern-.5\wd0}}

\def\dashint{\Xint-}

\newdimen\tableauside\tableauside=1.0ex
\newdimen\tableaurule\tableaurule=0.4pt
\newdimen\tableaustep
\def\phantomhrule#1{\hbox{\vbox to0pt{\hrule height\tableaurule width#1\vss}}}
\def\phantomvrule#1{\vbox{\hbox to0pt{\vrule width\tableaurule height#1\hss}}}
\def\sqr{\vbox{%
  \phantomhrule\tableaustep
  \hbox{\phantomvrule\tableaustep\kern\tableaustep\phantomvrule\tableaustep}%
  \hbox{\vbox{\phantomhrule\tableauside}\kern-\tableaurule}}}
\def\squares#1{\hbox{\count0=#1\noindent\loop\sqr
  \advance\count0 by-1 \ifnum\count0>0\repeat}}
\def\tableau#1{\vcenter{\offinterlineskip
  \tableaustep=\tableauside\advance\tableaustep by-\tableaurule
  \kern\normallineskip\hbox
    {\kern\normallineskip\vbox
      {\gettableau#1 0 }%
     \kern\normallineskip\kern\tableaurule}%
  \kern\normallineskip\kern\tableaurule}}
\def\gettableau#1{\ifnum#1=0\let\next=\null\else
\squares{#1}\let\next=\gettableau\fi\next}
\subheader{\hfill
{\small{\textsf{CERN-PH-TH-2014-110}}}
}

\title{Resurgent Transseries and the Holomorphic Anomaly: Nonperturbative Closed Strings in Local $\BC\BP^2$}

\author[a,b]{Ricardo~Couso-Santamar\'\i a,}
\affiliation[a]{Theory Division, Department of Physics, CERN,\\ 
CH--1211 Gen\`eve 23, Switzerland}
\affiliation[b]{Departamento de F\'\i sica de Part\'\i culas and IGFAE, Universidade de Santiago de Compostela,\\ E--15782 Santiago de Compostela, Spain}
\emailAdd{ricardo.couso@usc.es}

\author[b,c]{Jos\'e~D.~Edelstein,}
\affiliation[c]{Centro de Estudios Cient\'\i ficos, CECs, Casilla 1469, Valdivia, Chile}
\emailAdd{jose.edelstein@usc.es}

\author[a,d]{Ricardo~Schiappa,}
\affiliation[d]{CAMGSD, Departamento de Matem\'atica, Instituto Superior T\'ecnico, Universidade de Lisboa,\\ Av. Rovisco Pais 1, 1049--001 Lisboa, Portugal}
\emailAdd{schiappa@math.tecnico.ulisboa.pt}

\author[e]{Marcel~Vonk}
\affiliation[e]{Institute for Theoretical Physics, University of Amsterdam,\\ Science Park 904, 1090--GL Amsterdam, The Netherlands\\}
\emailAdd{m.l.vonk@uva.nl}

\abstract{
The holomorphic anomaly equations describe B-model closed topological strings in Calabi--Yau geometries. Having been used to construct perturbative expansions, it was recently shown that they can also be extended past perturbation theory by making use of resurgent transseries. These yield \textit{formal} nonperturbative solutions, showing integrability of the holomorphic anomaly equations at the nonperturbative level. This paper takes such constructions one step further by working out in great detail the specific \textit{example} of topological strings in the mirror of the local $\BC\BP^2$ toric Calabi--Yau background, and by addressing the associated (resurgent) large-order analysis of both perturbative and multi-instanton sectors. In particular, analyzing the asymptotic growth of the perturbative free energies, one finds contributions from three different instanton actions related by $\BZ_3$ symmetry, alongside another action related to the K\"ahler parameter. Resurgent transseries methods then compute, from the extended holomorphic anomaly equations, higher instanton sectors and it is shown that these precisely control the asymptotic behavior of the perturbative free energies, as dictated by resurgence. The asymptotic large-order growth of the one-instanton sector unveils the presence of resonance, \textit{i.e.}, each instanton action is necessarily joined by its symmetric contribution. The structure of different resurgence relations is extensively checked at the numerical level, both in the holomorphic limit and in the general nonholomorphic case, always showing excellent agreement with transseries data computed out of the nonperturbative holomorphic anomaly equations. The resurgence relations further imply that the string free energy displays an intricate multi-branched Borel structure, and that resonance must be properly taken into account in order to describe the full transseries solution.
}


\arxivnumber{1407.4821}

\begin{document}

%%%%%%%%%%%%%%%%%%%%%%%%%%%%%%%%%%%%%%%%%%%%%%%%%%%%%%%%%%%%%%%%%
%%%%%%%%%%%%%%%%%%%%%%%%%%%%%%%%%%%%%%%%%%%%%%%%%%%%%%%%%%%%%%%%%
\maketitle
%%%%%%%%%%%%%%%%%%%%%%%%%%%%%%%%%%%%%%%%%%%%%%%%%%%%%%%%%%%%%%%%%
%%%%%%%%%%%%%%%%%%%%%%%%%%%%%%%%%%%%%%%%%%%%%%%%%%%%%%%%%%%%%%%%%

\vfill

\eject

\allowdisplaybreaks

%%%%%%%%%%%%%%%%%%%%%%%%%%%%%%%%%%%%%%%%%%%%%%%%%%%%%%%%%%%%%%%%%
%%%%%%%%%%%%%%%%%%%%%%%%%%%%%%%%%%%%%%%%%%%%%%%%%%%%%%%%%%%%%%%%%
\section{Introduction and Summary}
%%%%%%%%%%%%%%%%%%%%%%%%%%%%%%%%%%%%%%%%%%%%%%%%%%%%%%%%%%%%%%%%%
%%%%%%%%%%%%%%%%%%%%%%%%%%%%%%%%%%%%%%%%%%%%%%%%%%%%%%%%%%%%%%%%%

A perturbative expansion lies at the very root of closed string theory; the familiar topological expansion in Riemann surfaces of given genus. Perhaps due to this birth out of perturbation theory, the question of how to adequately formulate closed string theory from a nonperturbative standpoint has been mostly open for over 40 years. One powerful approach to this problem has been to exploit the existence of large $N$ gauge theoretic duals \cite{m97, agmoo99}. This approach becomes particularly transparent and amenable to exact calculations when addressing closed topological strings and their matrix model  duals \cite{dv02, dv02a, emo07}. In these cases where the gauge theory can be reduced to a matrix model, large $N$ duality implies that one may take the matrix integral itself as the nonperturbative definition of the theory from the string point-of-view (also see, \textit{e.g.}, the recent proposals \cite{m08, em08, hmmo13, km13, hw14}). But could there be an alternative approach to this problem, where one tries to obtain information concerning the nonperturbative sectors of some given topological closed string theory without relying on the existence of a large $N$ dual?

As we shall try to motivate in the following, and proceeding with the analysis we started in our earlier paper on this topic \cite{cesv13}, the answer to the previous question is yes. The one (very reasonable) assumption one nonetheless has to make is that, being initially constructed via an asymptotic perturbative expansion whose coefficients grow factorially fast with genus \cite{s90, gp88}, the free energy of closed topological string theory \textit{is} a resurgent function\footnote{In particular this implies that adequate resummation procedures must be considered when trying to extract numerical information out of these asymptotic series and corresponding transseries, see, \textit{e.g.}, \cite{as13, gmz14}.}. In fact, the general theory of resurgence \cite{e81} precisely tries to make sense out of functions whose perturbative expansions in some adequate parameter (in this case, the string coupling) produce asymptotic series with vanishing radius of convergence (see, \textit{e.g.}, \cite{cnp93, ss03} for introductory reviews). Being ubiquitous across theoretical physics, it is of course no surprise that one has to properly address asymptotic series also within string theoretic problems. Let us then consider a perturbative expansion in some small parameter which is asymptotic, \textit{i.e.}, whose coefficients grow factorially fast with the perturbative order. Resurgence first shows that the origin of this factorial growth lies within the existence of nonperturbative sectors of the original problem under consideration (a fact which was of course already noticed long ago, albeit somewhat empirically, within the instanton literature, \textit{e.g.}, \cite{bw73, z81}). But resurgence goes beyond this prediction for the leading large-order behavior of the perturbative expansion, by making the relation between perturbative and nonperturbative sectors, and even the relations between the many distinct nonperturbative sectors themselves, quantitatively \textit{precise}. Indeed, resurgence provides a whole network of (large-order) relations between \textit{all} possible perturbative and nonperturbative sectors of the theory one is addressing, constraining it considerably and, at the same time, presenting a mechanism through which it is possible to check our analytic nonperturbative computations (which may be done to very precise numerical accuracy, by the use of large-order analysis). Let us further note that resurgence has been steadily applied to string theoretic problems over the last few years \cite{m06, msw07, m08, msw08, gm08, ps09, mpp09, gikm10, kmr10, dmp11, asv11, sv13, cesv13, gmz14}\footnote{Also, recently, resurgence has been applied to gauge theoretic problems; see, \textit{e.g.}, \cite{au12a, au12b, du12a, du12b}.}, albeit mostly having some form of large $N$ duality in mind. We refer the reader to the introduction of \cite{asv11} for a brief overview of these recent developments, and to the excellent review \cite{m10} for a more detailed introduction to these topics.

It is important to point out that one is facing distinct starting scenarios, when comparing the approach carried through in the aforementioned string theoretic references with the particular calculation framework we shall address in the present case of closed topological string theory. In fact, one way or another, the references above mostly deal with matrix models, where orthogonal polynomial methods yield finite-difference equations (the so-called ``string equations'') which may be solved perturbatively to compute genus-$g$ free energies \cite{biz80}, and where their double-scaling limits are described by nonlinear ordinary differential equations associated to integrable hierarchies, see, \textit{e.g.}, \cite{fgz93}. The main point is that these are equations in the string coupling itself, and one may then use these nonlinear finite-difference or differential equations in order to compute nonperturbative data via resurgent transseries techniques. This was for instance carried out in the examples of the Painlev\'e I equation describing 2d quantum gravity \cite{gikm10, asv11}, the Painlev\'e II equation describing 2d quantum supergravity \cite{m08, sv13}, and both one \cite{m08, asv11} and two-cut \cite{sv13} solutions to the quartic matrix model. Generically, such ``string equations'' do not exist in the topological string case (and are usually only available when explicitly considering the matrix model large $N$ dual). In these cases, one method of computing the string free energy is to solve the holomorphic anomaly equations \cite{bcov93b, bcov93, o94} (which have also been rather systematically used to produce very high genus perturbative data, see, \textit{e.g.}, \cite{hkq06, gkmw07, hkr08, kmr10, dmp11}). But the one important point to have in mind is that the holomorphic anomaly equations are recursive equations (in genus) for the string free energy, and they are \textit{not} equations in the string coupling. This means that the strategy of solution is now necessarily \textit{different}, and we have developed the general theory of resurgent transseries solutions to the holomorphic anomaly equations within this context in \cite{cesv13} (and in fact we refer the reader at the very least to the introduction of that paper, for an overview of the main ideas behind such an endeavor).

To be more specific, in \cite{cesv13} we have shown that the B-model holomorphic anomaly equations may be rewritten in a way that allows for solutions expressible as transseries, \textit{i.e.}, perturbative expansions both in the string coupling, $g_s$, and in the nonanalytic instanton factor, $\exp \left( -A/g_s \right)$ (see, \textit{e.g.}, \cite{e0801} for an introduction to transseries). Here $A$ is of course the instanton action. In \cite{cesv13} we have shown that this instanton action is holomorphic in the complex structure moduli, so that it may retain its geometrical interpretation as a period of the Calabi--Yau geometry. In other words, although the holomorphic anomaly equations cannot determine the instanton action (it is in some sense part of the holomorphic ambiguities), because toric Calabi--Yau threefolds may be essentially described by their mirror curves \cite{hv00}, one may use the accumulated (spectral curve) matrix model experience, \textit{e.g.}, \cite{d91, d92, kk04, msw07, ps09}, to pinpoint the topological-string instanton action as a combination of Calabi--Yau periods \cite{dmp11}. In \cite{cesv13} we have further shown how higher instanton free-energies could be computed and, in some cases, how their holomorphic ambiguities could be fixed. However, some important features of the transseries solution, such as the number of its parameters (equivalent to the number of different instanton actions explicitly appearing in the transseries), the starting powers of the string coupling $g_s$ in the asymptotic series associated to the many (nonperturbative) sectors, or, as already mentioned, the instanton actions themselves; all these have to be obtained from a resurgence study of the perturbative free energy. This (apparent) lack of computational power lies in the fact that the holomorphic anomaly equations are equations in the B-model complex structure moduli, instead of ``string equations'' in the string coupling; and this sums up to the fact that, in this class of problems, the resurgent analysis itself must provide information which will later be checked against analytically computed expressions! In summary, when a ``string equation'' is available, it determines a transseries which may then be validated by resurgence. When no such equation is available, as in our present scenario, resurgence plays a more prominent role as classes of transseries solutions may be put forward \cite{cesv13} but it will have to be resurgence itself that actually guides us in the construction of the final transseries. While these ideas were of course already introduced in \cite{cesv13}, it is perhaps only now, where we work out an explicit example in great detail, that it will also become fully transparent how one should proceed in practice.

In this paper we will address the paradigmatic example of closed topological string theory in the B-model mirror of the local $\BC\BP^2$ toric Calabi--Yau background. This is one of the simplest---albeit nontrivial---examples one may consider as the moduli space of this geometry is one-dimensional and thus the integration of the holomorphic anomaly equations becomes simpler (see \cite{kz99, hkr08} for the integration of these equations in the perturbative case). These simplifications allow for a somewhat easier and explicit computation of many perturbative and nonperturbative free energies, both analytically or numerically, which we shall later use to perform very high precision checks on the resurgent transseries structure of the nonperturbative free energy. As expected, we find that the large-order growth of the perturbative sector is precisely controlled by the one-instanton sector, which we compute directly from the nonperturbative holomorphic anomaly equations. In particular, there are several instanton actions one must consider, leading to an intricate structure of nonperturbative sectors at the level of the free energy transseries. Analysis of the asymptotic growth of the one-instanton free energies gives us access to both a two-instanton sector, and to a nonperturbative sector characterized by an instanton action along with its \textit{symmetric} contribution. This effect explicitly shows the appearance of resonance in the topological string setting---a phenomenon which was previously studied in the contexts of minimal strings and matrix models in \cite{gikm10, asv11, sv13}---hinting at the fact that the full string theoretic transseries may also include logarithmic sectors, or perhaps even more complicated nonanalytic structures. However, it is important to notice that while resonance and the (nonperturbative) logarithmic sectors it generates may be well understood in matrix model examples \cite{gikm10, asv11, sv13}, this is only the case because one has a ``string equation'' available, which does not occur within our topological string set-up as we already discussed. This means that in the string theoretic context a question remains open on what the full transseries consequences of resonance are. All these analyses, analytical and numerical, may be performed in full generality, \textit{i.e.}, considering both holomorphic and nonholomorphic dependences of the (nonperturbative) closed string free energies. Of course many simplifications take place when considering the holomorphic limit and we also study how resurgence is modified in this case. We perform many high precision numerical calculations which validate the structure we have proposed in \cite{cesv13} and which we further unveil in the present work. These numerical tests also clearly show that the holomorphic anomaly equations provide a powerful tool to compute nonperturbative effects, \textit{if} and \textit{when} conveniently guided by resurgence relations.

This paper is organized as follows. We begin with a review of the geometry of local $\BC\BP^2$ to set the stage, in section \ref{sec:CYgeo}. We then start our analysis in section \ref{sec:PertF}, where we describe the computation of the perturbative free energies in the language of the (holomorphic anomaly) propagators---a computation which we have carried out up to genus $114$, generating sufficiently precise large-order data. Having computed enough perturbative coefficients, we can then thoroughly analyze their large-order behavior, and extract the relevant dominant instanton action, which we do in section \ref{sec:largeorderanalysisoftheperturbativeexpansion}. Depending on the regime, there are other subdominant instanton actions which we also describe, as they will play important roles in the subsequent analyses. Furthermore, already at this level it will become clear that the string free energy has a multi-branched Borel structure, with singularities moving between Riemann sheets as we vary the moduli. We start section \ref{sec:resurgentanalysisofthetransseriessolution} reviewing the nonperturbative extension of the holomorphic anomaly equations we proposed in \cite{cesv13}. The rest of the section analyzes in detail the large-order growth of both perturbative and one-instanton free energies in our example of local $\BC\BP^2$, finding an excellent agreement between numerical results and expressions we compute from the nonperturbative holomorphic anomaly equations, as dictated by resurgence. The direct integration of these nonperturbative equations alongside the associated subtle issue of fixing the holomorphic ambiguities is considered in parallel with the large-order analysis. In section \ref{sec:resummation} we then address exponentially subleading contributions to the original perturbative growth. In particular, this involves a Borel--Pad\'e resummation of the one-instanton free energies, in order to access both one-instanton free energies associated to the other (subleading) instanton actions, and also the naturally expected two-instanton sectors. Note that both sections \ref{sec:resurgentanalysisofthetransseriessolution} and \ref{sec:resummation} include a very large number of high-precision numerical checks on the validity of our proposals. Furthermore, they include a discussion of how the resurgence results imply that the string free energy has an intricate Borel singularity structure, and how extra resonant sectors (which, however, we have not addressed in full generality) may still have an important role to play in the structure of the full transseries. We end with some conclusions and an outlook in section \ref{sec:conclusions}. In an appendix, we include a schematic description of the nonperturbative free energies we computed for local $\BC\BP^2$.

%%%%%%%%%%%%%%%%%%%%%%%%%%%%%%%%%%%%%%%%%%%%%%%%%%%%%%%%%%%%%%%%%
%%%%%%%%%%%%%%%%%%%%%%%%%%%%%%%%%%%%%%%%%%%%%%%%%%%%%%%%%%%%%%%%%
\section{Calabi--Yau Geometry and Local $\mathbb{C}\mathbb{P}^2$}\label{sec:CYgeo}
%%%%%%%%%%%%%%%%%%%%%%%%%%%%%%%%%%%%%%%%%%%%%%%%%%%%%%%%%%%%%%%%%
%%%%%%%%%%%%%%%%%%%%%%%%%%%%%%%%%%%%%%%%%%%%%%%%%%%%%%%%%%%%%%%%%

In this section we shall review the relevant facts concerning the geometry of the local $\mathbb{CP}^2$ toric Calabi--Yau threefold. For generalities concerning topological strings in Calabi--Yau geometries, we refer the reader to, \textit{e.g.}, the reviews \cite{m05, m04, nv04, v05, a12}. Herein, we shall mainly follow the conventions in \cite{hkr08} and will try to give a short and self-contained presentation. After a brief description of the toric geometry on the A-side, we construct the mirror by using standard techniques and present the associated Picard--Fuchs equations, which allow us to compute the periods of the geometry (they will play a role later as instanton actions). For the moment, these periods of course combine to yield both the mirror map, and the derivative of the genus-zero free energy. The Yukawa coupling is computed immediately afterwards.

%%%%%%%%%%%%%%%%%%%%%%%%%%%%%%%%%%%%%%%%%%%%%%%%%%%%%%%%%%%%%%%%%
\subsection{Toric Geometries and their Mirrors}
%%%%%%%%%%%%%%%%%%%%%%%%%%%%%%%%%%%%%%%%%%%%%%%%%%%%%%%%%%%%%%%%%

The three-dimensional complex manifold denoted by local $\mathbb{CP}^2$ may be described as the canonical line-bundle over the complex compact surface $\mathbb{CP}^2$, that is, $\mathcal{O}(-3)\rightarrow \mathbb{CP}^2$. Because the Chern class of the $\mathbb{CP}^2$ cancels that of the line bundle, the total space is a noncompact (or local) Calabi--Yau manifold, of toric type. These threefold geometries are described as quotients of $\mathbb{C}^n$, with some points removed, over the action of a group, $G$. Our example, local $\mathbb{CP}^2$, is given by
\be
\left( \mathbb{C}^4 - \{ x_1=x_2=x_3=0 \})\right) / G,
\ee
\noindent
where $x_0, \ldots, x_3$ are the usual coordinates on $\mathbb{C}^4$ and the group is $G = \{(t^{-3},t,t,t)\in (\mathbb{C}^\star)^4 \} \simeq \mathbb{C}^\star$ acting by component-wise multiplication. Notice that the projection $\pi: \mathbb{C}^4 \to \mathbb{C}^3$, onto the last three components, maps the Calabi--Yau to its base $\mathbb{CP}^2 = \left( \mathbb{C}^3-\{0,0,0\} \right)/\mathbb{C}^\star$, whereas the first component keeps the information concerning the fiber. In general, toric varieties can be described in terms of combinatorial data. For our description of the dual geometry we only need to know the vector $Q = (-3,1,1,1)$ which characterizes the group action. The condition for vanishing Chern class is then equivalent to $\sum_{i=0}^3 Q_i =0$, which is the case in the example of local $\BC\BP^2$.

For local toric Calabi--Yau manifolds one can apply the Hori--Vafa construction \cite{hv00} in order to find the B-side mirror geometry. Define variables $w^+, w^- \in \mathbb{C}$ and $X_0,\ldots, X_3 \in \mathbb{C}^\star$, with the homogeneity condition $X_i \sim \lambda X_i$, $i=0,\ldots,3$, $\forall \lambda \in \mathbb{C}^\star$. Then, the mirror geometry is defined by the equations
\bea
\label{eq:wwX}
w^+ w^- &=& \sum_{i=0}^3 X_i, \\
z &=& \prod_{i=0}^3 X_i^{Q_i}.
\label{eq:zXQ}
\eea
\noindent
Here $z$ is the complex structure modulus of the manifold (the K\"ahler modulus in the A-geometry was not made explicit in the above description). We can fix the ambiguity in the $X_i$ coordinates by choosing $X_1\equiv 1$, or defining $x_i = \frac{X_i}{X_1}, \; i=0,2,3$. Using \eqref{eq:zXQ} we can solve for, say, $x_2$, and after renaming $x\equiv x_0$ and $y\equiv x_3$, \eqref{eq:wwX} simply becomes
\be
w^+ w^- = 1+x+z\, \frac{x^3}{y}+y\equiv H.
\ee
\noindent
Computations on this Calabi--Yau manifold can sometimes be expressed in terms of the geometry of the Riemann surface (in this case, a punctured torus) described by the zero-locus equation $H = 0$. This is for instance the case for the periods of the threefold geometry, that is, integrals over three-cycles. These cycles are based on one-cycles of the embedded Riemann surface, where the rest of the geometry (the fiber) has degenerated into a pair of lines $w^+ w^- = 0$ which can be integrated out. The modular structure associated with the presence of a torus, with a parameter $z$ that controls its complex structure, is reflected in the modular properties of the perturbative free energies. Certain values of $z$ describe particularly interesting geometries. The case $z=0$ corresponds to the large radius, or large volume, limit of the mirror geometry, where the K\"ahler parameter goes to infinity. Another special point, $z= -1/27$, is recognized by looking at the discriminant of the Riemann surface, $\Delta = 1+27z$. At this so-called conifold point, the torus becomes singular (alternatively, one can see that the $j$-invariant of the torus has a singularity at this point). There is a third special point that will be found when we study the Picard--Fuchs equation, the orbifold point, $z^{-1} = 0$, where the geometry locally becomes $\BC^3/\BZ_3$.

%%%%%%%%%%%%%%%%%%%%%%%%%%%%%%%%%%%%%%%%%%%%%%%%%%%%%%%%%%%%%%%%%
\subsection{Periods, Mirror Map and Yukawa Couplings}
\label{subsection:periods}
%%%%%%%%%%%%%%%%%%%%%%%%%%%%%%%%%%%%%%%%%%%%%%%%%%%%%%%%%%%%%%%%%

Mirror symmetry establishes a correspondence between the moduli spaces of two (mirror) manifolds: on the A-side one considers a manifold along with its K\"ahler moduli space, while on the B-side one finds another manifold, this time around along with its complex-structure moduli space. Natural coordinates in the latter moduli space are given, in general, by periods of the nowhere vanishing (holomorphic) $(3,0)$-form $\Omega$ present in every Calabi--Yau threefold. Written in the local coordinates above,
\be
\Omega = \frac{\rmd H \wedge \rmd x \wedge \rmd y}{H x y}.
\ee
\noindent
As we anticipated earlier, in the toric case these periods descend to period integrals over cycles of the Riemann surface. At the end of the day, for our example, we have both the usual A and B cycles from the torus, and an extra C-cycle which appears because the geometry is noncompact (and which has a constant value). The two nontrivial periods may be written as
\be
T = \int_\text{A} \lambda, \qquad T_D = \int_\text{B} \lambda,
\ee
\noindent
where $\lambda = \log y\, \frac{\rmd x}{x}$ is the one-form on the zero-locus Riemann surface $H=0$, obtained from $\Omega$. Just as in the general case, $T_D$ is redundant as a coordinate in the complex moduli space, and in fact it may be written, up to a conventional factor, as a derivative of the prepotential:
\be
T_D \propto \frac{\partial F_0}{\partial T}.
\ee
\noindent
The prepotential, $F_0$, is the genus-zero free energy in the B-model topological string theory of this target space. Instead of trying to calculate the above integrals directly, it is much simpler to solve the Picard--Fuchs equation associated to the geometry under consideration. This is a differential equation in the complex modulus, $z$, defined by a differential operator which annihilates the periods of the geometry and which has the general form 
\be
\CCD_z = \prod_{i\, \left|\, Q_i>0 \right.} \left(Q_i \theta_z \right)_{Q_i} - z \prod_{i\, \left|\, Q_i<0 \right.} \left(Q_i \theta_z \right)_{-Q_i}.
\ee
\noindent
Here we are using the notation $(x)_n := x \left(x-1\right) \cdots \left(x-n+1\right)$ and $\theta_z := z\, \frac{\rmd}{\rmd z}$. For local $\mathbb{CP}^2$ this operator reads
\be
\label{eq:PFoperator}
\CCD_z = \theta_z^3 + 3 z \theta_z \left( 3\theta_z+1 \right)\left( 3\theta_z+2 \right).
\ee
\noindent
From the standard theory of linear differential equations, one can immediately see that \eqref{eq:PFoperator} has two regular singular points at finite distance, $z=0$ and $z=-1/27$, and another one at infinity, $z^{-1}=0$. Due to the existence of these points, one has to solve the equation locally around each of them, using, for instance, the Frobenius method, and then extend the solution analytically past the singularities. This is what we shall do in the following. Since the equation is of third order, we should find three independent solutions. The easiest one is the constant solution, which is associated to the C-cycle we mentioned before. The other two are to be identified with the A and B periods. Furthermore, the description of the regular singular points is dependent upon the coordinate we choose to use for the complex modulus. For example, if we use a new coordinate (which will become relevant very soon), $\psi$, defined by
\be
\label{eq:psidefinition}
\psi^{-3}=-27z,
\ee
\noindent
then the Picard--Fuchs equation reads
\be
\label{eq:PFinpsicoordinate}
f'''(\psi) - \frac{3\psi^2}{1-\psi^3}\, f''(\psi) - \frac{\psi}{1-\psi^3}\, f'(\psi) = 0.
\ee
\noindent
In this coordinate one clearly sees that the three cubic-roots of unity are regular singular points. One can also check that $\psi^{-1}=0$ is a regular singular point.

\begin{figure}[t]
\begin{center}
\begin{subfigure}[b]{0.325\textwidth}
\includegraphics[width=\textwidth]{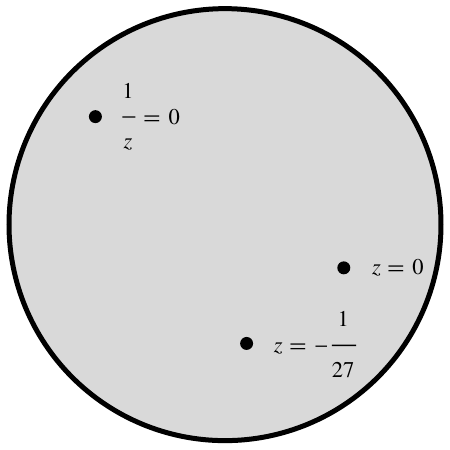}
\caption{The $z$ moduli space.}
\label{fig:zmodulispace}
\end{subfigure}
~
$\qquad\qquad$
~
\begin{subfigure}[b]{0.325\textwidth}
\includegraphics[width=\textwidth]{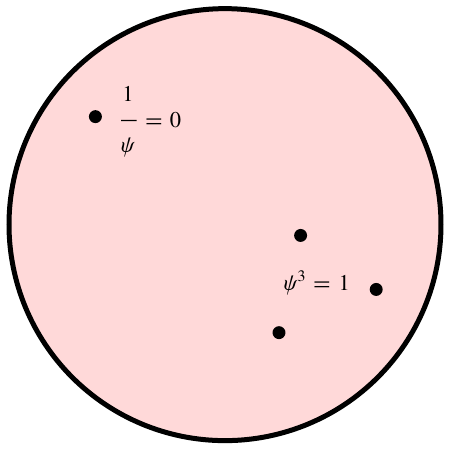}
\caption{The $\psi$ moduli space.}
\label{fig:psimodulispace}
\end{subfigure}
\end{center}
\vspace{-1\baselineskip}
\caption{Complex-structure moduli space of local $\BC\BP^2$ in different coordinates.}
\label{fig:modulispaces}
\end{figure}

To summarize, in the $z$-description we have three special points: large-radius ($z=0$), conifold ($z=-1/27$), and orbifold ($z^{-1}=0$); whereas in the $\psi$-description we have one large-radius point ($\psi^{-1}=0$), and three conifold points ($\psi^3-1=0$). Note, however, that the orbifold point, $\psi = 0$, is no longer a singular point. Figures \ref{fig:zmodulispace} and \ref{fig:psimodulispace} depict the complex structure moduli-space in $z$ and $\psi$ coordinates, respectively. A thorough account of the solutions to the above Picard--Fuchs equation is given in, \textit{e.g.}, \cite{dg99}. Let us consider these points in turn.

%%%%%%%%%%%%%%%%%%%%%%%%%%%%%%%%%%%%%%%%%%%%%%%%%%%%%%%%%%%%%%%%%
%%%%%%%%%%%%%%%%%%%%%%%%%%%%%%%%%%%%%%%%%%%%%%%%%%%%%%%%%%%%%%%%%

We begin the description of the solutions to the Picard--Fuchs equations by considering the large-radius point, $z=0$, around which the mirror map and the derivative of the prepotential are given by\footnote{The proportionality factor $-9$ in \eqref{eq:dTF0} is chosen in order to obtain standard Gromov--Witten invariants.}
\bea
\label{eq:periodT}
T &=& X^{(1)}, \\
-9\, \p_T F^{[\textrm{LR}]}_0 =\,\, T_D &=& \frac{3}{2} X^{(1,1)} - \frac{3}{2} X^{(1)} + \frac{3}{4}, \label{eq:dTF0}
\eea
\noindent
where
\bea
X^{(1)} &=& \frac{1}{2\pi\rmi} \left( \log z + \sigma_1 (z) \right), \\
X^{(1,1)} &=& \frac{1}{(2\pi\rmi)^2} \left( \log^2 z + 2 \sigma_1 (z)\, \log z + \sigma_2 (z) \right), \\
\sigma_1 (z) &=& \sum_{n=1}^{+\infty} 3\, \frac{(3n-1)!}{(n!)^3} \left(-z\right)^n = -6z+ 45z^2 - 560z^3 + \cdots, \\
\sigma_2 (z) &=& \sum_{n=1}^{+\infty} 18 \left( \sum_{k=n+1}^{3n-1}\frac{1}{k} \right) \frac{(3n-1)!}{(n!)^3} \left(-z\right)^n = -18z + \frac{423}{2}z^2 + \cdots. 
\eea
\noindent
From these expressions we can already calculate the Yukawa couplings by simply taking the third derivative of the prepotential. One finds,
\bea
C_{TTT} &=& \left(2\pi\rmi\right)^3 \frac{\partial^3 F_0^{[\text{LR}]}}{\partial T^3} = - \frac{1}{3} + \mathcal{O}(z), \\
C_{zzz} &=& \left( \frac{\partial T}{\partial z} \right)^3 C_{TTT} = - \frac{1}{3z^3(1+27z)}.
\label{eq:Czzz}
\eea
\noindent
Note how the last expression has a singularity at the conifold locus; this is a general feature of the Yukawa couplings. As we shall see, in the holomorphic anomaly equations the information concerning the genus-zero free energy will enter only through this rational function.

As we turn to the conifold point, we can make use of the centered coordinate $\Delta = 1+27z$. Around this point, the mirror map is given by the regular\footnote{Regular at the conifold point, thus unaffected by the logarithms present in the large-radius expression.} solution of the Picard--Fuchs equation, and it is given by the analytic continuation of $T_D$ above, \eqref{eq:dTF0}, up to normalization. The appropriate coordinates are now $t_\text{c}$, and its dual ${t_\text{c}}_D$, given by
\bea
t_\text{c} &=& \Delta + \frac{11}{18} \Delta^2 + \frac{109}{243} \Delta^3 + \cdots = -\frac{2\pi}{\sqrt{3}}\, T_D, \\
{t_\text{c}}_D &=& \partial_{t_c} F^{[\textrm{c}]}_0 = - T_D \log\Delta + \mathcal{O}(\Delta^0) = \frac{4\pi^2\rmi}{3\sqrt{3}}\, T.
\eea
\noindent
The coordinate $t_\text{c}$ is also called the flat coordinate around the conifold point, and we shall make extensive use of it in the following sections since it will turn out to be intimately related to the instanton actions controlling the large-order behavior of the perturbative sector (and of course also controlling the large-order behavior of higher instanton sectors).

Finally, around the orbifold point $z^{-1} = 0$, we shall use the coordinate $\psi$, defined in \eqref{eq:psidefinition}. It is now simple to check that the solutions of \eqref{eq:PFinpsicoordinate} are hypergeometric functions, and thus the mirror map and associated derivative of the prepotential are given by
\bea
\sigma &=& 3\alpha\psi\,\, {}_3 F_2 \left. \left( \frac{1}{3},\frac{1}{3},\frac{1}{3}; \frac{2}{3},\frac{4}{3}\, \right| \psi^3 \right), \\
\p_\sigma F^{[\textrm{orb}]}_0 &=& \frac{1}{6} \left( 3\alpha\psi\right)^2\, {}_3 F_2 \left. \left( \frac{2}{3},\frac{2}{3},\frac{2}{3}; \frac{4}{3},\frac{5}{3}\, \right| \psi^3 \right),
\eea
\noindent
where $\alpha = (-1)^{1/3}$. The orbifold point is not going to play a very significant role in the nonperturbative description of local $\BC\BP^2$, but we shall nonetheless express the conifold flat coordinate, $t_\text{c}$, in terms of the above hypergeometric functions. One can check that
\be
\label{eq:tcpsi}
t_\text{c} = \frac{2\pi}{\sqrt{3}} \left( \frac{3\psi}{\G\left(\frac{2}{3}\right)^3}\,\,  {}_3 F_2 \left. \left( \frac{1}{3},\frac{1}{3},\frac{1}{3}; \frac{2}{3},\frac{4}{3}\, \right| \psi^3 \right) - \frac{\frac{9}{2}\psi^2}{\G\left(\frac{1}{3}\right)^3}\,\, {}_3 F_2 \left. \left( \frac{2}{3},\frac{2}{3},\frac{2}{3}; \frac{4}{3},\frac{5}{3}\, \right| \psi^3 \right) - 1 \right).
\ee
\noindent
We mentioned earlier that the description of the complex structure moduli space may be equally carried out in terms of the coordinate $\psi$. In that picture, one finds three conifold points at the cubic roots of identity, while the orbifold point is no longer special, \textit{i.e.}, the Picard--Fuchs equation is regular at $\psi = 0$ and consequentially the solutions have no nontrivial monodromy around it. Later on we will find that the instanton actions that control the large-order behavior of the perturbative free energies are precisely related to these three conifold points!

%%%%%%%%%%%%%%%%%%%%%%%%%%%%%%%%%%%%%%%%%%%%%%%%%%%%%%%%%%%%%%%%%
%%%%%%%%%%%%%%%%%%%%%%%%%%%%%%%%%%%%%%%%%%%%%%%%%%%%%%%%%%%%%%%%%
\section{The Perturbative String Free Energy}\label{sec:PertF}
%%%%%%%%%%%%%%%%%%%%%%%%%%%%%%%%%%%%%%%%%%%%%%%%%%%%%%%%%%%%%%%%%
%%%%%%%%%%%%%%%%%%%%%%%%%%%%%%%%%%%%%%%%%%%%%%%%%%%%%%%%%%%%%%%%%

In this section, we shall review in some detail the computation of the higher genus \textit{perturbative} free energies \cite{hkr08} using the holomorphic anomaly equations \cite{bcov93} (we also refer the reader to \cite{cesv13} where a short introduction to the holomorphic anomaly equations was presented, essentially including all one needs to know in the following). Due to the recursive nature of this calculation, one has to begin with the genus-one free energy: we will review how to explicitly calculate its holomorphic limit, and then show its fundamental relation to the so-called propagator. For higher genera $g\geq 2$ one simply makes use of the aforementioned recursive nature of the holomorphic anomaly equations. The one detail to fix is that before starting the integration process one has to write every ingredient of the equations in terms of the propagator, and this still requires fixing some ambiguities associated with it. Finally, we will exemplify the process of solving the holomorphic anomaly equations with the explicit computation of the perturbative genus-two contribution, $F^{(0)}_2$, and we shall further recall how to fix the holomorphic ambiguities that naturally appear in the process of integration.

%%%%%%%%%%%%%%%%%%%%%%%%%%%%%%%%%%%%%%%%%%%%%%%%%%%%%%%%%%%%%%%%%
\subsection{The Genus-One Free Energy}
%%%%%%%%%%%%%%%%%%%%%%%%%%%%%%%%%%%%%%%%%%%%%%%%%%%%%%%%%%%%%%%%%

The genus-one free energy satisfies a holomorphic anomaly equation of its own \cite{bcov93b}, which describes how it fails to be a holomorphic quantity; given by
\be
\label{eq:F1HAE}
\p_{\bar{j}}\p_i F^{(0)}_1 = \frac{1}{2}C_{ijk}\, {\bar{C}_{\bar{j}}}^{\ jk} - \left( \frac{\chi}{24} - 1 \right) G_{i\bar{j}},
\ee
\noindent
and where barred indices naturally denote antiholomorphic variables. Here, the $C_{ijk}$ are the Yukawa couplings, and ${\bar{C}_{\bar{j}}}^{\ \ell k} = \bar{C}_{\bar{j}\bar{\ell}\bar{k}}\, G^{\ell\bar{\ell}}\, G^{k\bar{k}}\, \rme^{2K}$, where $G_{i\bar{j}} = \p_i\p_{\bar{j}}K$ is the K\"ahler metric on the complex-structure moduli space. As for the Euler characteristic $\chi$ of the Calabi--Yau threefold, we do not need its actual value. Equation \eqref{eq:F1HAE} can be integrated in general \cite{bcov93b} and its solution is
\be
F^{(0)}_1 = \log\left( \rme^{\frac{K}{2} \left(3 + h^{1,1} -\frac{\chi}{12} \right)} \left( \det G \right)^{-\frac{1}{2}} \left|f_1(z)\right|^2 \right),
\ee
\noindent
where the explicit value of $h^{1,1}$ is not relevant for us either. Here $f_1(z)$ is a holomorphic ambiguity which appears in the integration process. In order to fix this holomorphic ambiguity, one considers the special points of the geometry, near which the behavior of the free energy $F^{(0)}_1$ is known. With the \textit{ansatz} $f_1(z) = \Delta^r\, z^b$, where $\Delta = 1 + 27z$ and $r$ and $b$ yet to be determined, one first finds that the value of $r$ is associated with the conifold point, $z = -1/27$, and has the universal value of $-\frac{1}{12}$. Near the large-radius point, the holomorphic limit\footnote{A word on notation: while $F$ denotes general closed-string free energies, $\CF$ denotes their holomorphic limit.} of $F^{(0)}_1$ has to satisfy \cite{bcov93b}
\be
\lim_{z\to 0} \CF^{[\textrm{LR}](0)}_1 = \lim_{z\to 0} -\frac{1}{24}\, 2\pi\rmi T \int_M c_2 J,
\ee
\noindent
where, for local $\BC\BP^2$, $\int_M c_2 J = -2$. One thus finds $b = -\frac{7}{12}$, resulting in
\be
\mathcal{F}^{\textrm{[LR]}(0)}_1 = -\frac{1}{2} \log \frac{\partial T}{\partial z} -\frac{1}{12} \log z^7 \left(1+27z\right).
\label{eq:holoF1}
\ee
\noindent
Here we have used the fact that, in the holomorphic limit, the metric $G_{z\bar{z}}$ is proportional to $\p_z T$. Since later on we shall only be interested in the derivative of the genus-one free energy, herein we have also omitted any possible additive constants.

At this point, let us comment on the nature of the perturbative free energies when considering their holomorphic limit. As for many other Calabi--Yau threefolds, the perturbative free energies of local $\BC\BP^2$ may be written in terms of modular forms, in such a combination that any $F^{(0)}_g$ is in fact a modular \textit{function}, that is, has modular weight zero \cite{abk06}. The modular parameter turns out to be a function of the complex modulus, whose origin is rooted in the torus (the mirror curve) embedded in the Calabi--Yau geometry. In these modular functions, the nonholomorphic dependence is naturally implemented by the nonholomorphic extension of the quasimodular form $E_2$ (this is the second Eisenstein series). This also implies that the holomorphic $\CF^{(0)}_g$ are quasimodular forms. In this way, modularity is then present at the cost of nonholomorphicity. Now, taking this holomorphic limit introduces the notion of a \textit{frame}, as modular transformations on the aforementioned quasimodular free energies turn out to be equivalent to a change of frame. This means that whenever one takes the holomorphic limit, one has to specify a frame, and that is why throughout this paper we use a label for the frame. Special points of moduli space turn out to have preferred frames associated with them, so the labels $\text{[LR]}$, $\text{[c]}$ or $\text{[orb]}$ will make reference to those. In general topological string theories the identification of the modular properties of the model may not be a simple task but general approaches do exist, \textit{e.g.}, \cite{asyz13}.

Having discussed the large-radius point above, we may now turn to the other frames, either conifold or orbifold, where one finds analogous expressions to \eqref{eq:holoF1} with $T$ replaced by the corresponding flat coordinate, either $t_\text{c}$ or $\sigma$. The conifold frame will be important in the remainder of the paper,
\be
\mathcal{F}^{\textrm{[c]}(0)}_1 = -\frac{1}{2} \log \frac{\partial t_c}{\partial z} -\frac{1}{12} \log z^7 \left(1+27z\right).
\label{eq:holoF1c}
\ee
\noindent
As to the derivative of the full nonholomorphic free energy, it will be directly related to the propagator, \textit{i.e.}, to an antiholomorphic variable which is conveniently used to integrate the holomorphic anomaly equations---see \eqref{eq:derivativeF1propagator} below.

%%%%%%%%%%%%%%%%%%%%%%%%%%%%%%%%%%%%%%%%%%%%%%%%%%%%%%%%%%%%%%%%%
\subsection{The Holomorphic Anomaly Equations}
%%%%%%%%%%%%%%%%%%%%%%%%%%%%%%%%%%%%%%%%%%%%%%%%%%%%%%%%%%%%%%%%%

The holomorphic anomaly equations of \cite{bcov93} essentially describe the failure of a given fixed-genus (perturbative) free energy to be holomorphic, by (differentially) relating this failure to lower-genus (perturbative) free energies. They read
\be
\p_{\bar{i}} F^{(0)}_g = \frac{1}{2} {\bar{C}_{\bar{i}}}^{\ jk}\left( D_j\p_k F^{(0)}_{g-1} + \sum_{h=1}^{g-1} \p_j F^{(0)}_h\, \p_k F^{(0)}_{g-h} \right).
\label{eq:orginalHAEs}
\ee
\noindent
These equations allow for a recursive integration of the free energies in the antiholomorphic variables but, at the same time, they naturally introduce an undetermined holomorphic function as an ``integration constant'', the aforementioned holomorphic ambiguity, which needs to be fixed for each genus. The choice of antiholomorphic variables can be important in order to achieve a well-organized expression for the computed free energies. A particularly useful choice are the so-called propagator variables, $S^{ij}$. In our particular example of local $\mathbb{CP}^2$ where the complex-structure moduli space is one-dimensional, there is only one\footnote{With several complex-structure moduli, $S^{ij}$ is a symmetric matrix whose rank equals the dimension of the moduli space.} propagator, $S^{zz}$. Along with $S^{zz}$ there are also other propagators, $S^{z}$, $S$ and $K_z$, which nevertheless can be chosen to vanish in the holomorphic limit. In fact, since the dependence of the free energies on all these propagators is algebraic, we can consider them as formal independent external parameters that we can turn off if we so wish. In order to simplify the computations, we shall do so for all of them except, of course, $S^{zz}$ (but see \cite{asyz13} for a thorough discussion of propagators and their connection to modularity). The propagator satisfies, by definition,
\be
\label{eq:propagatordef}
\partial_{\bar{z}} S^{zz} = {\bar{C}_{\bar{z}}}^{\ zz},
\ee
\noindent
where ${\bar{C}_{\bar{z}}}^{\ zz}$ explicitly appears in the holomorphic anomaly equations \eqref{eq:orginalHAEs}, and
\be
{\bar{C}_{\bar{z}}}^{\ zz} = C_{\bar{z}\bar{z}\bar{z}} \left( G^{z\bar{z}} \right)^2 \rme^{2K}.
\ee
\noindent
The nonholomorphic dependence of ${\bar{C}_{\bar{z}}}^{\ zz}$ is complicated, thus preventing us from straightforwardly integrating the equations \eqref{eq:orginalHAEs}. The role of the propagator, $S^{zz}$, is precisely to overcome this obstruction: by using the chain rule and \eqref{eq:propagatordef}, one finds a simpler version of the holomorphic anomaly equations as
\be
\partial_{S^{zz}} F^{(0)}_g = \frac{1}{2} \left( D_z \partial_z F^{(0)}_{g-1} + \sum_{h=1}^{g-1} \partial_z F^{(0)}_h\, \partial_z F^{(0)}_{g-h} \right),
\label{eq:propHAEs}
\ee
\noindent
in which ${\bar{C}_{\bar{z}}}^{\ zz}$ is no longer present, and where
\be
D_z \partial_z F^{(0)}_g = \partial_z^2 F^{(0)}_g - \Gamma^z_{zz}\, \partial_z F^{(0)}_g.
\ee
\noindent
Here, $\Gamma^z_{zz}$ is the Christoffel symbol associated with the metric in the complex-structure moduli space. Perhaps the most important property that the propagator satisfies is that, when acted upon by the covariant derivative, it produces no antiholomorphic dependence other than itself \cite{yy04, al07, alm08}. Indeed, from special geometry and the definition of the propagator one can show that
\bea
\label{eq:covariantderivativepropagator}
D_z S^{zz} &=& - C_{zzz} \left(S^{zz}\right)^2 + f^{zz}_z, \\
\Gamma^z_{zz} &=& - C_{zzz}\, S^{zz} + \tilde{f}^z_{zz}.
\label{eq:christoffelpropagator}
\eea
\noindent
Here, $f^{zz}_z (z)$ and $\tilde{f}^z_{zz}(z)$ are two holomorphic functions that appear due to the definition of the propagator in \eqref{eq:propagatordef} being given in terms of its antiholomorphic derivative; we shall fix them below. Before that, it can be shown that the derivative of the genus-one free energy is linear in the propagator
\be
\partial_z F^{(0)}_1 = \frac{1}{2} C_{zzz}\, S^{zz} + \alpha_z,
\label{eq:derivativeF1propagator}
\ee
\noindent
because its holomorphic anomaly equation \eqref{eq:F1HAE} involves ${\bar{C}_{\bar{z}}}^{\ zz}$. In equation \eqref{eq:derivativeF1propagator}, $\alpha_z = -\frac{1}{2} \tilde{f}^z_{zz}+\partial_z \log f_1$ (recall that $f_1(z)$ was determined in the previous subsection). This equation will be important to compute the holomorphic limit of the propagator, which in fact is \textit{not} zero.

In order to fix  $f^{zz}_z (z)$ and $\tilde{f}^z_{zz}(z)$ a conventional choice must be made, and following \cite{hkr08} we will impose $\alpha_z = 0$. From this, it immediately follows that
\be
\label{eq:ftilde}
\tilde{f}^z_{zz}(z) = - \frac{7+216z}{6z \left(1+27z\right)}.
\ee
\noindent
Finally, to calculate $f^{zz}_z (z)$, we simply take the holomorphic limit of \eqref{eq:covariantderivativepropagator} and use that $\left(\Gamma^z_{zz}\right)_{\textrm{hol}} = \left( \frac{\partial T}{\partial z} \right)^{-1} \frac{\partial^2 T}{\partial z^2}$. This results in
\be
f^{zz}_z(z) = - \frac{z}{12 \left(1+27z\right)}.
\ee
\noindent
Via \eqref{eq:derivativeF1propagator} the propagator is now fully determined (and we shall be more explicit below).

Having fixed the ambiguities associated with the propagator, one can start integrating the holomorphic anomaly equations. The first equation, for $g=2$, is now
\be
\partial_{S^{zz}} F^{(0)}_2 = \frac{1}{2} \left( D_z^2 F^{(0)}_1 + \left(\partial_z F^{(0)}_1 \right)^2 \right).
\ee
\noindent
Using \eqref{eq:covariantderivativepropagator}, \eqref{eq:christoffelpropagator} and \eqref{eq:derivativeF1propagator}, and integrating with respect to $S^{zz}$, it is very simple to obtain
\be
\label{eq:pertF2unfixed}
F^{(0)}_2 = C_{zzz}^2 \left( \frac{5}{24} \left(S^{zz}\right)^3 - \frac{3z^2}{16}\left(S^{zz}\right)^2 + \frac{z^4}{16}\, S^{zz} \right) + f^{(0)}_2(z),
\ee
\noindent
where $f^{(0)}_2(z)$ is the genus-two perturbative holomorphic ambiguity. To fix it, one first notices that the holomorphic limit of the free energy \eqref{eq:pertF2unfixed} diverges at the conifold point, since the Yukawa coupling does so (recall \eqref{eq:Czzz}). This is actually perfectly compatible with the universal behavior of the holomorphic limit of the free energies near this special point (where $t_\text{c} = 0$), which is described by the gap condition \cite{bcov93, gv95, agnt95, gv98a, gv98c, hk06, hkq06}
\be
\mathcal{F}^{\text{[c]}(0)}_g = \frac{\mathfrak{c}^{g-1}\, B_{2g}}{2g \left(2g-2\right) t_\text{c}^{2g-2}}+\mathcal{O}(t_\text{c}^0), \qquad g\geq 2.
\label{eq:gapcondition}
\ee
\noindent
Here $B_{2g}$ are the Bernoulli numbers, and, for the case of local $\mathbb{CP}^2$, $\mathfrak{c}=3$. The requirement that there is only \textit{one} singular term followed by regular terms (the ``gap''), restricts the form of the ambiguity almost completely. Using the \textit{ansatz}
\be
f^{(0)}_g = \frac{p_{g,0}(z)}{\Delta^{2g-2}},
\ee
\noindent
and imposing that $\mathcal{F}^{(0)}_g$ is regular at every other point in moduli space, we obtain that $p_{g,0}(z)$ has to be a polynomial of degree $2g-2$. The gap condition will thus fix $2g-2$ out of the $2g-1$ coefficients in $p_{g,0}$. The last one is calculated from the behavior of the free energy near the large-radius point, $z = 0$, \cite{bcov93, mm98, gv98a, fp98, gv98c},
\be
\mathcal{F}^{\text{[LR]}(0)}_g = \frac{(-1)^{g-1}\, \chi\, B_{2g-2}B_{2g}}{4g \left(2g-2\right) \left(2g-2\right)!} + \mathcal{O}(z), \qquad g\geq 2,
\label{eq:aroundlargeradiusconstantmap}
\ee
\noindent
where the Euler characteristic is $\chi = 3$ for our example. This first term in \eqref{eq:aroundlargeradiusconstantmap} is the so-called constant map contribution. It is important to note that, in order to calculate the value of the holomorphic free energy near either conifold or large-radius points, one still has to use the appropriate expression for the holomorphic limit of the propagator, obtained from the corresponding expression for $\mathcal{F}^{(0)}_1$. Recall that the holomorphic limit requires the choice of a frame, which we have been denoting with the symbol $[\text{f}]$. Then, from \eqref{eq:derivativeF1propagator},
\be
\label{eq:generalholoS}
S^{zz}_{\text{[f]},\hol} = \frac{2}{C_{zzz}}\, \p_z \CF^{\text{[f]}(0)}_1,
\ee
\noindent
alongside \eqref{eq:holoF1} and its analogue in the conifold frame, we can calculate
\bea
S^{zz}_{\textrm{[LR]},\hol} &=& \frac{2}{C_{zzz}} \left( \frac{1}{12z\left(1+27z\right)} - \frac{{}_2F_1 \left. \left( \frac{2}{3}, \frac{4}{3}, 1\, \right| -27z \right)}{6z\, {}_2F_1 \left. \left( \frac{1}{3}, \frac{2}{3}, 1\, \right| -27z \right)} \right), \label{eq:LRholoS} \\
S^{zz}_{\textrm{[c]},\hol} &=& \frac{z^2}{2} \left( -1-54z+2 \frac{\pi\, P_{2/3}(1+54z) + 2\sqrt{3}\, Q_{2/3}(1+54z) }{\pi\, P_{-1/3}(1+54z) + 2\sqrt{3}\, Q_{-1/3}(1+54z)} \right),
\eea
\noindent
where $P_\nu(x)$ and $Q_\nu(x)$ are Legendre functions (see, \textit{e.g.}, \cite{olbc10}). An analogous calculation can be carried through for the orbifold frame,
\be
S^{zz}_{\textrm{[orb]},\hol} = \frac{z}{54} \left( -27z + \left(1+27z\right) \frac{{}_2F_1 \left. \left( \frac{4}{3}, \frac{4}{3}, \frac{5}{3}\, \right| -\frac{1}{27z} \right)}{{}_2F_1 \left. \left( \frac{1}{3}, \frac{1}{3}, \frac{2}{3}\, \right| -\frac{1}{27z} \right)} \right).
\ee
\noindent
In this way, we can finally compute
\be
f^{(0)}_2(z) = C_{zzz}^2\, z^6\, \frac{729z^2 +162 z -11}{1920}.
\ee

We thus (very explicitly) conclude that $F^{(0)}_2$ is a polynomial of degree three in the propagator, with rational functions of $z$ as coefficients. In doing the next computation for $F^{(0)}_3$ one finds a degree six polynomial, and, in general, one has that $F^{(0)}_g$ is a polynomial of degree $3g-3$ in the propagators \cite{al07}. Finally, the $S^{zz}$-independent term is always given by the holomorphic ambiguity, which still has to be computed at each genus as we reviewed above.

%%%%%%%%%%%%%%%%%%%%%%%%%%%%%%%%%%%%%%%%%%%%%%%%%%%%%%%%%%%%%%%%%
%%%%%%%%%%%%%%%%%%%%%%%%%%%%%%%%%%%%%%%%%%%%%%%%%%%%%%%%%%%%%%%%%
\section{Large-Order Analysis of the Perturbative Expansion}
\label{sec:largeorderanalysisoftheperturbativeexpansion}
%%%%%%%%%%%%%%%%%%%%%%%%%%%%%%%%%%%%%%%%%%%%%%%%%%%%%%%%%%%%%%%%%
%%%%%%%%%%%%%%%%%%%%%%%%%%%%%%%%%%%%%%%%%%%%%%%%%%%%%%%%%%%%%%%%%

Having understood how to produce high genus data, we may now start unveiling the resurgent structure of the topological string free-energy in local $\BC\BP^2$. The first thing we shall check is that the perturbative sector is indeed asymptotic and that at leading order the free energies grow factorially fast as $\left( 2g \right)!$, where $g$ is the genus. This fact immediately demands for the need of nonperturbative sectors in the free energy, \textit{e.g.}, \cite{s90}. The quantity that controls the growth at next-to-leading order is the dominant instanton action. We shall see that, depending on the value of the modulus $\psi$, one of three possible instanton actions becomes the dominant one. One of them is actually expected and universal: it is the instanton action arising from the constant-map contribution. Another instanton action has its origin at the conifold point. However, at this point the nonperturbative structure actually becomes a bit subtle. It turns out that, beyond this first, dominant, conifold instanton action, there are two other instanton actions associated with the two conifold points in the $\psi$-plane which are not $\psi = 1$, namely, the points $\psi = \rme^{+2\pi\rmi/3}$ and $\psi = \rme^{-2\pi\rmi/3}$ (recall the discussion in subsection \ref{subsection:periods}). Because of this phenomenon, it will be necessary to change coordinate from $z$ to $\psi$, in order to properly describe the \textit{three} contributing conifold instanton actions and their interrelations---and already in the present section we shall obtain clear evidence that those actions exist and are all relevant. Finally, once having factored out the constant-map contribution, a third instanton action needs to be considered which takes dominance near the large radius point. In the computations that follow the propagator is present but, as shown in \cite{cesv13}, the instanton actions are holomorphic, so the role of the propagator in this section turns out to be somewhat irrelevant. 

%%%%%%%%%%%%%%%%%%%%%%%%%%%%%%%%%%%%%%%%%%%%%%%%%%%%%%%%%%%%%%%%%
\subsection{The Dominant Instanton Action}
%%%%%%%%%%%%%%%%%%%%%%%%%%%%%%%%%%%%%%%%%%%%%%%%%%%%%%%%%%%%%%%%%

\begin{figure}[ht!]
\begin{center}
\includegraphics[scale=0.7]{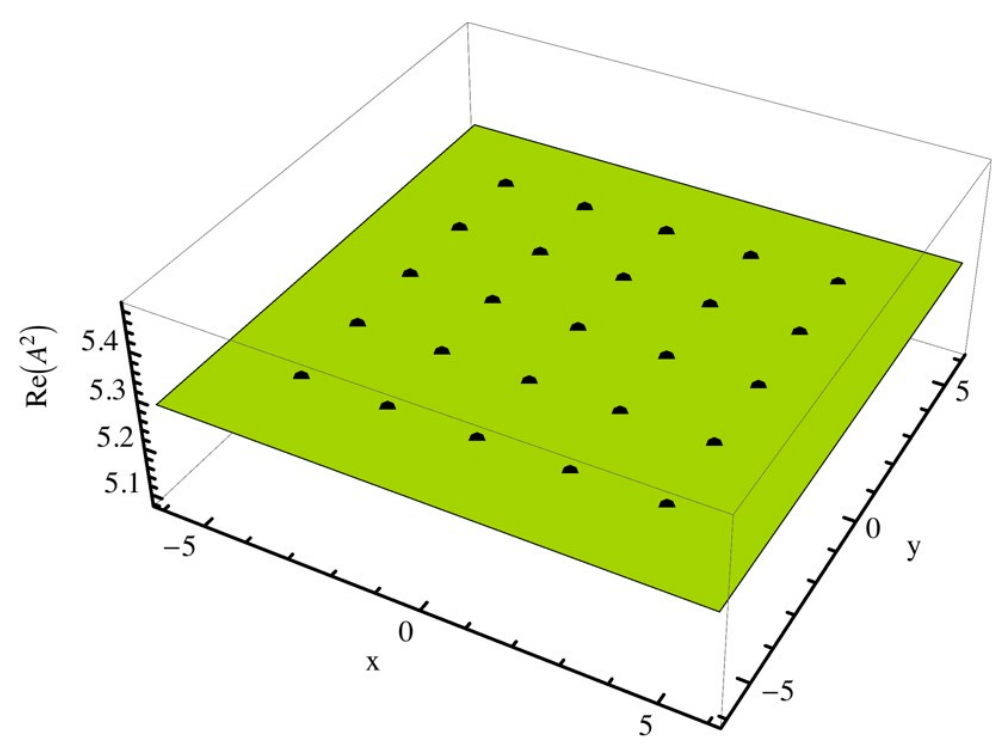} \hspace{0.5cm}
\includegraphics[scale=0.7]{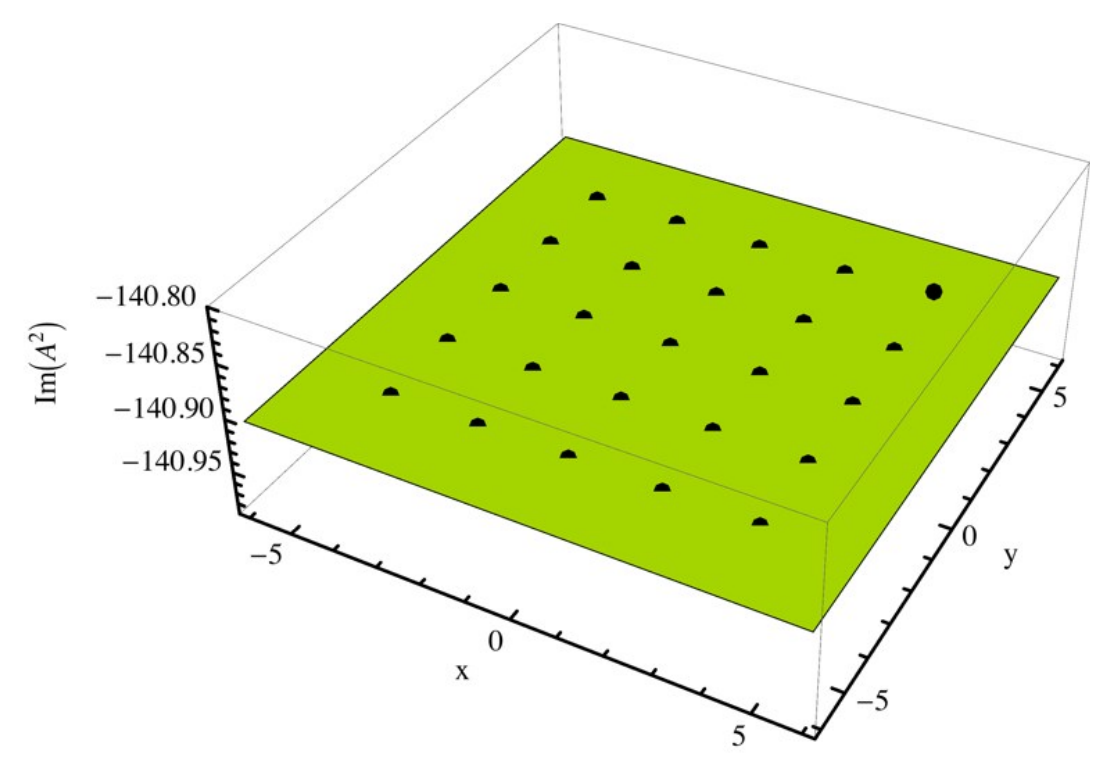}\\
\vspace{0.5cm}
\includegraphics[width=\textwidth]{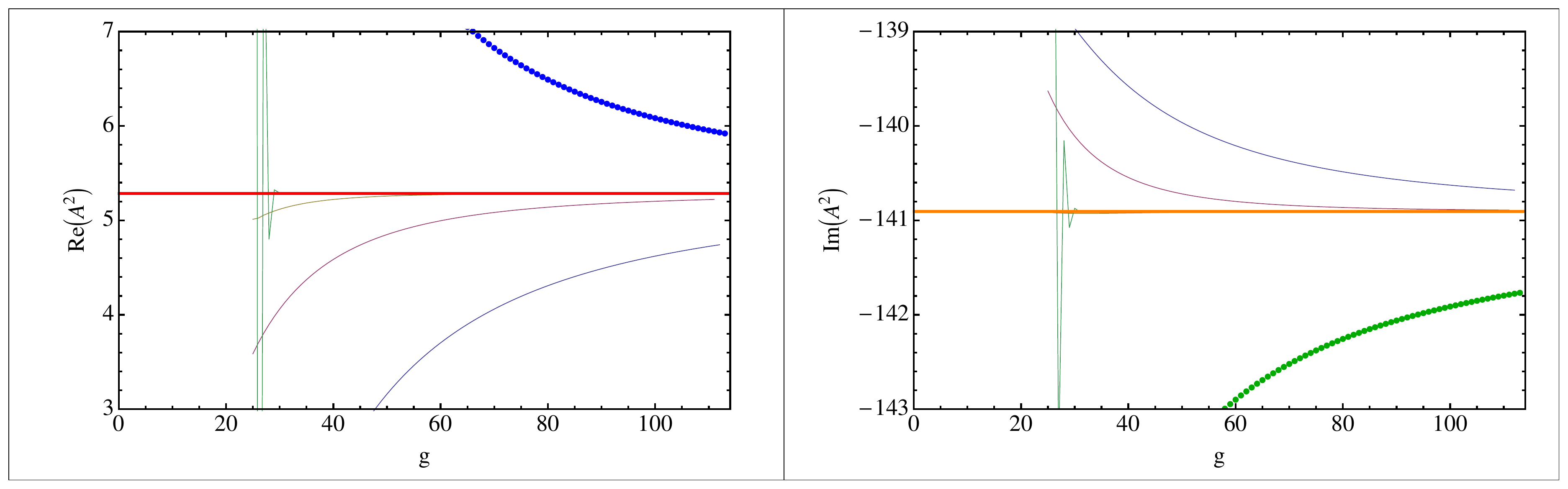} 
{\scriptsize
\bea
A^2 &=& 5.285\,620\,821\,373 - 140.904\,851\,596\,926 \, \rmi  \nonumber\\
\text{14 Richardson Transforms} &=& 5.285\,620\,821\,381 - 140.904\,851\,596\,903 \, \rmi. \nonumber
\eea
}
\end{center}
\captionsetup{singlelinecheck=off}
\caption[.]{The two images on top show the holomorphicity of the instanton action at fixed $\psi=2\,\rme^{\rmi\pi/6}$, and with varying $S^{zz} = S^{zz}_\hol \cdot \left( 1 + x - \rmi\, y \right)$. We display both real and imaginary components of $A^2$ and in both cases all (numerical) points intersect the (theoretical) constant-height surface of $A^2$. In the two images below, we show the real and imaginary parts of \eqref{eq:Asquareratio} for the particular value of $(x,y) = (4,-4)$, along with several Richardson transforms. We can compare the numerical results after fourteen Richardson transforms, and find that they agree with the predicted value of the instanton action achieving a precision as high as $\sim 10^{-10}$. If $S^{zz}$ is too large the convergence towards $A^2$ starts only at higher values of the genus, and the precision is consequently lower. The ``theoretical'' value of the instanton action is given by \eqref{eq:Atc} and marked with a horizontal line.}
\label{fig:AisholomorphicRT}
\end{figure}

The perturbative free energies obtained in the previous section (and of which we have computed over one hundred using a \textit{Mathematica} code, to generate sufficiently precise large-order data), grow factorially fast with the genus,
\be
F^{(0)}_g (z,S^{zz}) \sim \frac{\G(2g-1)}{A_\textrm{dom}(z)^{2g-1}} + \cdots.
\label{eq:sketchylargeoder}
\ee
\noindent
This expression is a bit schematic: there is also a function, of both $z$ and $S^{zz}$ but independent of $g$, multiplying the right-hand-side (the one-loop one-instanton contribution), and also additive subleading terms (higher loops and higher instanton numbers), but we shall worry about all those later. For the moment let us focus on leading and next-to-leading contributions to large order. In this case, in the above expression, $A_\textrm{dom}(z)$ is a holomorphic function given by the smallest of the instanton actions (in absolute value) at the specific point in moduli space. If there is factorial growth, we should be able to calculate the instanton action numerically using the limit
\be
A_\textrm{dom}^2 = \lim_{g\to\infty} 4g^2\, \frac{F^{(0)}_g}{F^{(0)}_{g+1}}. \label{eq:Asquareratio}
\ee
\noindent
Of course if the growth were milder than factorial this limit would go to infinity, and this would be clearly seen in the numerics. The fact that the instanton action is holomorphic was shown to be a consequence of the holomorphic anomaly equations in \cite{cesv13}. But we can now explicitly show that this is the case for local $\BC\BP^2$ by selecting some value of $\psi$ in moduli space and then varying the propagator. Our results are displayed in figure \ref{fig:AisholomorphicRT}, fully validating  our theoretical expectations (let us note that the high accuracy of the numerical limit is obtained by making use of accelerated convergence techniques such as Richardson transforms; see, \textit{e.g.}, \cite{bo78, msw07}).

%%%%%%%%%%%%%%%%%%%%%%%%%%%%%%%%%%%%%%%%%%%%%%%%%%%%%%%%%%%%%%%%%
%%%%%%%%%%%%%%%%%%%%%%%%%%%%%%%%%%%%%%%%%%%%%%%%%%%%%%%%%%%%%%%%%

\begin{figure}[t!]
\begin{center}
\begin{subfigure}[b]{0.48\textwidth}
\includegraphics[width=\textwidth]{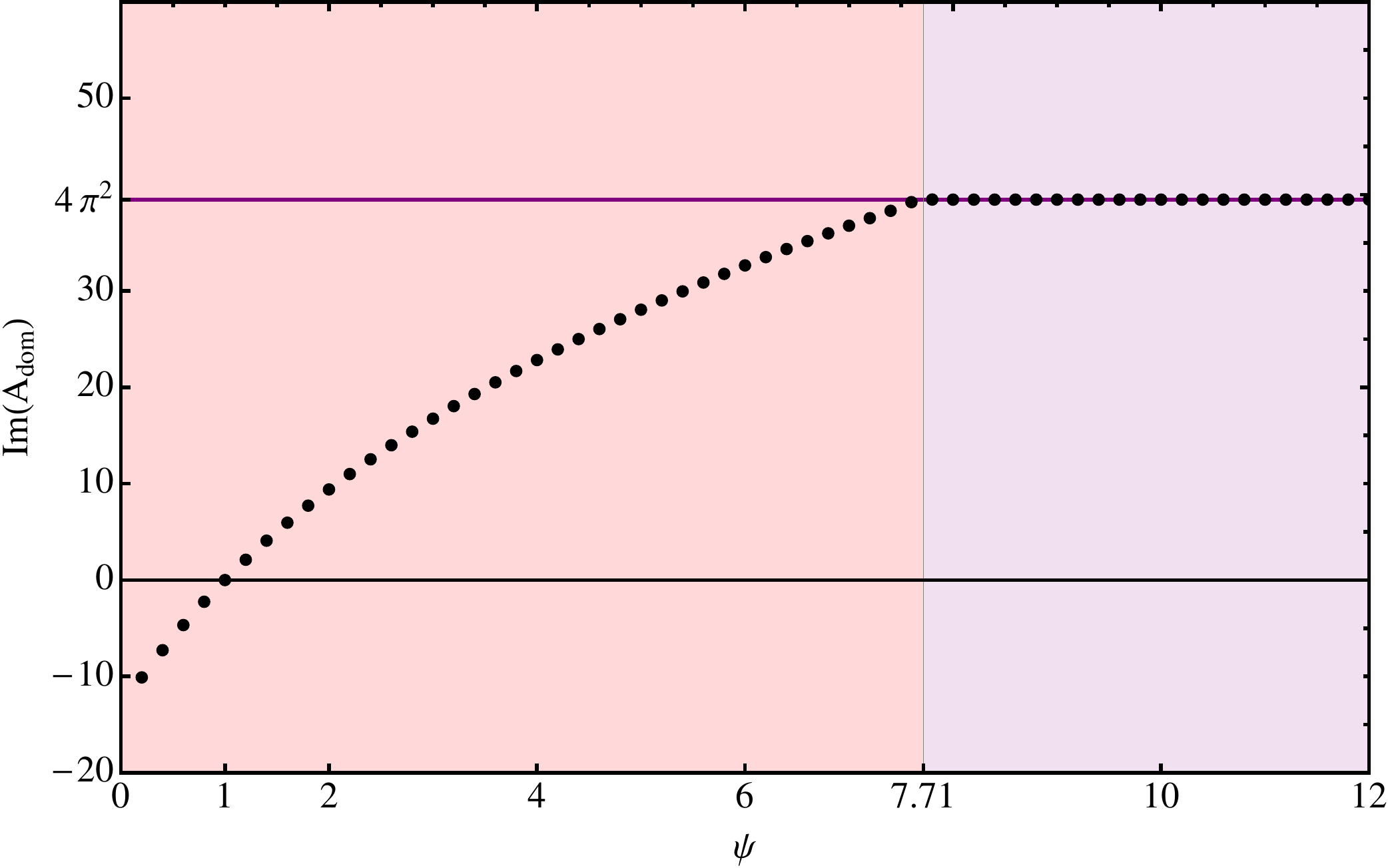}
\caption{With constant map.}
\label{fig:withCMdominantinstantonaction}
\end{subfigure}
~
\begin{subfigure}[b]{0.48\textwidth}
\includegraphics[width=\textwidth]{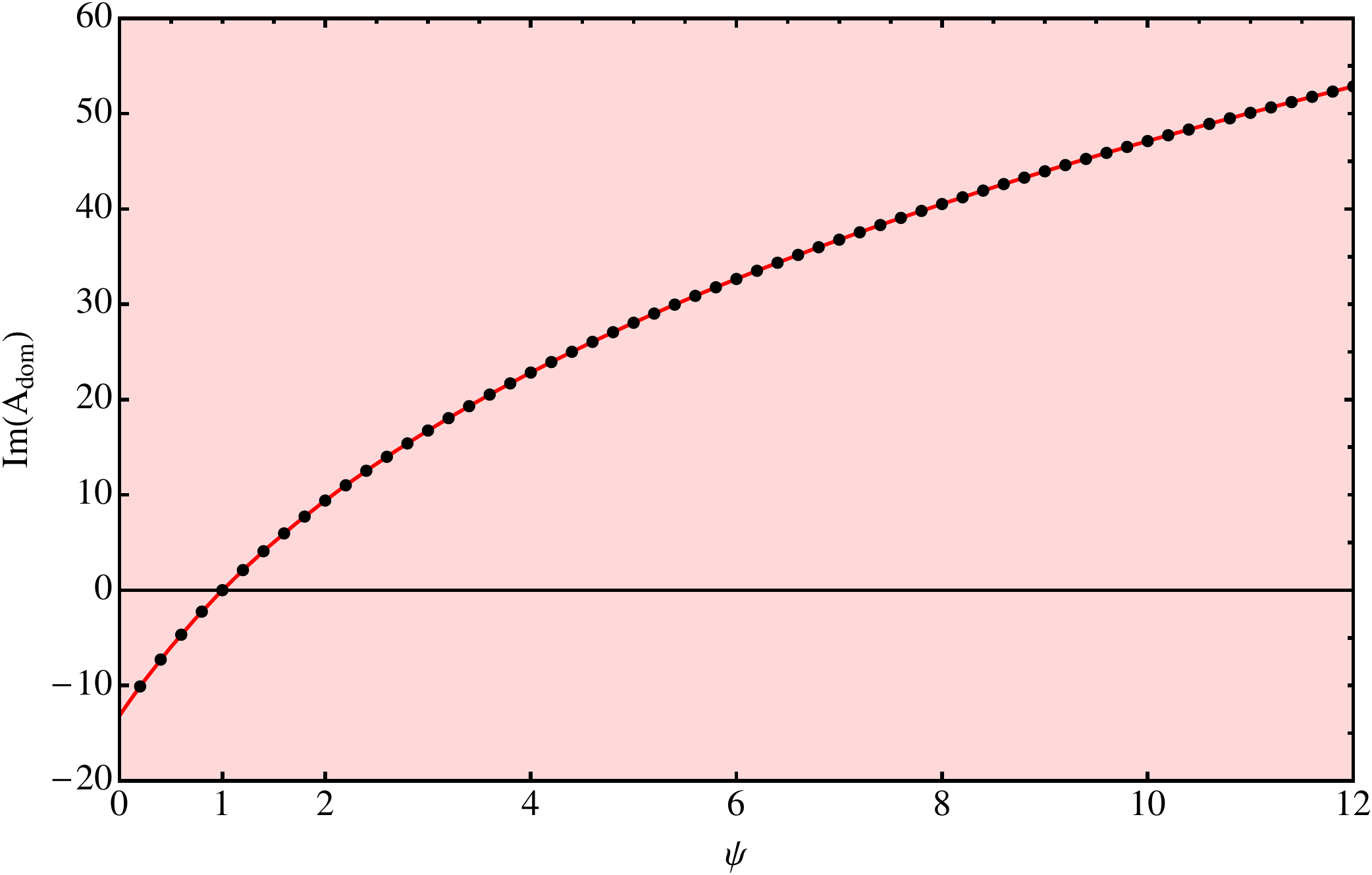}
\caption{Without constant map.}
\label{fig:withoutCMdominantinstantonaction}
\end{subfigure}
\end{center}
\vspace{-1\baselineskip}
\caption{The dominant instanton action as obtained from large-order \eqref{eq:Asquareratio}, with (left) and without (right) the constant map contribution \eqref{eq:removeCM}. In the plots we have chosen $\psi$ real and positive (or $z$ real and negative), so that the conifold point $\psi = 1$ is clearly shown (note that the instanton action vanishes at this point). Also, in this case of $\arg(\psi)=0$, the instanton action is purely imaginary. The continuous red line that fits the data is given by \eqref{eq:Atc}. Four Richardson transforms on the numerical data were enough to give an agreement of about one part in $10^9$. Precision is lower close to the transition point $\psi = 7.71$, on the left plot, because the two instanton actions become of the same order at this point.}
\label{fig:dominantinstantonaction}
\end{figure}

Having shown the holomorphicity of $A_{\textrm{dom}}$ we can start looking at its dependence across moduli space next. To this end we choose to look at negative values of $z$ in the real line, corresponding to positive real $\psi$ (recall \eqref{eq:psidefinition}). The results are essentially the same should we choose other lines in the complex moduli-space parametrized by $\psi$, except for $z\in\BR^+$ (we shall later see that on this line two instanton actions become dominant at the same time, and they thus combine to give an oscillatory behavior in $g$, which prevents us from being able to use Richardson transforms). Anticipating future conclusions, and in order to make the plots easier to read, it will be better to use the $\psi$ coordinate from now on---and we shall do so in the following. Since these coordinates are related by a cubic root via \eqref{eq:psidefinition}, we shall momentarily restrict to a wedge in the $\psi$-plane with $\arg(\psi) \in (-\pi/3,+\pi/3)$. We see in figure \ref{fig:withCMdominantinstantonaction} that there are two distinct instanton actions, which become dominant in different regions of moduli space. The one that dominates near the large-radius point ($\psi^{-1} = 0$) is constant and equal to $4\pi^2\rmi$. This is the instanton action that arises directly from the constant map contribution \eqref{eq:aroundlargeradiusconstantmap}, due to the factorial growth of the Bernoulli numbers, and on its own it was originally addressed in \cite{ps09}. Since this instanton action is always going to be present in any model, and it is rather well understood, we can remove it in order to focus solely on the particular features of local $\BC\BP^2$. Redefine
\be
F^{(0)}_g \longrightarrow F^{(0)}_g - \frac{(-1)^{g-1}\, 3\, B_{2g-2}B_{2g}}{4g \left(2g-2\right) \left(2g-2\right)!}.
\label{eq:removeCM}
\ee
\noindent
If we now recompute the plot yielding the dominant instanton action, we find that there is only one instanton action which dominates, at least for values of the modulus up to $\psi \sim 12$, shown in figure  \ref{fig:withoutCMdominantinstantonaction}. This instanton action is given by 
\be
A_{\textrm{dom}} (\psi) = \frac{2\pi\rmi}{\sqrt{3}}\, t_\text{c} (\psi),
\label{eq:Atc}
\ee
\noindent
where $t_\text{c}$ is given in \eqref{eq:tcpsi} and it is clearly a rather nontrivial function of the modulus $\psi$. This result was already anticipated in section 7 of \cite{cesv13}, based upon the universal behavior of the free energies at the conifold point, that is, the gap condition \eqref{eq:gapcondition}. As a simple consistency check, do note how the instanton action vanishes at the conifold point $\psi = 1$. As argued in \cite{dmp11}, we should in general expect vanishing instanton actions at points in moduli space where the free energies blow up. This may be understood from the diverging behavior of \eqref{eq:sketchylargeoder} and \eqref{eq:gapcondition} at the conifold point. Note that the Calabi--Yau geometry is pinched at that point, since the period $t_c$, which is an integral over a cycle, is zero.

%%%%%%%%%%%%%%%%%%%%%%%%%%%%%%%%%%%%%%%%%%%%%%%%%%%%%%%%%%%%%%%%%
%%%%%%%%%%%%%%%%%%%%%%%%%%%%%%%%%%%%%%%%%%%%%%%%%%%%%%%%%%%%%%%%%

%%%%%%%%%%%%%%%%%%%%%%%%%%%%%%%%%%%%%%%%%%%%%%%%%%%%%%%%%%%%%%%%%
\subsection{Resurgent Transseries and Large-Order Relations}
\label{sec:largeorderfromresurgence}
%%%%%%%%%%%%%%%%%%%%%%%%%%%%%%%%%%%%%%%%%%%%%%%%%%%%%%%%%%%%%%%%%

Before discussing what other instanton actions play a role in the resurgent structure of the free energy, it will prove useful to put forward some guidelines on how to proceed---essentially amounting to what the general theory of resurgence tells us about the large-order growth of the perturbative free energies (we have already discussed this issue a bit in \cite{cesv13}, and two of us further discussed it in great detail in \cite{asv11}, and we refer the reader to those discussions). In the most ``favorable'' situation, one starts off with a transseries describing the full nonperturbative solution to our problem, and then resurgence tells us very precisely what the large-order behavior of any chosen sector in this transseries is, in terms of data concerning its other sectors. In some sense, resurgence ``validates'' the original transseries structure. In order to write down these resurgence relations, one needs to know the action on every possible sector of the (pointed) alien derivatives, $\dot{\Delta}_\omega$, which are the fundamental differential operators of resurgence theory \cite{e81} (also see, \textit{e.g.}, \cite{asv11} for very explicit formulae). They precisely capture the singularities, $\omega$, of (the Borel transform of) an asymptotic series, which at the end of the day are responsible for its divergence. To make a long story short, these most ``favorable'' cases are those in which one explicitly knows the form of the transseries and the \textit{bridge equation}, relating alien and usual calculus,
\be
\dot{\Delta}_\omega \Phi \propto \p_\bfs\Phi
\label{eq:standardbridgeequation}
\ee
\noindent
(see \eqref{eq:Ftransseries} just below for notation). In this case, the action of the alien derivatives on a given sector may be very simply related to other (known) sectors present in the transseries. Now, the explicit computation of transseries solutions is possible when we have ``string equations'' and, in some of theses cases, also the existence of a standard bridge equation such as \eqref{eq:standardbridgeequation} is guaranteed. Let us then start by considering the situation where a transseries is known and a bridge equation exists, in order to set some notation.

Transseries are formal objects, the simplest of which are constructed perturbatively in the monomials $g_s$ and $\rme^{-\alpha/g_s}$ (with $\alpha$ to be identified with the instanton action in physical problems). If considering a multi-parameter transseries, then there are several instanton actions that we may collect into a vector $\bfa = (\alpha_1,\ldots,\alpha_q)$. Such transseries are generically of the form
\bea
\Phi(\bfs, g_s) &=& \sum_{\bfk\in\BN^q} \bfs^\bfk\, \rme^{-\bfk \cdot \bfa/g_s}\, \Phi^{(\bfk)}(g_s),
\label{eq:Ftransseries} \\
\Phi^{(\bfk)}(g_s) &\simeq& \sum_{g = 0}^{+\infty} g_s^{g+b^{(\bfk)}} \phi^{(\bfk)}_g.
\label{eq:perturbativephik}
\eea
\noindent
Here, the vector notation $(\bfk) = (k_1 | \cdots |Êk_q)$ denotes a nonperturbative sector with total instanton action $\bfk \cdot \bfa = \sum_{i=1}^q k_i \, \alpha_i$. The transseries parameters $\bfs^\bfk = \prod_{i=1}^q \sigma_i^{k_i}$ keep track of this nonperturbative sector, around which there are asymptotic series in the string coupling as shown above. The perturbative sector is obviously the one for which $\bfk = \bf0$. In a general sector $(\bfk)$, $b^{(\bfk)}$ will be the starting power of the perturbative expansion. Do note that one sometimes needs to consider more complicated transseries, which may include further (nonanalytic) monomials in $g_s$. An example is $\log g_s$, and in that case we say that (nonperturbative) logarithmic sectors are present (see, \textit{e.g.}, \cite{gikm10, asv11, sv13}). Having a transseries of the form \eqref{eq:Ftransseries}, the large-order growth of the perturbative coefficients $\phi^{(\bf0)}_g$ in \eqref{eq:perturbativephik} can then be obtained from a resurgent analysis of the bridge equation (see \cite{asv11} for a very explicit computation), as
\be
\phi^{(\bf0)}_g \simeq \sum_{i=1}^q \sum_{k=1}^{+\infty} \frac{(S_i)^k}{2\pi\rmi}\, \sum_{h=0}^{+\infty} \frac{\G\left( g+b^{(\bf0)}-b^{(k \bfep_i)}-h \right)}{\left( k \alpha_i \right)^{g+b^{(\bf0)}-b^{(k \bfep_i)}-h}}\, \phi^{(k \bfep_i)}_h.
\label{eq:largeoderperturbativephi}
\ee
\noindent
There are three sums in \eqref{eq:largeoderperturbativephi}. The first is over the different parameters or instanton actions present in the transseries. The second runs over ``pure'' instanton sectors $(k\bfep_i) :=(0 | \cdots | k | \cdots | 0)$, with $k$ in the $i$-th position. The last sum is over all the perturbative, loop coefficients of a given sector $(k\bfep_i)$. The factorial growth of the perturbative coefficients is very explicit in \eqref{eq:largeoderperturbativephi}. The instanton actions appear in the denominator multiplied by the instanton sector number, $k$, and raised to the power $g$. This means that the leading contribution to the large-order will be controlled by the instanton sector with \textit{smallest} instanton action, $k=1$ and $h=0$. Note how different sectors of the transseries $\phi^{(k\bfep_i)}_h$ will contribute at different orders to the growth of the perturbative coefficients. Finally, the coefficients $S_i$ in \eqref{eq:largeoderperturbativephi} are the Stokes constants which first appear in the expanded version of the bridge equation \eqref{eq:standardbridgeequation} as proportionality constants between alien and regular derivatives (see, \textit{e.g.}, \cite{asv11}). Equivalently, they are associated to the singularities of the Borel transform of each asymptotic series in the transseries, $\Phi^{(\bfk)}$.

The resurgence equation \eqref{eq:largeoderperturbativephi} has allowed for many high precision tests in string theoretic examples, \textit{e.g.}, \cite{msw07, m08, ps09, gikm10, asv11, sv13}. Interestingly enough, for all of these problems the perturbative sector has a topological genus expansion in the string coupling, that is,
\be
\Phi^{(\bf0)} = F^{(\bf0)} \simeq \sum_{g=0}^{+\infty} g_s^{2g-2} F^{(\bf0)}_g,
\ee
\noindent
and we would like to see this $g_s^2$ structure emerge as a \textit{consequence} of the resurgence properties of the free energy, rather than being imposed from the start. Referring back to \eqref{eq:perturbativephik}, this means that $\phi^{(\bf0)}_{\text{odd}} = 0$, $\phi^{(\bf0)}_{2g} = F^{(\bf0)}_g$, and $b^{(\bf0)} = -2$. In particular, there must be a cancellation on the right-hand-side of \eqref{eq:largeoderperturbativephi} when $g$ is odd. The simplest way this is realized is to have instanton actions appearing in pairs of symmetric signs,
\be
\label{eq:organizationoftheinstantonactions}
\bfa = (+A_1,-A_1,+A_2,-A_2,\ldots).
\ee
\noindent
Moreover, for the required cancellation to work, we must have
\bea
\left( S_{2i-1} \right)^k\, \phi^{(k\bfep_{2i-1})}_h &=& \left(-1\right)^{-b^{(\bf0)}+b^{(k\bfep_{2i})}+h} \left( S_{2i} \right)^k\, \phi^{(k\bfep_{2i})}_h, \\
b^{(\bfep_{2i-1})} &=& b^{(\bfep_{2i})},
\eea
\noindent
which is actually shown to be the case in Painlev\'e and matrix model examples \cite{asv11, sv13}. For topological string theories a general argument was provided in section 5 of \cite{cesv13} (but see also the present appendix \ref{ap:structure} for an explicit example). Once we implement this cancellation, we find
\be
F^{(\bf0)}_{g} \simeq \sum_{i=1}^q \sum_{k=1}^{+\infty} \frac{\left( S_{1,i} \right)^k}{\rmi\pi}\, \sum_{h=0}^{+\infty} \frac{\G\left( 2g-2-b^{(k \bfe_i)}-h \right)}{\left( k A_i \right)^{2g-2-b^{(k \bfe_i)}-h}}\, F^{(k \bfe_i)}_h.
\label{eq:generalperturbativelargeorder}
\ee
\noindent
Here we have introduced the notation $S_{1,i}$ for $S_{2i-1}$. Further standard notation (see \cite{asv11}) is to write $\widetilde{S}_{-1,i}$ for $S_{2i}$. We have also renamed $F^{(\bfk)}_h := \phi^{(\bfk)}_h$ and $(\bfe_i) = (\bfep_{2i-1})$. Let us further mention that, just as one can derive \eqref{eq:generalperturbativelargeorder}, it is possible to compute the large-order growth of all higher-instanton sector coefficients; they involve many other sectors of the transseries creating a whole network of nonperturbative relations \cite{asv11}.

As we outlined above, our final equation \eqref{eq:generalperturbativelargeorder} is valid when we both have a well-motivated transseries \textit{ansatz} for our solution, such as, \textit{e.g.}, \eqref{eq:Ftransseries}; and also have a bridge equation allowing us to extract resurgence relations. But this need not always be the case and, in general, we should expect generalizations of these situations, where more complicated transseries structures may emerge, or where more involved bridge equations may be at play. Of course resurgence will still bind the nonperturbative sectors in the transseries together, but one may now reasonably expect that large-order relations will differ (possibly even considerably) from \eqref{eq:generalperturbativelargeorder}. In these cases, one should reverse the logic where resurgence ``validates'' the proposed transseries structure, into having resurgence \textit{uncover} the underlying transseries structure we will be searching for. As such, the numerical large-order analyses of perturbative and higher instanton sectors of the theory will now play a rather prominent role in the complete understanding of both the resurgence properties of the transseries, and of the transseries itself. For our example of topological strings in local $\BC\BP^2$ we precisely find that we are in this general situation. Nonetheless, many features of the standard case still remain the same (the factorial growths, the instanton actions of symmetric signs, instanton sectors contributing at different orders, and so on). We have already seen that the perturbative coefficients grow like \eqref{eq:sketchylargeoder}. We shall very explicitly see in the next sections that there are one and two (and higher) instanton sectors involved in that growth. Our main point is that \textit{all} these sectors can be computed by making use of the nonperturbative extension of the holomomorphic anomaly equations we proposed in \cite{cesv13} and describe in subsection \ref{sec:thenonperturbativeHAEs}. The numerical analysis of different asymptotic growths will then be our guiding principle in understanding both the resurgence relations and the transseries sectors which should appear in the topological string free energy. This process will also include the fixing of the holomorphic ambiguities appearing after each integration.

%%%%%%%%%%%%%%%%%%%%%%%%%%%%%%%%%%%%%%%%%%%%%%%%%%%%%%%%%%%%%%%%%
\subsection{Finding Several Instanton Actions}
\label{sec:instantonactions}
%%%%%%%%%%%%%%%%%%%%%%%%%%%%%%%%%%%%%%%%%%%%%%%%%%%%%%%%%%%%%%%%%

\begin{figure}[ht!]
\begin{center}
\includegraphics[scale=0.5]{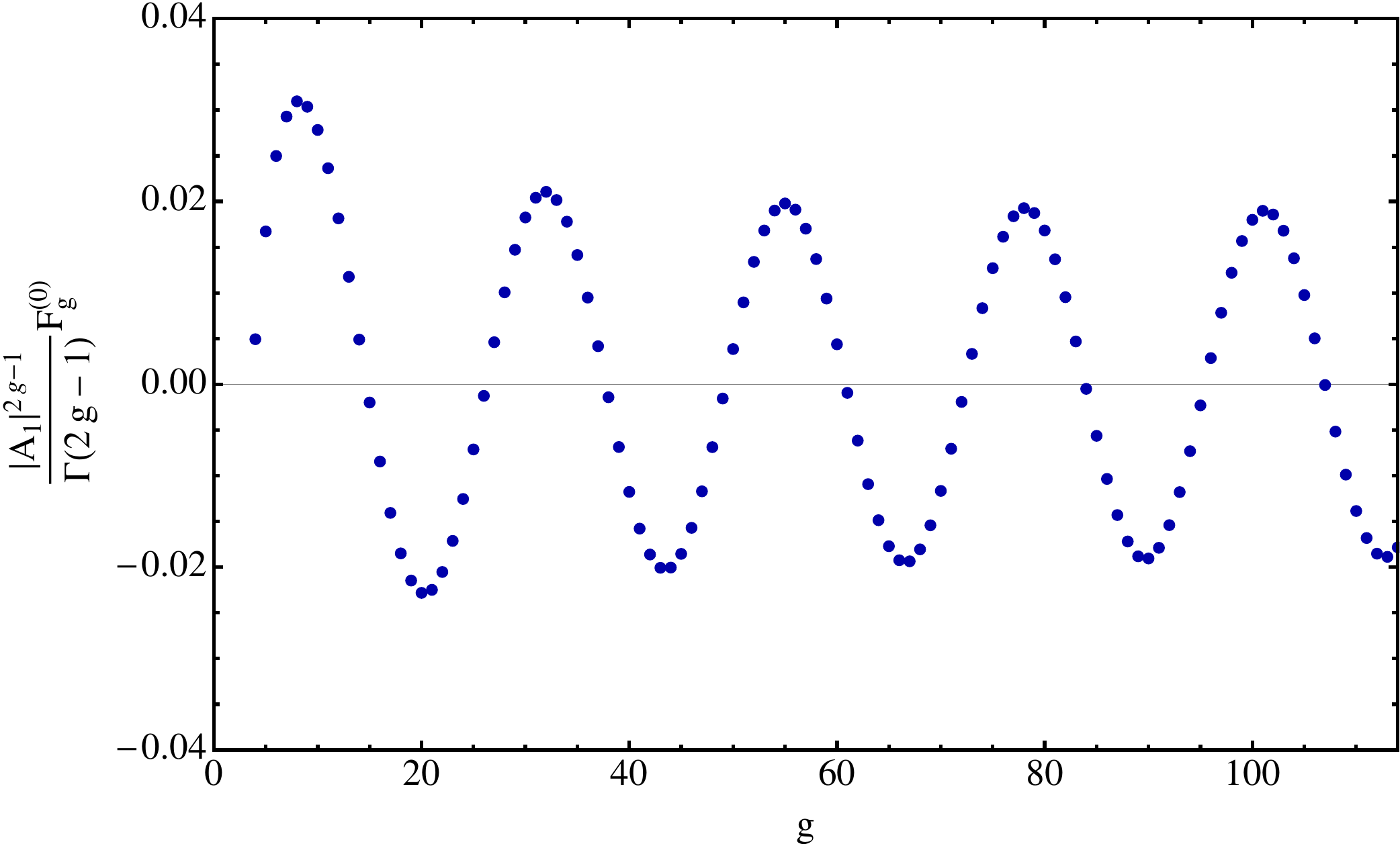}
\end{center}
\vspace{-1\baselineskip}
\caption{Oscillatory behavior of the perturbative sector, due to complex contributions of both $A_1$ and $A_2$, for values of the modulus $\psi = 1.25\, \rme^{\rmi\pi/3}$ and  propagator $S^{zz} = 10^{-5} \sim 0.15\,|S^{zz}_{1,\hol}|$.}
\label{fig:oscillatoryplotonlydata}
\end{figure}

Having understood some basic properties concerning multi-parameter resurgent transseries, let us return to the study of the instanton actions present in local $\BC\BP^2$. At this stage we have gathered extensive numerical evidence that, removing the constant map contribution and restricting ourselves to values of the modulus $z\notin \BR^+$ and $\psi$ not too large, there is a single dominant instanton action given by \eqref{eq:Atc}. However, at points $z\in\BR^+$ in moduli space, one finds new (oscillatory) large-order behavior of the perturbative sector, as illustrated in figure \ref{fig:oscillatoryplotonlydata}. Now it is important to note that, on the one hand, we have used a real value for the propagator leading to \textit{real} perturbative free energies. On the other hand, here the instanton action \eqref{eq:Atc} is actually \textit{complex}. This immediately implies that \eqref{eq:sketchylargeoder} necessarily has to be supplemented with its complex conjugate, in order for large order to produce an overall real combination; see, \textit{e.g.}, \cite{msw07, dmp11}. Physically, what is happening here is that one of the subdominant instanton actions in \eqref{eq:generalperturbativelargeorder} actually \textit{becomes} the complex conjugate of the (originally) dominant one. In fact, if one considers points in moduli space with $\arg(\psi) = +\pi/3$, or real positive $z$ approached from above the axis, then the instanton action associated to the conifold point at $\psi = \rme^{2\pi\rmi/3}$, say $A_2$, precisely becomes the complex conjugate of the instanton action associated to the ``original'' conifold point $\psi = 1$, that is $A_1$. Since they now have the same absolute value, they are both equally leading in the large-order growth of \eqref{eq:generalperturbativelargeorder}, in fact combining to provide for an oscillatory dependence in $g$, as shown in figure \ref{fig:oscillatoryplotonlydata}. Similarly, there is one further third instanton action associated to the conifold point $\psi = \rme^{-2\pi\rmi/3}$ that influences the large-order behavior for $\arg(\psi) = -\pi/3$. Later on in this paper, once we have computed the one-instanton sector via the (nonperturbative) holomorphic anomaly equations, we shall present a very precise fit of the numerical data, matching both amplitude and frequency of the oscillations in figure \ref{fig:oscillatoryplotonlydata}.

%%%%%%%%%%%%%%%%%%%%%%%%%%%%%%%%%%%%%%%%%%%%%%%%%%%%%%%%%%%%%%%%%
%%%%%%%%%%%%%%%%%%%%%%%%%%%%%%%%%%%%%%%%%%%%%%%%%%%%%%%%%%%%%%%%%

\begin{figure}[ht!]
\begin{center}
\includegraphics[width=\textwidth]{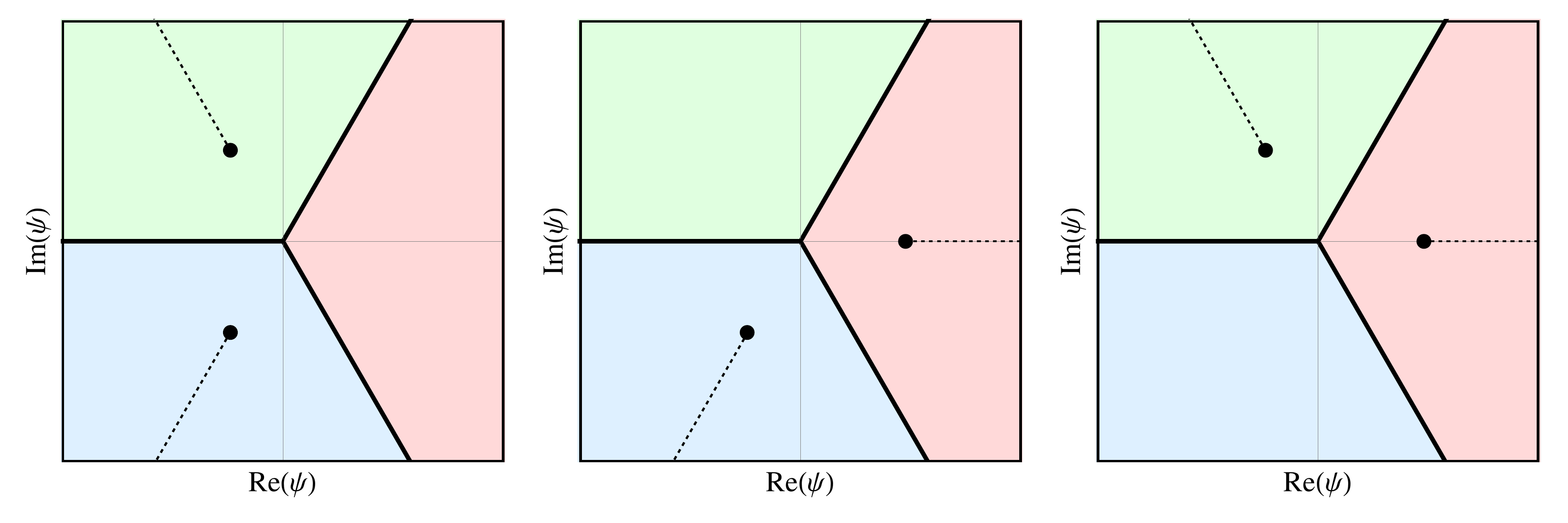}
\end{center}
\vspace{-1.5\baselineskip}
\caption{Branch points and cuts of the instanton actions $A_1$ (left), $A_2$ (center) and $A_3$ (right), in the complex $\psi$ plane. Each wedge of angle $2\pi/3$ is in correspondence with one full complex $z$ plane. The rightmost wedge, in red, includes the first conifold point, $\psi = 1$; the upper wedge, in green, includes the second conifold point, $\psi = \rme^{2\pi \rmi/3}$; and the lower wedge, in blue, includes the third conifold point, $\psi = \rme^{-2\pi \rmi/3}$.}
\label{fig:psiplane}
\end{figure}

\begin{figure}[ht!]
\begin{center}
\includegraphics[scale=0.45]{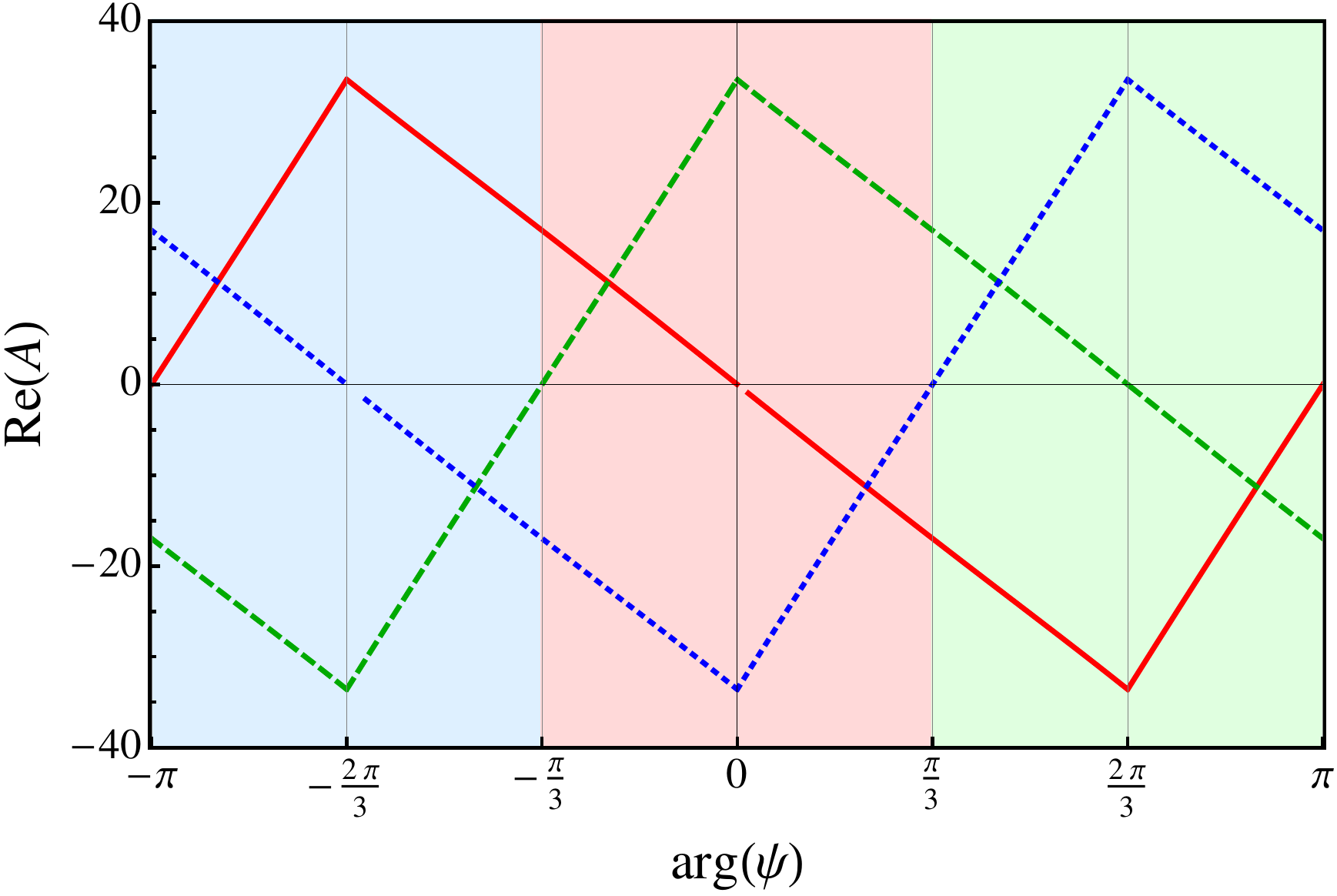} 
\includegraphics[scale=0.45]{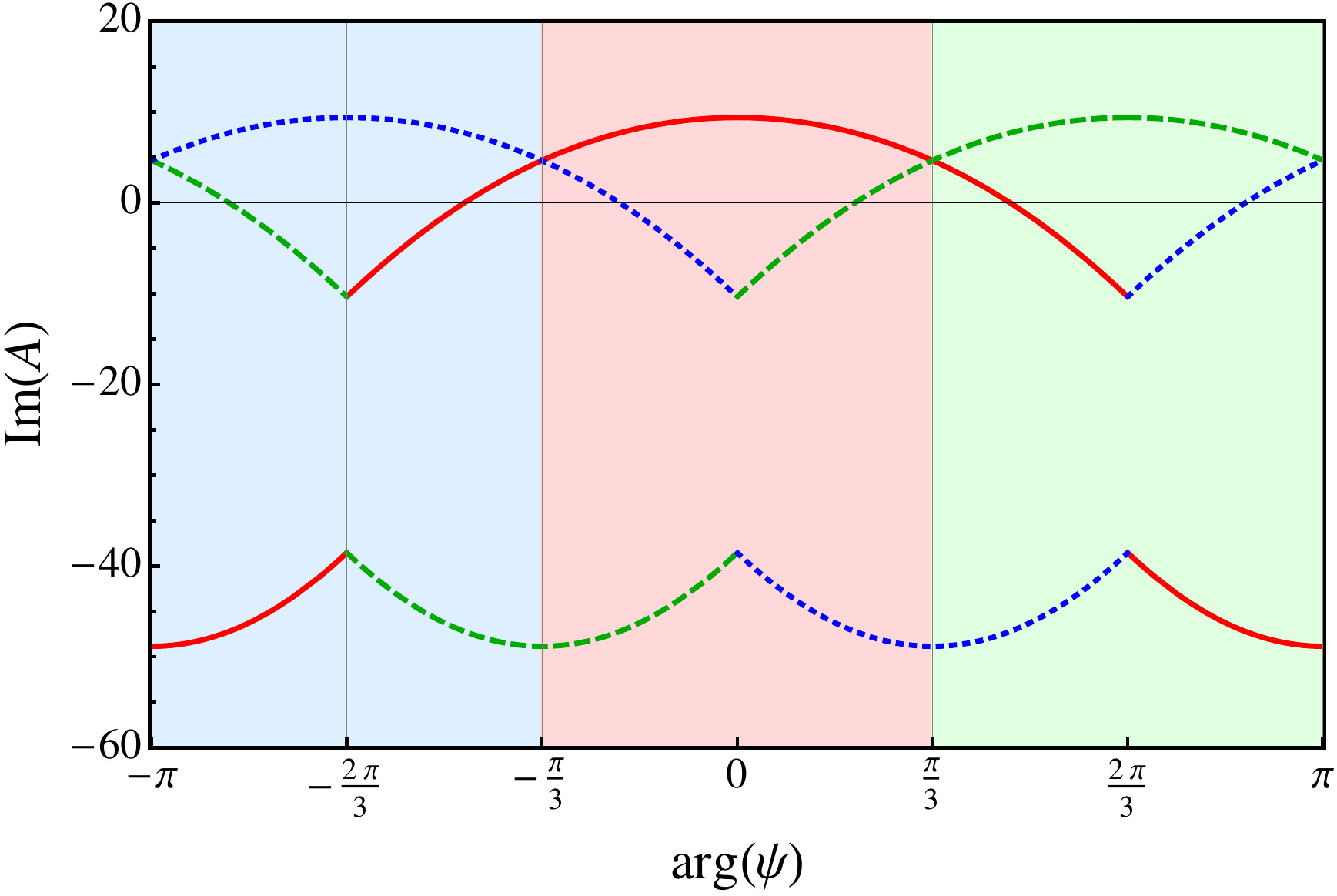} \\
\includegraphics[scale=0.45]{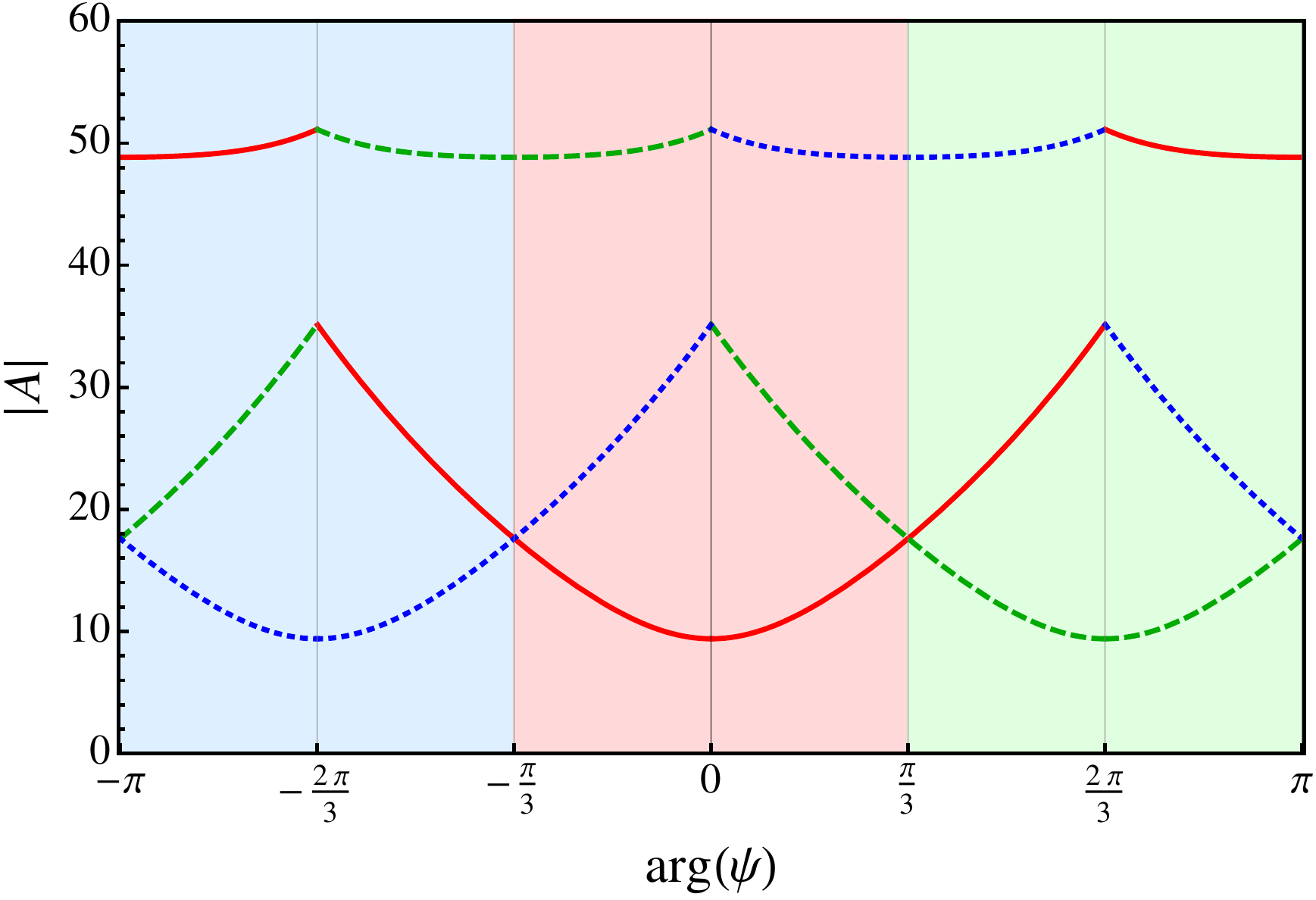}
\includegraphics[scale=0.45]{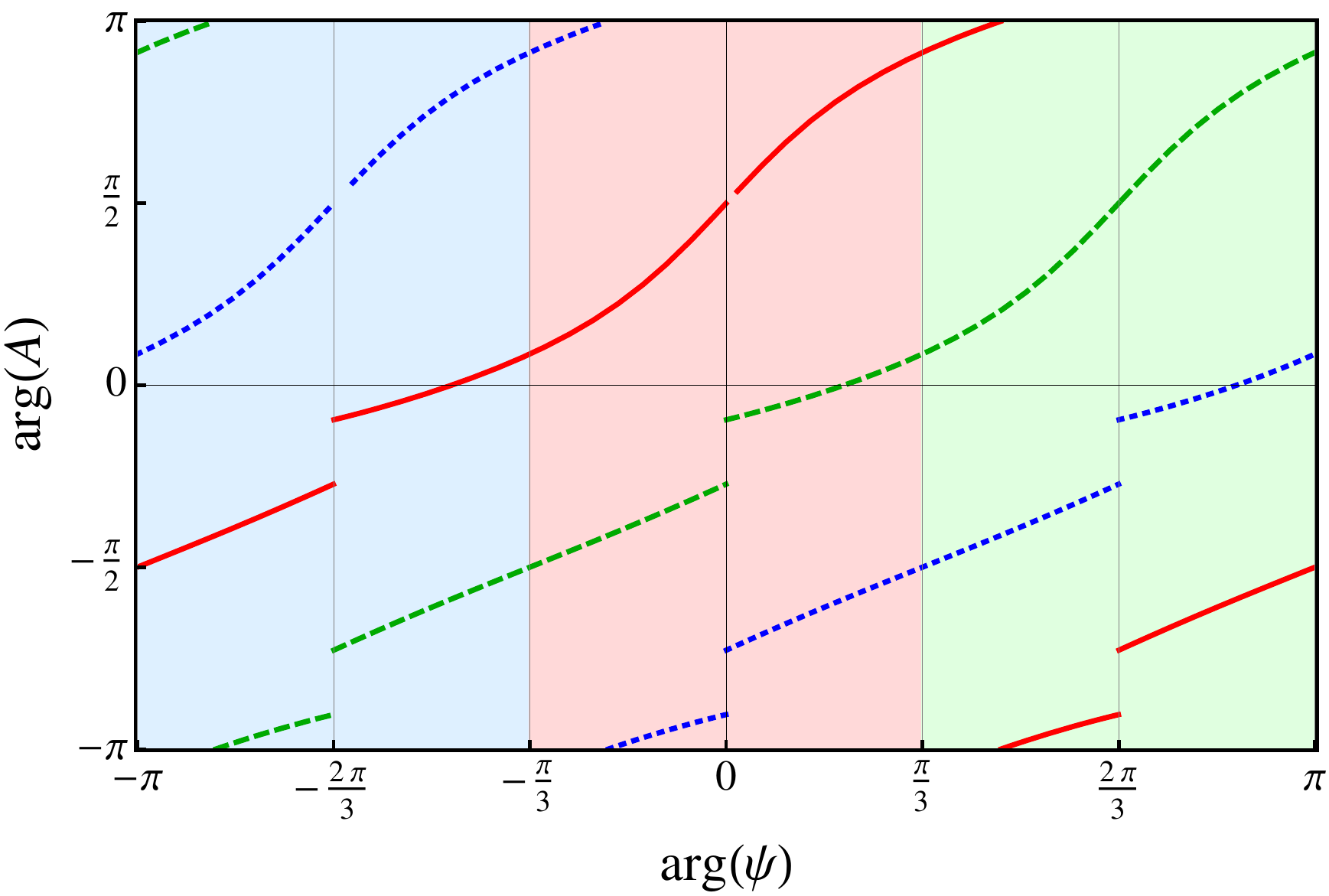}Ê\\
\includegraphics[scale=0.45]{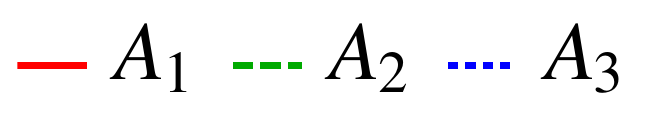}
\end{center}
\vspace{-1\baselineskip}
\caption{Real parts, imaginary parts, absolute values, and arguments of the three instanton actions for fixed absolute value $| \psi | = 2$ and varying argument $\arg (\psi)$. The colors are in precise correspondence with those in figure \ref{fig:psiplane}. Do note that a complete picture should also include the symmetric instanton actions, $-A_i$, $i=1,2,3$ (see the main text).
}
\label{fig:instantonactionsplots}
\end{figure}

\begin{figure}[ht!]
\begin{center}
\includegraphics[scale=0.39]{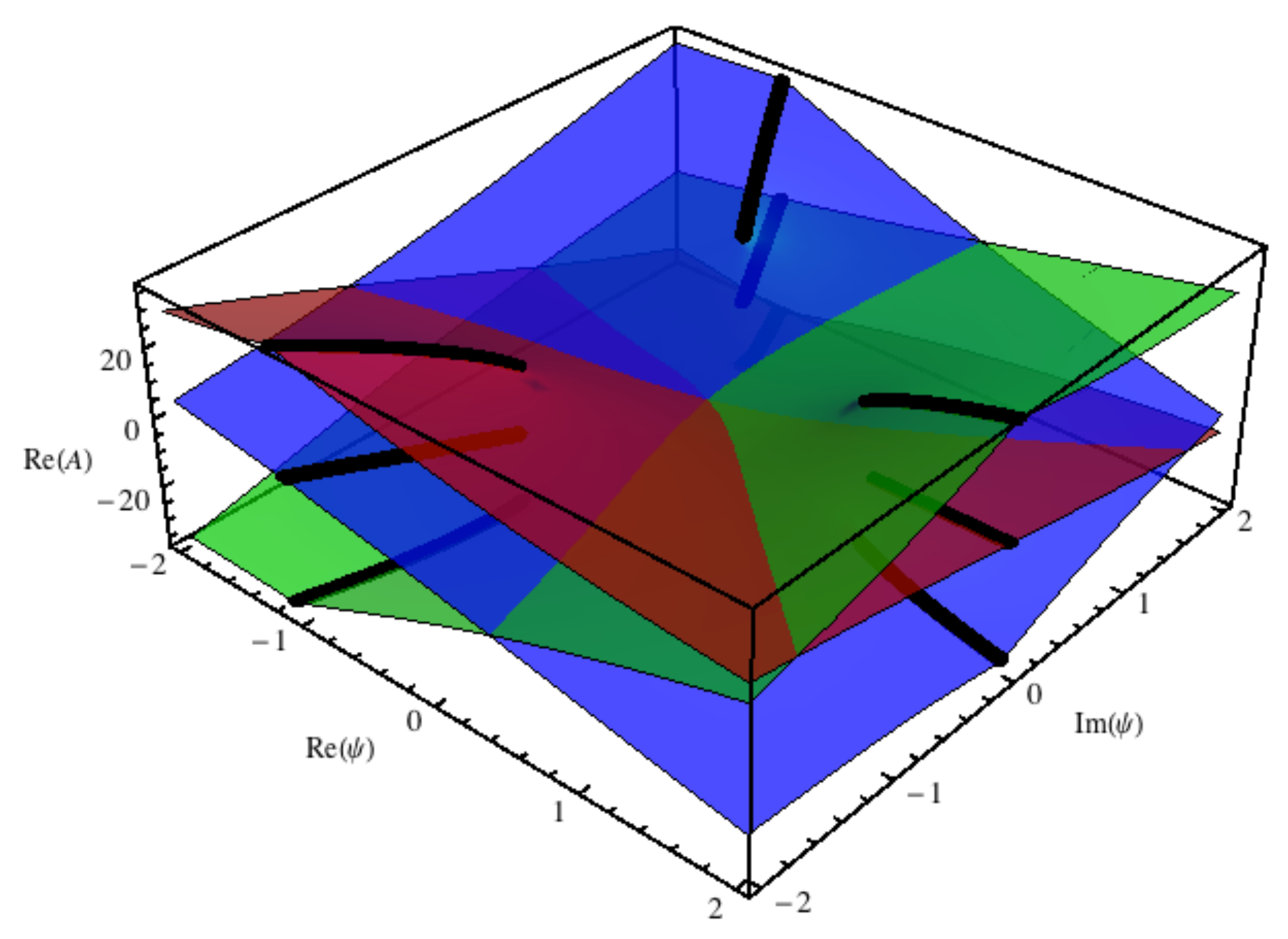}
\hspace{0.3cm}
\includegraphics[scale=0.39]{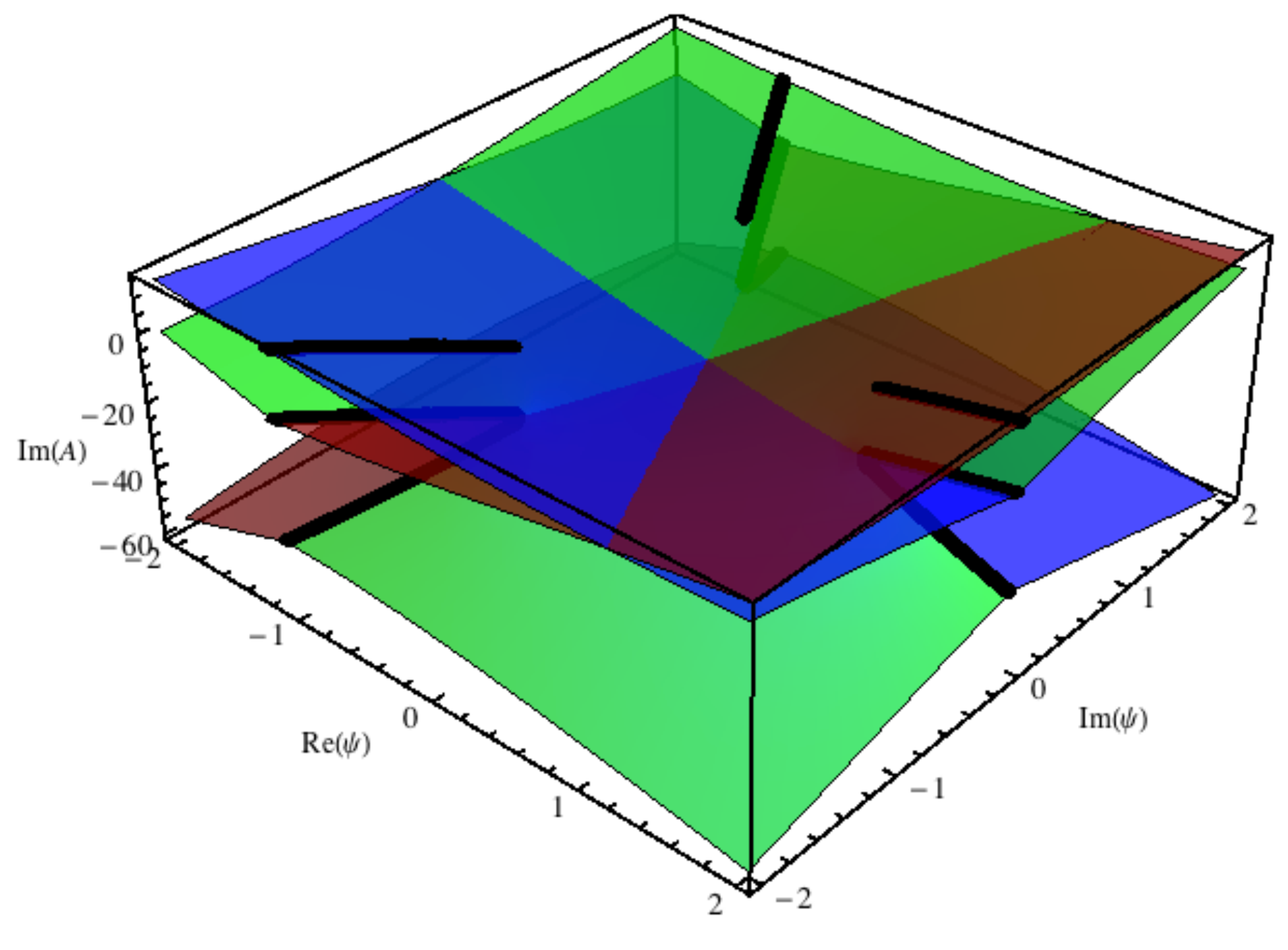}\\
\includegraphics[scale=0.39]{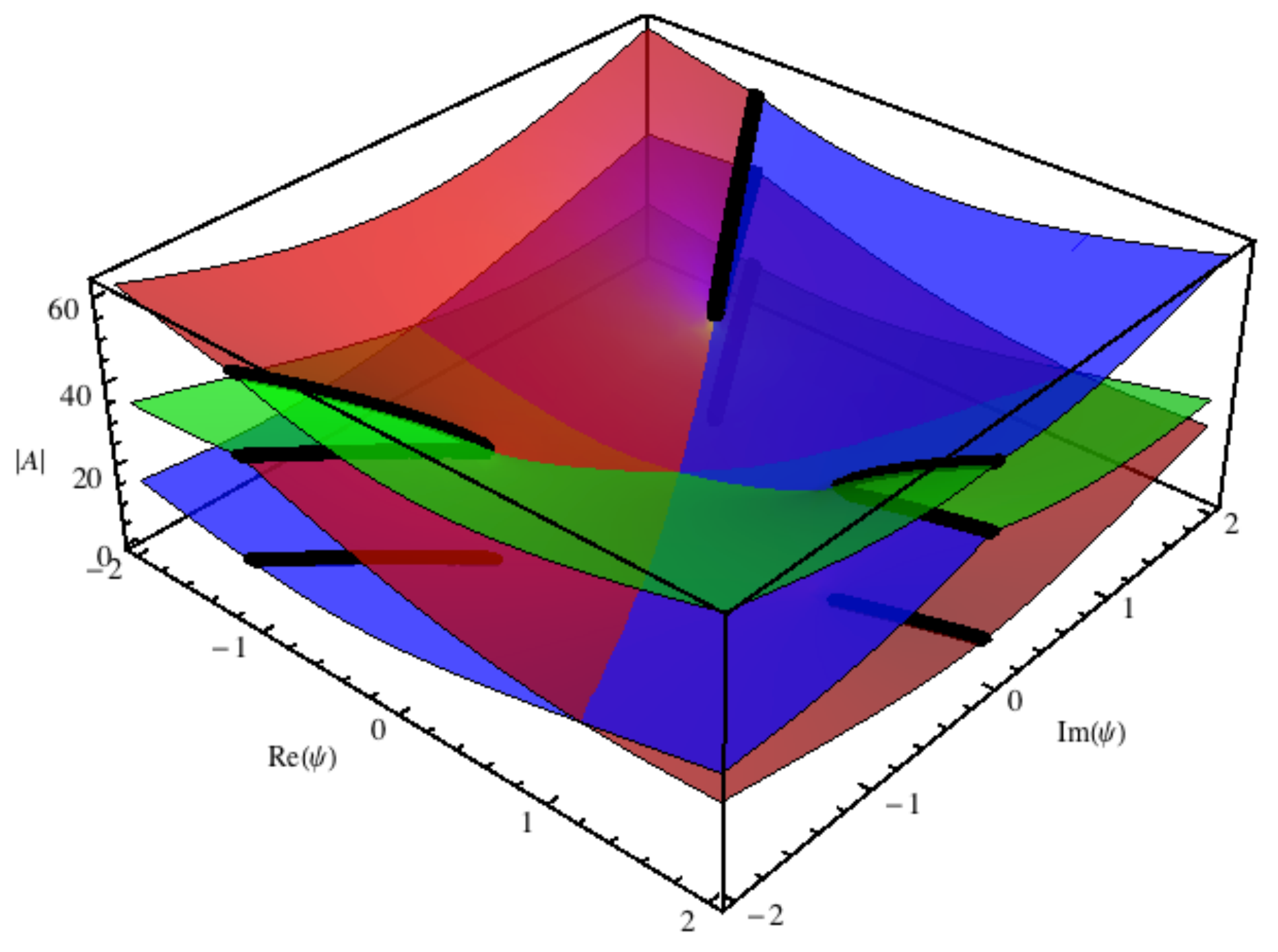} 
\end{center}
\vspace{-1\baselineskip}
\caption{The images on top show the real and imaginary parts of the three instanton actions, over the complex $\psi$ plane, while in the image below we plot their absolute values. As usual, the colors red, green and blue denote the instanton actions $A_1$, $A_2$ and $A_3$, respectively. The black lines represent the possible branch cuts.}
\label{fig:instantonactionsplots3d}
\end{figure}

Let us be more explicit with respect to these ``new'' conifold instanton actions. Since each point $\psi = 1, \rme^{+2\pi\rmi/3}, \rme^{-2\pi\rmi/3}$ is a conifold point in its own right, the instanton actions are still going to be given by \eqref{eq:Atc}, but where $t_\text{c}$ should now have a label $i\in \{1,2,3\}$, respectively. Moreover, since the various conifold points are related by a rotation of $2\pi/3$ in the $\psi$ plane, we can just conclude
\be
A_i (\psi)= \frac{2\pi\rmi}{\sqrt{3}}\, t_{\text{c},i}(\psi), \qquad i=1,2,3,
\label{eq:Aitci}
\ee
\noindent
where
\bea
t_{\text{c},1}(\psi) &=& t_\text{c} (\psi), \\
t_{\text{c},2}(\psi) &=& t_\text{c} (\rme^{-2\pi\rmi/3}\,\psi), \\
t_{\text{c},3}(\psi) &=& t_\text{c} (\rme^{+2\pi\rmi/3}\,\psi).
\eea
\noindent
The two hypergeometric functions that compose the instanton actions via \eqref{eq:tcpsi} have a branch-cut along the line $(1,+\infty)$ of their arguments. Since this argument is actually $\psi^3$, there may be three branch-cuts for the instanton actions along the rays $(1,+\infty)$, $\rme^{+2\pi\rmi/3}\, (1,+\infty)$ and $\rme^{+2\pi\rmi/3}\,  (1,+\infty)$ in the complex plane. As it turns out, $A_1$ is continuous on the first of these lines but has jumps of its imaginary part on the other two. The other instanton actions, $A_2$ and $A_3$, have the precise same structure but rotated by one third of a full turn in one or the other direction. This means that $A_2$ and $A_3$ have jumps at $\psi \in (1,+\infty)$. The structure of the complex $\psi$-plane, including the choice of branch cuts for the instanton actions, is depicted in figure \ref{fig:psiplane}, while the jumps are represented graphically in figures \ref{fig:instantonactionsplots} and \ref{fig:instantonactionsplots3d}. Note that the presence of these discontinuities might be a relevant issue to address within the context of the resummation of the transseries, but such an issue is beyond the scope of the present work. On what does concern us, the resurgence relations and their associated large-order, we have found no new phenomena related to the existence of these branch cuts.

%%%%%%%%%%%%%%%%%%%%%%%%%%%%%%%%%%%%%%%%%%%%%%%%%%%%%%%%%%%%%%%%%
%%%%%%%%%%%%%%%%%%%%%%%%%%%%%%%%%%%%%%%%%%%%%%%%%%%%%%%%%%%%%%%%%

%%%%%%%%%%%%%%%%%%%%%%%%%%%%%%%%%%%%%%%%%%%%%%%%%%%%%%%%%%%%%%%%%
%%%%%%%%%%%%%%%%%%%%%%%%%%%%%%%%%%%%%%%%%%%%%%%%%%%%%%%%%%%%%%%%%

%%%%%%%%%%%%%%%%%%%%%%%%%%%%%%%%%%%%%%%%%%%%%%%%%%%%%%%%%%%%%%%%%
%%%%%%%%%%%%%%%%%%%%%%%%%%%%%%%%%%%%%%%%%%%%%%%%%%%%%%%%%%%%%%%%%

One important aspect to retain is that the three instanton actions are not all independent, since they are all given by periods and, as discussed, the Picard--Fuchs equation only has two nontrivial independent solutions. It turns out that
\be
A_1 + A_2 + A_3 = -4\pi^2\rmi,
\label{eq:leadstoresonance}
\ee
\noindent
where $4\pi^2\rmi$ is precisely the constant-map instanton action \cite{ps09}, which becomes dominant at the large-radius point. Without going into the details that we reserve for later sections, this already tells us that the full topological string transseries will be rather intricate. One starts off with three relevant instanton actions, $A_1$, $A_2$ and $A_3$, which must be accompanied by their symmetric values in order to guarantee that the perturbative free energy has a topological genus expansion (as explained in the previous subsection, but see also \cite{gikm10, kmr10, dmp11, asv11, sv13, as13, cesv13, bdu13} for further discussions concerning these ``symmetric'' instanton actions). The fact that the transseries includes instanton actions satisfying, in pairs, dependence relations such as the (trivial) $A_i + (- A_i) = 0$ leads to the appearance of resonance in the transseries \cite{gikm10, asv11, sv13, cesv13} and, at least in the matrix model context, to the inclusion of nonperturbative logarithmic sectors. The expression above further says that the three conifold instanton actions will intertwine with the large-radius instanton action and this will most likely enlarge the number of transseries parameters and, at the same time, induce new resonant sectors we have still to fully unveil. Finally, note that the large-order of the perturbative sector just implies that we must construct the transseries in the complex $\psi$ plane, in order to accommodate for the required instanton actions $A_2$ and $A_3$. In retrospect, the fact that we find three instanton actions is related to the $\BZ_3$ symmetry that is present at the orbifold point, where the B-side geometry becomes $\BC^3/\BZ_3$.

\begin{figure}[t!]
\begin{center}
\includegraphics[scale=0.44]{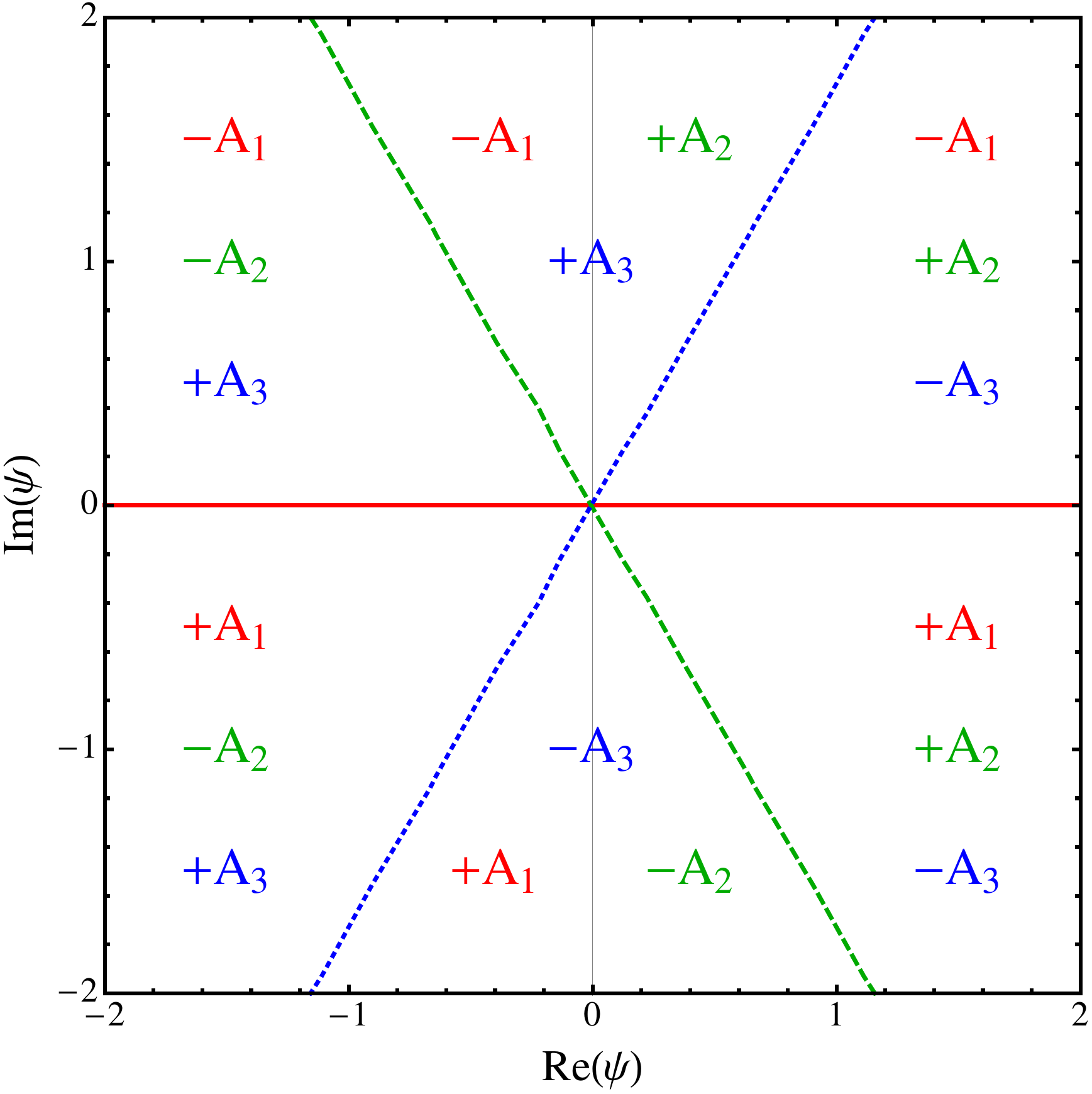}
\hspace{0.3cm}
\includegraphics[scale=0.44]{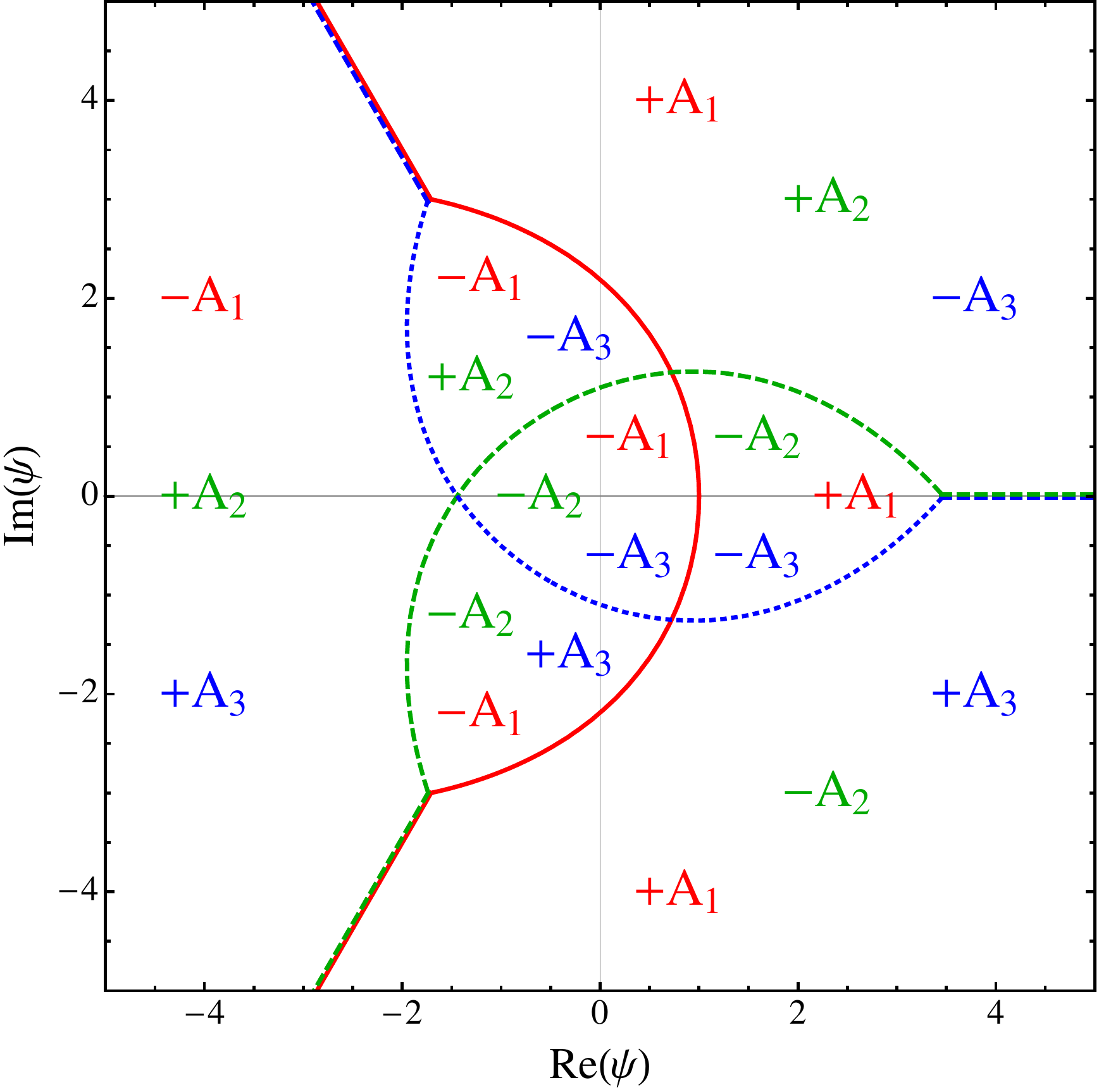}
\end{center}
\vspace{-1\baselineskip}
\caption{Phase diagrams for local $\BC\BP^2$. On the left $g_s\in\BR^+$, while on the right $g_s\in\rmi\BR^+$. The anti-Stokes phase boundaries satisfy $\re \left( A(\psi)/g_s \right) = 0$ and in the plots we mark which instanton actions satisfy $\re\left( A(\psi)/g_s \right) > 0$, in each region of the complex $\psi$ plane. Straight double lines on the right plot denote a branch-cut jump in $\re \left( A(\psi)/g_s \right)$ from positive to negative value.}
\label{fig:phasediagrams}
\end{figure}

To end the description of the instanton actions, let us briefly discuss the phase diagram of the local $\BC\BP^2$ nonperturbative free-energy. In order to have a proper transseries along some direction in the string-coupling complex-plane\footnote{Of course this is usually (small) real positive string coupling.}, \textit{i.e.}, a transseries which is amenable to resummation, it must be the case that all multi-instanton sectors are exponentially suppressed as compared to the perturbative sector along the chosen direction. This essentially translates to the fact that $\re\left( A(\psi)/g_s \right)$ must be positive\footnote{Sectors that do not satisfy this constraint will be turned off, by setting the corresponding transseries parameter $\sigma$ to zero in the appropriate region of coupling space. Of course the resurgence relations still involve all possible sectors, in spite of having some of them turned off in particular explicit forms of the transseries. In this way, Stokes lines will turn on parts of the transseries which might have not been present before, as the corresponding transseries parameters $\bfs$ jump at these lines; see, \textit{e.g.}, \cite{as13}.}, where at this stage $A$ can be any of the instanton actions appearing in the general transseries \eqref{eq:Ftransseries}. As one moves around moduli and coupling spaces, these exponential terms may change and, at some stage, may become of order one, \textit{i.e.}, they will no longer be exponentially suppressed but of the same order as the perturbative series. This happens at the anti-Stokes lines, where, due to the simultaneous contribution of all multi-instanton sectors which are no longer exponentially suppressed, a phase transition will take place. A full analysis of transseries-induced phase transitions of the topological-string free energy in local $\BC\BP^2$ is beyond the scope of this work, but we do illustrate this idea with some simple phase diagrams. In figure \ref{fig:phasediagrams}, on the left, we show the phase boundaries in the situation where $g_s\in\BR^+$; and on the right we further show the case where $g_s\in\rmi\BR^+$. In both plots we mark which sectors could be turned on, \textit{i.e.}, having $\re\left( A(\psi)/g_s \right) > 0$, in each region. The boundaries satisfy $\re\left( A(\psi)/g_s \right) = 0$ for some instanton action $A$. We leave a more detailed analysis of this phase structure for future work.

%%%%%%%%%%%%%%%%%%%%%%%%%%%%%%%%%%%%%%%%%%%%%%%%%%%%%%%%%%%%%%%%%
%%%%%%%%%%%%%%%%%%%%%%%%%%%%%%%%%%%%%%%%%%%%%%%%%%%%%%%%%%%%%%%%%

%%%%%%%%%%%%%%%%%%%%%%%%%%%%%%%%%%%%%%%%%%%%%%%%%%%%%%%%%%%%%%%%%
\subsection{Analysis at the Large-Radius Point}\label{sec:largeradiusactions}
%%%%%%%%%%%%%%%%%%%%%%%%%%%%%%%%%%%%%%%%%%%%%%%%%%%%%%%%%%%%%%%%%

As previously mentioned, in our present example of local $\BC\BP^2$ the Picard--Fuchs equation only has two nontrivial solutions. This may seem to imply that as one probes moduli space, not many novelties may occur compared to what we have already uncovered. Nonetheless, different combinations of these solutions may take over as the dominant instanton action at different points in moduli space, and these combinations may also enjoy some adequate geometrical interpretation helping us further understand the physics under scrutiny. One question one may ask is whether the conifold instanton action we have just addressed remains the dominant action as we move in moduli space to larger values of the modulus, approaching the large-radius point, or if some other action takes over\footnote{We are thankful to Alba Grassi and Szabolcs Zakany for raising this issue.}. Recall that at this large-radius point the mirror map was computed in \eqref{eq:periodT} and one may in fact explicitly write the K\"ahler parameter in terms of the complex-structure coordinate $\psi$ via a Meijer $G$-function
\be
\label{meijerkahlerparameter}
T(\psi) = - \frac{1}{2\pi \rmi}\frac{\sqrt{3}}{2\pi} G^{22}_{33}\left( \begin{array}{ccc|} \frac{1}{3} & \frac{2}{3} & 1 \\ 0 & 0 & 0 \end{array} \; - \frac{1}{\psi^{3}} \right).
\ee

\begin{figure}[t!]
\begin{center}
\includegraphics[width=\textwidth]{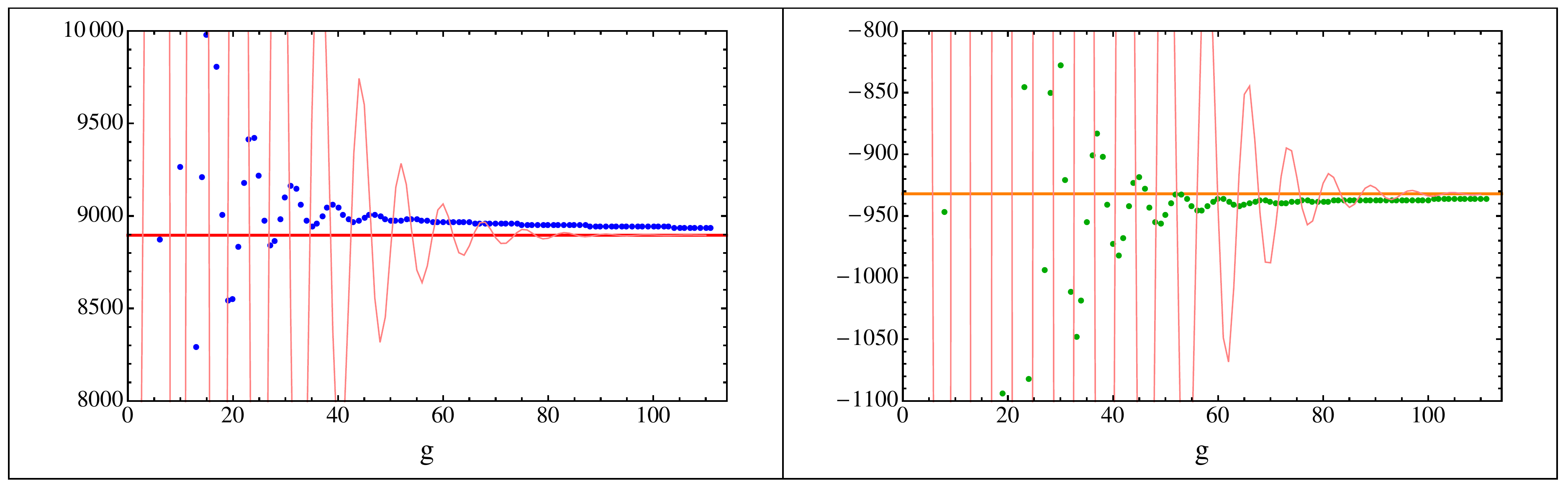}
\vspace{-1\baselineskip}
{\scriptsize
\bea
A_{\text{K}}^2 &=& 8\,896.1 - 932.2\,\rmi  \nonumber\\
\text{1 Richardson Transform} &=& 8\,888.8-929.8\,\rmi \nonumber
\eea
}
\end{center}
\vspace{-1\baselineskip}
\caption{Real and imaginary parts of the limit \eqref{eq:Asquareratio} for $\psi = 50\,  \rme^{\rmi \pi/4}$, along with one Richardson transform. For values of the modulus close to the large-radius point the limits converge slowly and with oscillations, which make the use of Richardson transforms limited. Nevertheless the agreement with the (predicted) analytic value of $A_{\text{K}}^2$ is around 0.1\%, which is still very good.}
\label{fig:AKlargeorder1RTpsi50}
\end{figure}

\begin{figure}[t!]
\begin{center}
\includegraphics[scale=0.415]{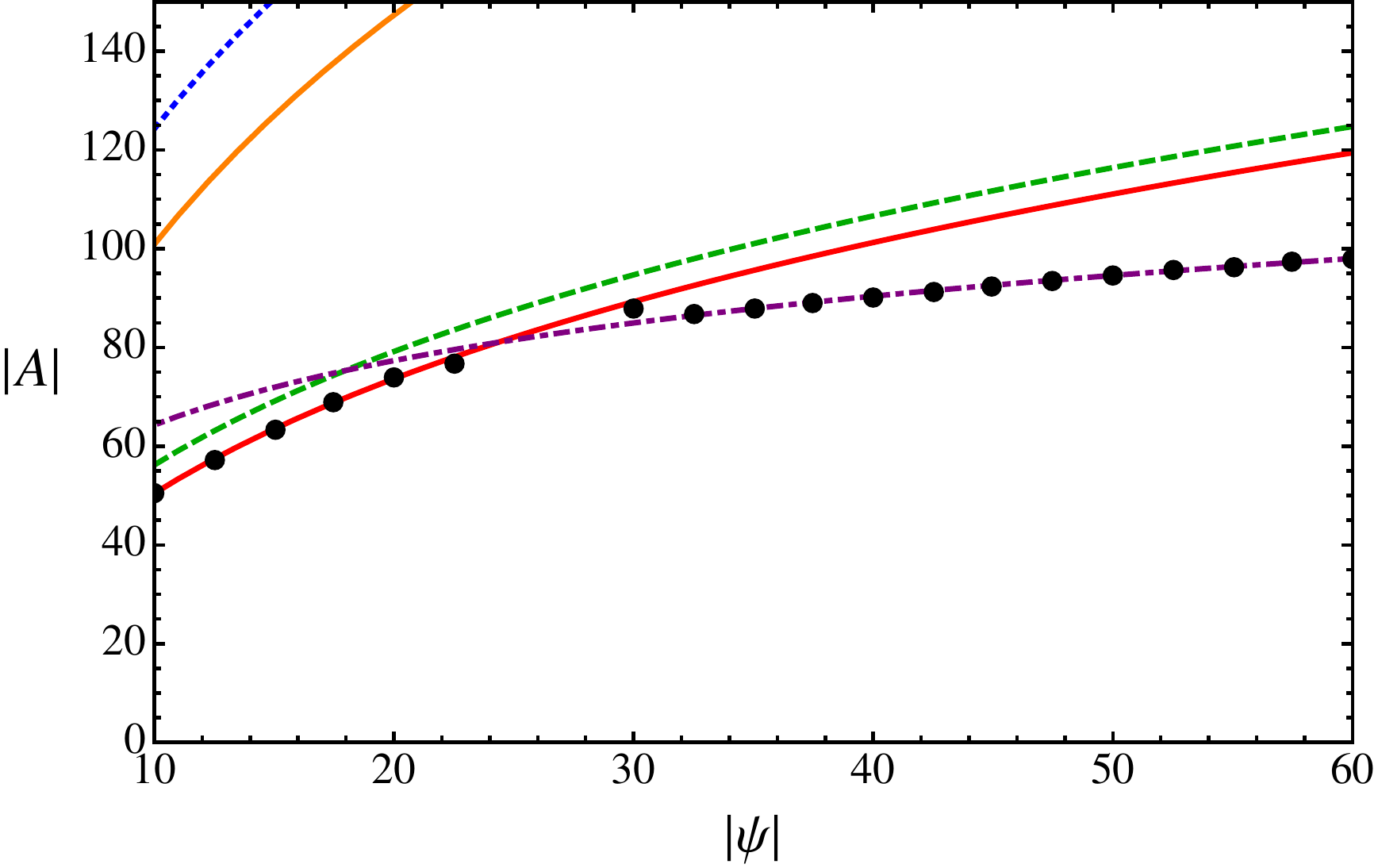}
$\,$
\includegraphics[scale=0.4]{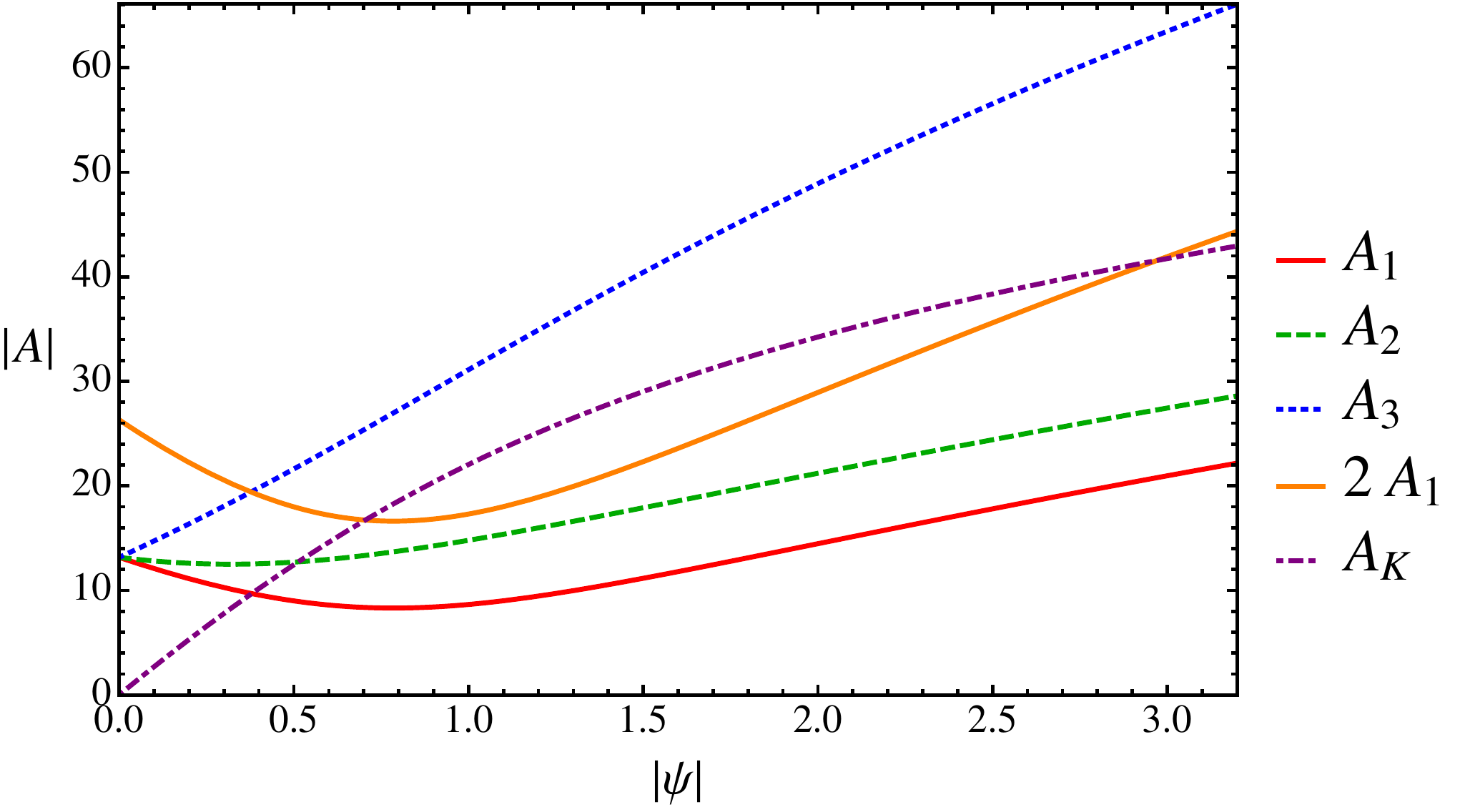}
\end{center}
\vspace{-1\baselineskip}
\caption{Absolute value of conifold, $A_i$, and large-radius, $A_{\text{K}}$, instanton actions. The smaller their value the closer they are to the origin in the complex Borel plane; with the smallest taking dominance in the leading large-order behavior. Here $\arg (\psi) = \frac{\pi}{4}$, and it is clear on the left plot that as one moves towards large-radius, there is a change of leading dominance between $A_1$ and $A_{\text{K}}$ around $\psi \sim 24$. The left plot includes numerical data (the dots) on top of the analytical expectations. The right plot seems to indicate that a similar change of dominance would take place at small values of $\psi$, but this is \textit{not} supported by the large-order numerics.}
\label{fig:kahleractionappears}
\end{figure}

A numerical study of the perturbative free energies at large values of the modulus reveals that, in the region of moduli space associated to the large-radius point, the dominant instanton action actually becomes
\be
\label{kahleraction}
A_{\text{K}} (\psi) = 4\pi^2 \rmi\, T (\psi).
\ee
\noindent
Of course this expression may be written as a difference of conifold actions---although due to the branched structure of these functions not the same difference across moduli space: for $\arg (\psi) >0$ it is $A_{\text{K}} = A_1-A_2$, while for $\arg (\psi) <0$ it is $A_{\text{K}} = A_3-A_1$. As such we shall explicitly work with \eqref{kahleraction} in the following. Numerical evidence showing that the dominant instanton action near the large-radius point is no longer $A_1$ may be found in figure \ref{fig:AKlargeorder1RTpsi50}, where the large-order numerics near this point clearly vindicate $A_{\text{K}}$ as the dominant action. With this result in mind, we have thus gathered extensive numerical evidence that, at small values of the complex-structure modulus, the conifold instanton action dominates the large-order behavior; while at large values of the complex-structure modulus, it is instead the large-radius action associated to the K\"ahler parameter which dominates large order. Where does the transition from one behavior to the next occur? Repeating the previous numerical exercise for different values of the absolute value of $\psi$, we are able to see the transition of dominance between the conifold and K\"ahler instanton actions, which is illustrated in the left plot of figure \ref{fig:kahleractionappears}. Note, however, that in the vicinity of the transition both instanton actions have comparable absolute value and, as such, dominate the large order equally. This reduces the precision of the usual numerical limit or makes it impossible altogether. Nonetheless, a bit further away from the takeover point we can clearly see how the transition occurs. Figure \ref{fig:kahleractionappears} also shows the absolute value of the several instanton actions we have to deal with, \textit{i.e.}, the distances of these singularities to the origin in the complex Borel plane, with the closest action being the dominating one. But for \textit{small} values of the modulus, as shown in the right plot of figure \ref{fig:kahleractionappears}, another crossover of dominance between $A_1$ and $A_{\text{K}}$ seems to take place, and this we know \textit{cannot} be the case from the large-order numerics as we have been discussing throughout this section. So how come the K\"ahler parameter is smaller than the conifold action at small values of $\psi$ and yet it does not take over controlling the large-order behavior of the perturbative free energy? As we will see next, the explanation for this apparent mismatch reveals an intricate multi-branched Borel structure of the topological string free-energy.

%%%%%%%%%%%%%%%%%%%%%%%%%%%%%%%%%%%%%%%%%%%%%%%%%%%%%%%%%%%%%%%%%
%%%%%%%%%%%%%%%%%%%%%%%%%%%%%%%%%%%%%%%%%%%%%%%%%%%%%%%%%%%%%%%%%

%%%%%%%%%%%%%%%%%%%%%%%%%%%%%%%%%%%%%%%%%%%%%%%%%%%%%%%%%%%%%%%%%
%%%%%%%%%%%%%%%%%%%%%%%%%%%%%%%%%%%%%%%%%%%%%%%%%%%%%%%%%%%%%%%%%

%%%%%%%%%%%%%%%%%%%%%%%%%%%%%%%%%%%%%%%%%%%%%%%%%%%%%%%%%%%%%%%%%
\begin{figure}[t!]
\begin{center}
\includegraphics[width=\textwidth]{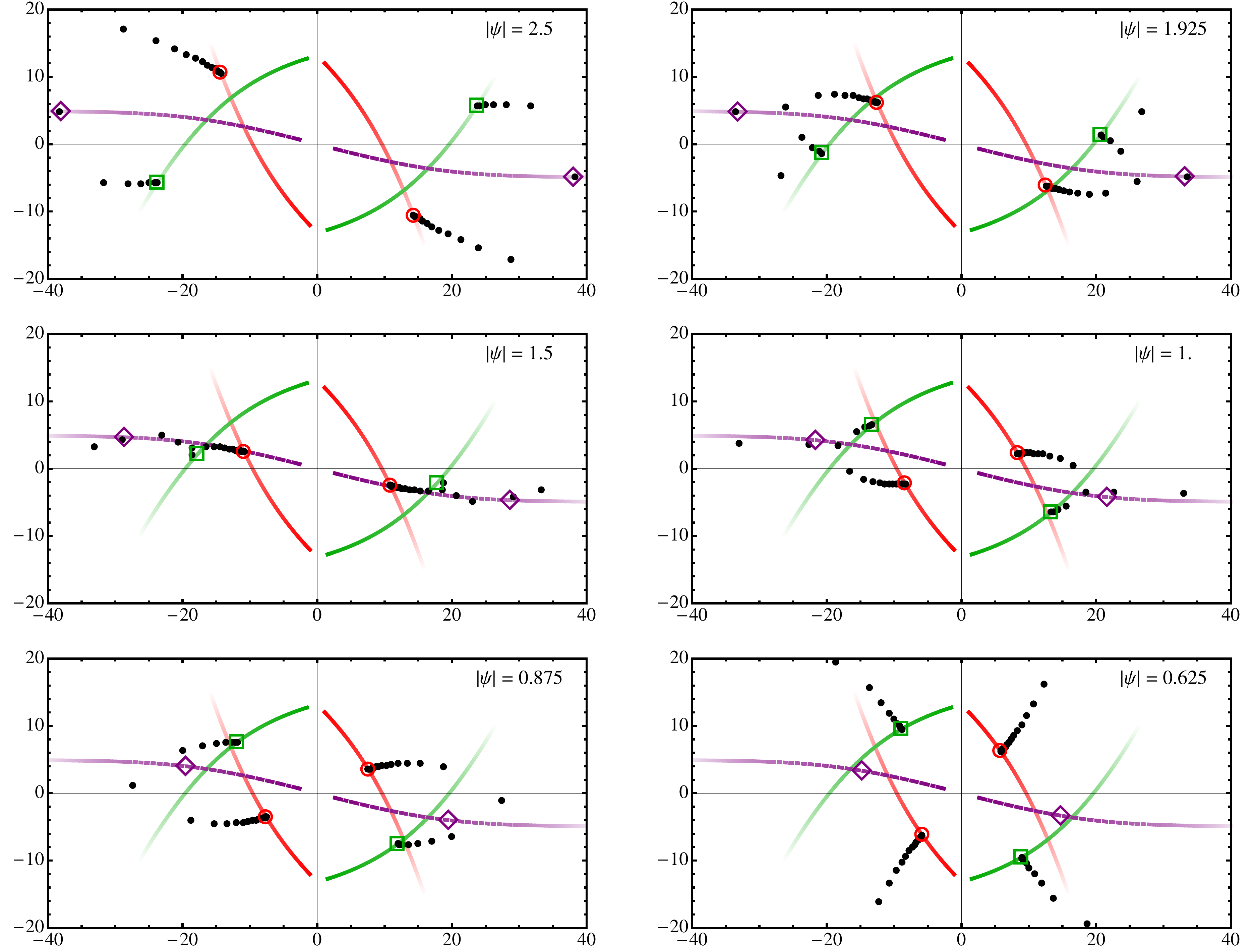}
\end{center}
\vspace{-1\baselineskip}
\caption{Pad\'e analysis of the Borel singularity structure. We illustrate snapshots of the Borel plane for different absolute values of the modulus $\psi$. The red circle shows the analytical value of the conifold action $A_1$, and the red line the trajectory it follows as the modulus is varied. Similarly for the green square and trajectory, associated to the conifold action $A_2$; and for the purple rhombus and trajectory, associated to the large-radius action $A_{\text{K}}$. The black dots are the Pad\'e poles of the Borel transform of the perturbative free energy, and their accumulation signals the onset of a branch cut. It can be clearly seen that around $|\psi| \sim 1$ the Pad\'e pole associated to the large-radius action disappears from the principal Riemann sheet of the perturbative sector.}
\label{fig:changingRiemannsheets}
\end{figure}
%%%%%%%%%%%%%%%%%%%%%%%%%%%%%%%%%%%%%%%%%%%%%%%%%%%%%%%%%%%%%%%%%

A safer way to understand which instanton actions contribute to the large-order behavior is, of course, to analyze the singularity structure of the Borel transform of the perturbative free energy. However, we only know finitely many coefficients in this expansion (up to genus $g=114$ as mentioned) preventing us from exactly evaluating this Borel transform. One thus turns to its Pad\'e approximant (see, \textit{e.g.}, \cite{bo78}), which yields a rational approximation to the required Borel transform. The interesting point is that the poles of this rational function help identify some of the Borel singularities, while their accumulations identify Borel branch cuts. A plot of the Pad\'e poles associated to the perturbative free energy is presented in figure \ref{fig:changingRiemannsheets}, for different (small) values of the complex-structure modulus. In the same plots we have identified the analytical values of the conifold actions $A_1$ and $A_2$, and also of the large-radius action $A_{\text{K}}$ (and their symmetric negatives). The singularity which is closer to the origin will dominate the large-order behavior, and we want to understand why $A_{\text{K}}$ does not take dominance at small $\psi$ as figure \ref{fig:kahleractionappears} seems to indicate. What we find is rather interesting. Both conifold actions, $A_1$ and $A_2$, are rather clear in the Pad\'e results, alongside the branch cuts we expect to find near these singularities (these would be logarithmic branch cuts if these are simple resurgence singularities; see, \textit{e.g.}, \cite{asv11}). As we vary $\psi$ these singularities and their associated branch cuts are \textit{always} visible. But the same does \textit{not} happen for the large-radius action $A_{\text{K}}$. Around $|\psi| \sim 1$, the $A_1$ and $A_2$ singularities cross in the complex Borel plane with their branch cuts effectively pulling the $A_{\text{K}}$ singularity \textit{away} from the principal sheet of the perturbative sector, into \textit{another} Riemann sheet of the Borel surface. It is rather clear in the figure that the Pad\'e pole associated to the $A_{\text{K}}$ singularity is no longer present once $|\psi|$ is smaller than one. This explains why while analytically the large-radius action is indeed smaller than all others (at small $\psi$), as shown in the analytic plots of figure \ref{fig:kahleractionappears}, it is still the case that it does \textit{not} contribute to the leading large-order as vindicated by the numerics: this singularity is now in another Riemann sheet and thus, effectively, further away from the origin in the principal sheet. At the same time this implies that the string free energy has an intricate \textit{multi-branched} Borel structure, with higher-order Stokes phenomena \cite{hlo04} taking an active role in the analysis.

In the remainder of this paper we will mostly focus on the conifold instanton action(s), as we  address subleading and multi-instanton corrections to the large-order growth. Of course the large-radius multi-instanton sectors should follow a very similar story, and it would thus be very interesting to explicitly address the large-order resurgent analysis of these sectors in future work.

Let us end this subsection with a curious remark concerning what takes place \textit{exactly} at the large-radius point. As discussed above, we have seen that the dominant instanton action controlling the large-order growth of the perturbative sector (with the constant-map contribution removed) grows arbitrarily large as one approaches the large-radius point, $\psi^{-1} = 0$ or $z = 0$. This seems to suggest that, at exactly this large-radius point, the perturbative series is no longer asymptotic as any possible singularity in the complex Borel plane has now gone off to infinity.

Let us discuss this point a little more closely. In the holomorphic limit with respect to the large-radius frame, where \eqref{eq:LRholoS} is the relevant expression to consider, the free energies $\CF^{\text{[LR]}(0)}_g$ actually go to zero as we approach the large-radius point, and everything seems rather consistent. Instead, in the nonholomorphic setting the limit $z\to 0$ is a bit subtler and it can only be taken after rescaling the propagator by $z^2$, or otherwise we would find a singular limit. In this way, let us work with a different variable, denoted by $\Sigma$ and given by $S^{zz}  =: z^2\, \Sigma$. This is also naturally motivated by the actual behavior of $S^{zz}_{\text{[LR]},\hol} = z^2\left(\frac{1}{2} + 9 z + \cdots \right)$. If we now study the large-order behavior of the ``rescaled'' free energies,
\be
H^{(0)}_g (\Sigma) := \lim_{z\to 0}F^{(0)}_g(z, z^2\, \Sigma),
\ee
\noindent
we find the following large-order growth
\be
H^{(0)}_g \simeq \frac{\G(g-1)}{A_H(\Sigma)^{g-1}}\, \mu(\Sigma) + \cdots \quad \text{ with } \quad A_H(\Sigma) = \frac{6}{\left( \Sigma - \frac{1}{2} \right)^3}. 
\ee
\noindent
There are a few interesting things to notice from this result. First, the leading growth is now somewhat milder, of the type $\sim g!$ rather than $\sim \left( 2g \right)!$. As to the formula for the ``instanton action'', it is found empirically out of numerics and we have no first-principles approach to it. Curiously enough, it now turns out to have a propagator dependence, something which does not happen anywhere else. The ``instanton factor'' $\mu(\Sigma)$ blows up for $\Sigma = \frac{1}{2}$ and is finite elsewhere. Finally, note that for $\Sigma = \frac{1}{2}$, $H^{(0)}_g = 0$ for all $g \geq 2$. This is because $H^{(0)}_g(\Sigma) = \left( \Sigma - \frac{1}{2} \right)^{2g-3} \, \text{Pol}(\Sigma; g)$, where $\text{Pol}(x; d)$ denotes a polynomial of degree $d$ in $x$.

%%%%%%%%%%%%%%%%%%%%%%%%%%%%%%%%%%%%%%%%%%%%%%%%%%%%%%%%%%%%%%%%%
%%%%%%%%%%%%%%%%%%%%%%%%%%%%%%%%%%%%%%%%%%%%%%%%%%%%%%%%%%%%%%%%%
\section{Resurgent Analysis and the Transseries Solution}
\label{sec:resurgentanalysisofthetransseriessolution}
%%%%%%%%%%%%%%%%%%%%%%%%%%%%%%%%%%%%%%%%%%%%%%%%%%%%%%%%%%%%%%%%%
%%%%%%%%%%%%%%%%%%%%%%%%%%%%%%%%%%%%%%%%%%%%%%%%%%%%%%%%%%%%%%%%%

In the previous section we learned about the existence of different instanton actions, and how they come to play leading roles in the large-order behavior of the perturbative string free-energy, as the modulus is varied. In this section we will dig deeper into the large-order resurgence relations (recall subsection \ref{sec:largeorderfromresurgence}) in order to explicitly see polynomially subleading effects due to the one-instanton sector. We shall do this by addressing the numerical analysis of two resurgence relations: the one describing the perturbative asymptotics, and the one describing the asymptotics of the one-instanton sector. For definiteness, we focus on the conifold action(s) and mainly address the sector associated to $A_1$, but we will also see a little about the ones associated to $A_2$ and $A_3$, which, while being exponentially subleading will be further studied and in greater detail in the next section. Then, having explicitly seen the one-instanton sector in the perturbative asymptotics, we can address it on its own and study its resurgent nature. In particular, we shall show that the asymptotic growth of this (conifold) one-instanton sector is controlled by nonperturbative free energies associated to sectors $(2\bfe_1) = (2|0\|\cdots)$ and $(\bfe_{1,1}) = (1|1\|0\cdots)$, both of which can be computed from the holomorphic anomaly equations. We start, therefore, by swiftly reviewing the main results of \cite{cesv13} on the latter.

%%%%%%%%%%%%%%%%%%%%%%%%%%%%%%%%%%%%%%%%%%%%%%%%%%%%%%%%%%%%%%%%%
\subsection{The Nonperturbative Holomorphic Anomaly Equations}
\label{sec:thenonperturbativeHAEs}
%%%%%%%%%%%%%%%%%%%%%%%%%%%%%%%%%%%%%%%%%%%%%%%%%%%%%%%%%%%%%%%%%

In \cite{cesv13} we described at length how the holomorphic anomaly equations of \cite{bcov93} can be generalized beyond the perturbative sector, in order to allow for transseries solutions. We also described the structure of the higher instanton free energies with respect to the propagators, in various generic situations. In particular, the propagator dependence of the free energies turns from just polynomial, in the perturbative sector, to a further exponential dependence, in the nonperturbative sectors. The starting point involves writing the holomorphic anomaly equations \eqref{eq:propHAEs} as a single equation for the perturbative free energy, $F^{(\bf0)} \simeq \sum_{g=0}^{+\infty} g_s^{2g-2}F^{(\bf0)}_g$, as
\be
\p_{S^{zz}}F^{(\bf0)} + U\, \p_z F^{(\bf0)} - \frac{1}{2} g_s^2 \left( D_z \p_z F^{(\bf0)} + \left( \p_z F^{(\bf0)} \right)^2 \right) = \frac{1}{g_s^2}\, W + V.
\label{eq:prenpHAE}
\ee
\noindent
Here, the definitions
\be
U = \p_z F^{(\bf0)}_0, \qquad V = \p_{S^{zz}} F^{(\bf0)}_1 - \frac{1}{2} D_z \p_z F^{(\bf0)}_0, \qquad W = \frac{1}{2} \left( \p_z F^{(\bf0)}_0 \right)^2,
\ee
\noindent
ensure that we precisely recover \eqref{eq:propHAEs} when inserting the perturbative \textit{ansatz}. Next, we promote \eqref{eq:prenpHAE} to be valid for the full nonperturbative free energy of topological string theory. We can thus construct nonperturbative solutions by plugging in a transseries of the form (but see \cite{cesv13} for broader classes of possible transseries solutions)
\be
F(\bfs,g_s;z,S^{zz}) = \sum_{g=0}^{+\infty} g_s^{2g-2} F^{(\bf0)}_g (z,S^{zz}) + \sum_{\bfn \neq \bf0}^{+\infty} \bfs^\bfn\, \rme^{-A^{(\bfn)}(z,S^{zz})/g_s}\, \sum_{g=0}^{+\infty} g_s^{g+b^{(\bfn)}} F^{(\bfn)}_g (z,S^{zz}),
\label{eq:Ftransseriesansatz}
\ee
\noindent
where we use the notation $A^{(\bfm)} = \sum_{\alpha=1}^q m_\alpha A_\alpha$ to denote the several instanton actions at play. Inserting back in \eqref{eq:prenpHAE}, we find the nonperturbative holomorphic anomaly equations \cite{cesv13}
\begin{eqnarray}
\p_{S^{zz}} A^{(\bfn)} \cdot F^{(\bfn)}_{g+1} &=& 0, \label{eq:holoAHAEs}\\
\left( \partial_{S^{zz}} - \frac{1}{2} \left( \partial_z A^{(\boldsymbol{n})} \right)^2 \right) F^{(\boldsymbol{n})}_g &=& - \sum_{h=1}^g \mathcal{D}^{(\boldsymbol{n})}_h F^{(\boldsymbol{n})}_{g-h} + \label{eq:NPeq}\\
&&
\hspace{-130pt} + \frac{1}{2}\, \sideset{}{'}\sum_{\boldsymbol{m}=\boldsymbol{0}}^{\boldsymbol{n}}\, \sum_{h=0}^{g-B(\boldsymbol{n},\boldsymbol{m})} \left( \partial_z F^{(\boldsymbol{m})}_{h-1} - \partial_z A^{(\boldsymbol{m})}\, F^{(\boldsymbol{m})}_h \right) \left( \partial_z F^{(\boldsymbol{n}-\boldsymbol{m})}_{g-1-B(\boldsymbol{n},\boldsymbol{m})-h} - \partial_z A^{(\boldsymbol{n}-\boldsymbol{m})}\, F^{(\boldsymbol{n}-\boldsymbol{m})}_{g-B(\boldsymbol{n},\boldsymbol{m})-h} \right). \nonumber
\end{eqnarray}
\noindent
Here we have defined $B(\bfn,\bfm) = b^{(\bfm)}+b^{(\bfn-\bfm)}-b^{(\bfn)}$ and the differential operators
\bea
\mathcal{D}^{(\bfn)}_1 &=& \frac{1}{2}\, D_z^2 A^{(\bfn)} + \partial_z A^{(\bfn)} \left( \p_z + \partial_z F^{(\bf0)}_1 \right), \label{eq:calD1}\\
\mathcal{D}^{(\bfn)}_2 &=& - \frac{1}{2}\, D_z^2 - \partial_z F^{(\bf0)}_1\, \p_z, \label{eq:calD2}\\
\mathcal{D}^{(\bfn)}_{2h-1} &=& \partial_z A^{(\bfn)}\, \partial_z F^{(\bf0)}_h, \qquad h = 2, 3, \ldots, \\
\mathcal{D}^{(\bfn)}_{2h} &=& -\partial_z F^{(\bf0)}_h\, \p_z, \qquad h = 2, 3, \ldots. \label{eq:calD2h}
\eea
\noindent
The first equation above, \eqref{eq:holoAHAEs}, implies that the instanton action is holomorphic. This was already extensively discussed in \cite{cesv13} and is now very explicitly seen in an example as we have shown in figure \ref{fig:AisholomorphicRT}. In the second equation above, \eqref{eq:NPeq}, the prime in the sum over $\bfm$ means that the sectors $\bfm=\bf0$ and $\bfm=\bfn$ are excluded from the range. This makes the equations recursive in both the instanton sector, $\bfm$, and the expansion index, $h$. The dependence on the perturbative sector is hidden in the operators $\mathcal{D}^{(\bfn)}_h$. For more details and generalizations see \cite{cesv13}. For our explicit example of local $\BC\BP^2$, we shall integrate the equations above for each relevant sector whenever it appears in the resurgent large-order analyses, and will discuss the fixing of the associated holomorphic ambiguities at the same time.

%%%%%%%%%%%%%%%%%%%%%%%%%%%%%%%%%%%%%%%%%%%%%%%%%%%%%%%%%%%%%%%%%
\subsection{Perturbative Large-Order: Generic Analysis}
\label{sec:largeorderoftheperturbativesector}
%%%%%%%%%%%%%%%%%%%%%%%%%%%%%%%%%%%%%%%%%%%%%%%%%%%%%%%%%%%%%%%%%

\begin{figure}[ht!]
\begin{center}
\includegraphics[scale=0.36]{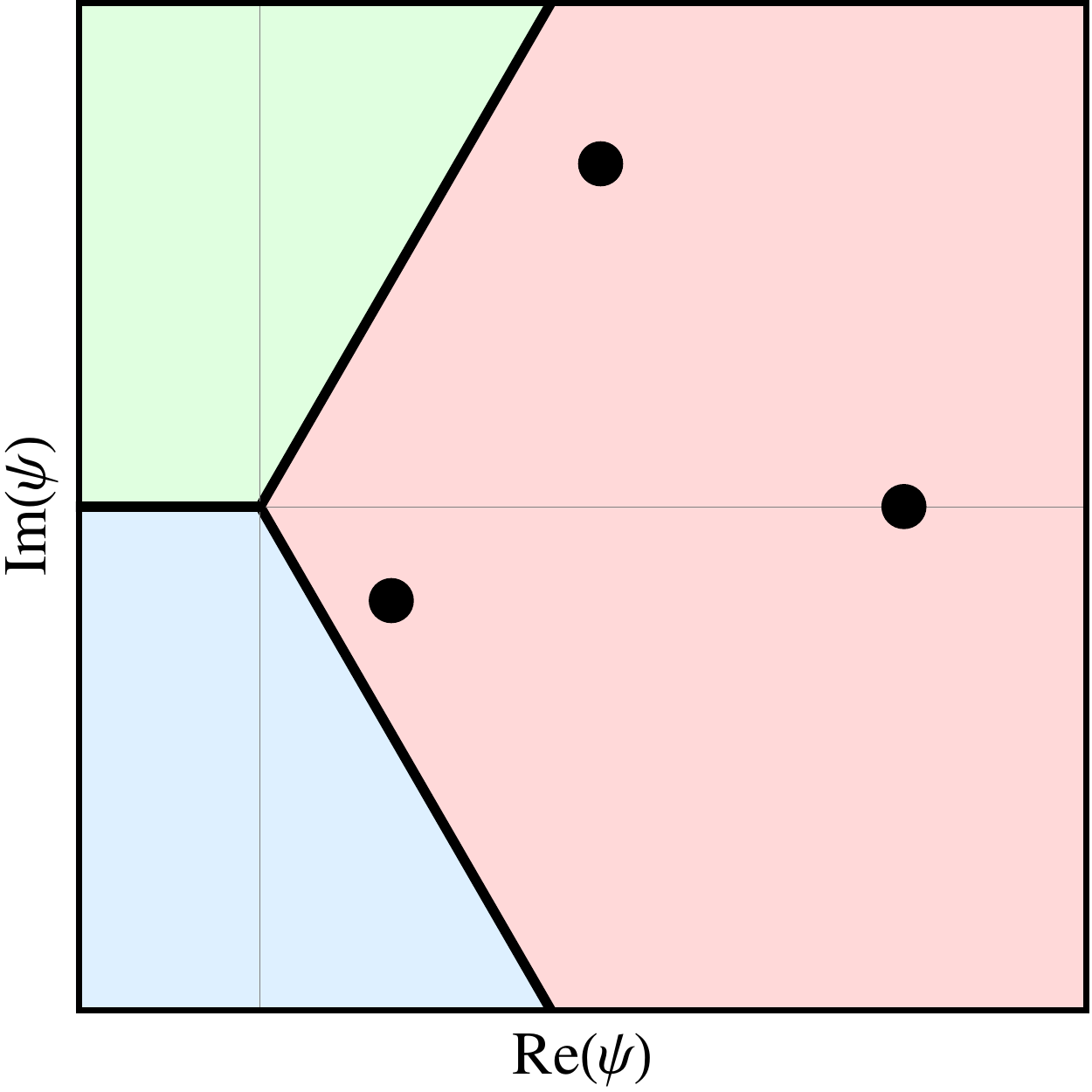}\\ \vspace{0.5cm}
\includegraphics[width=0.32\textwidth]{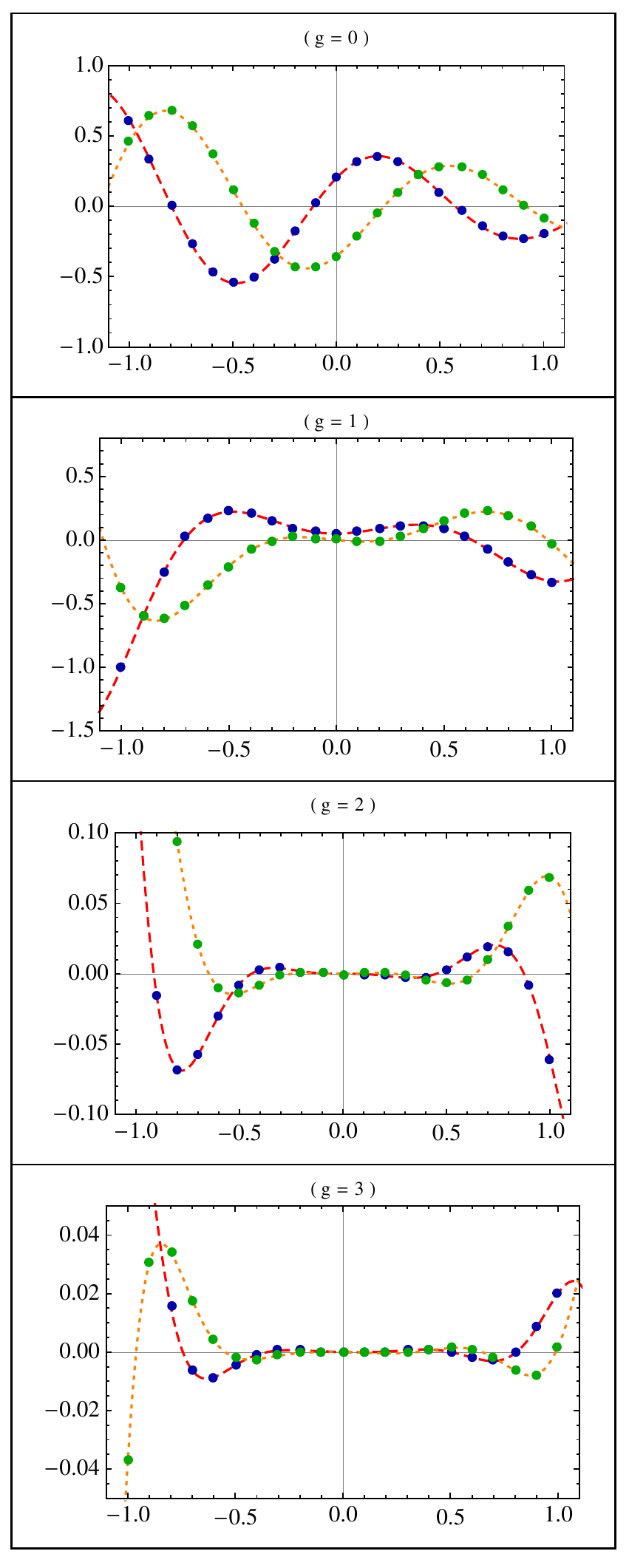}
\includegraphics[width=0.32\textwidth]{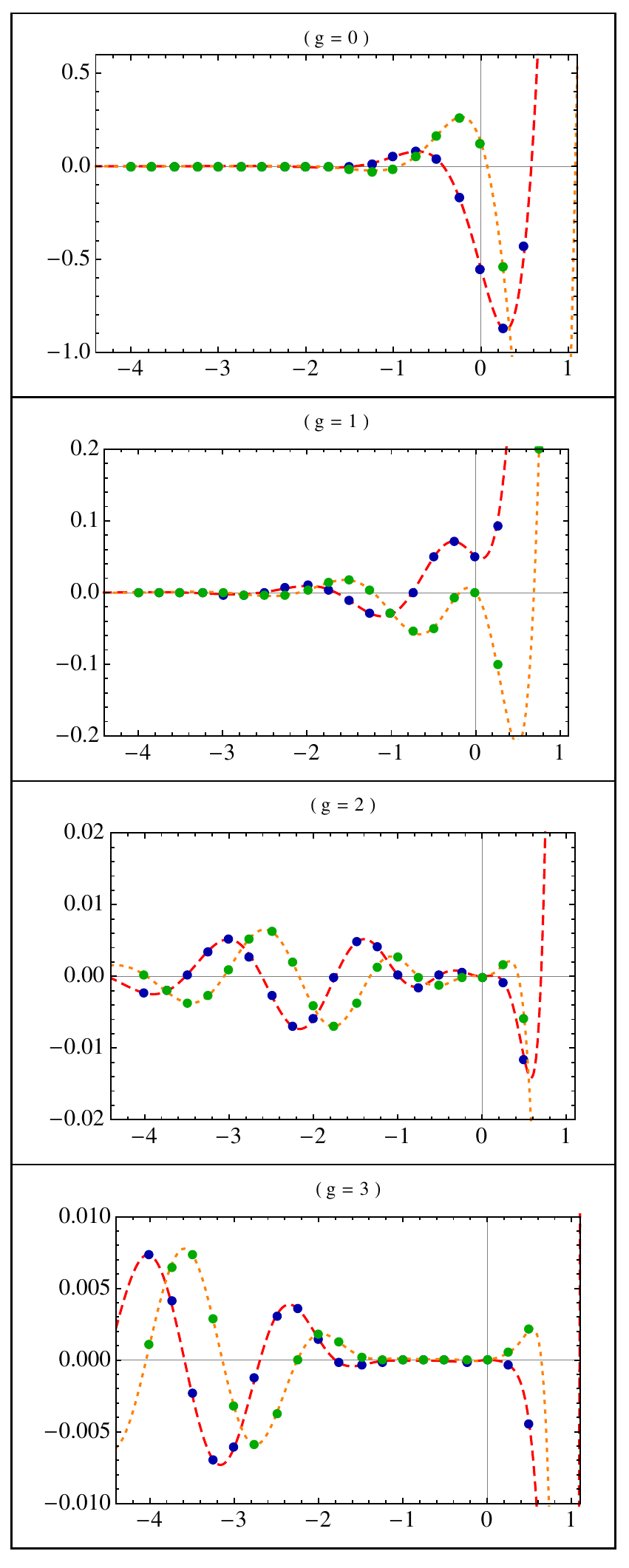}
\includegraphics[width=0.32\textwidth]{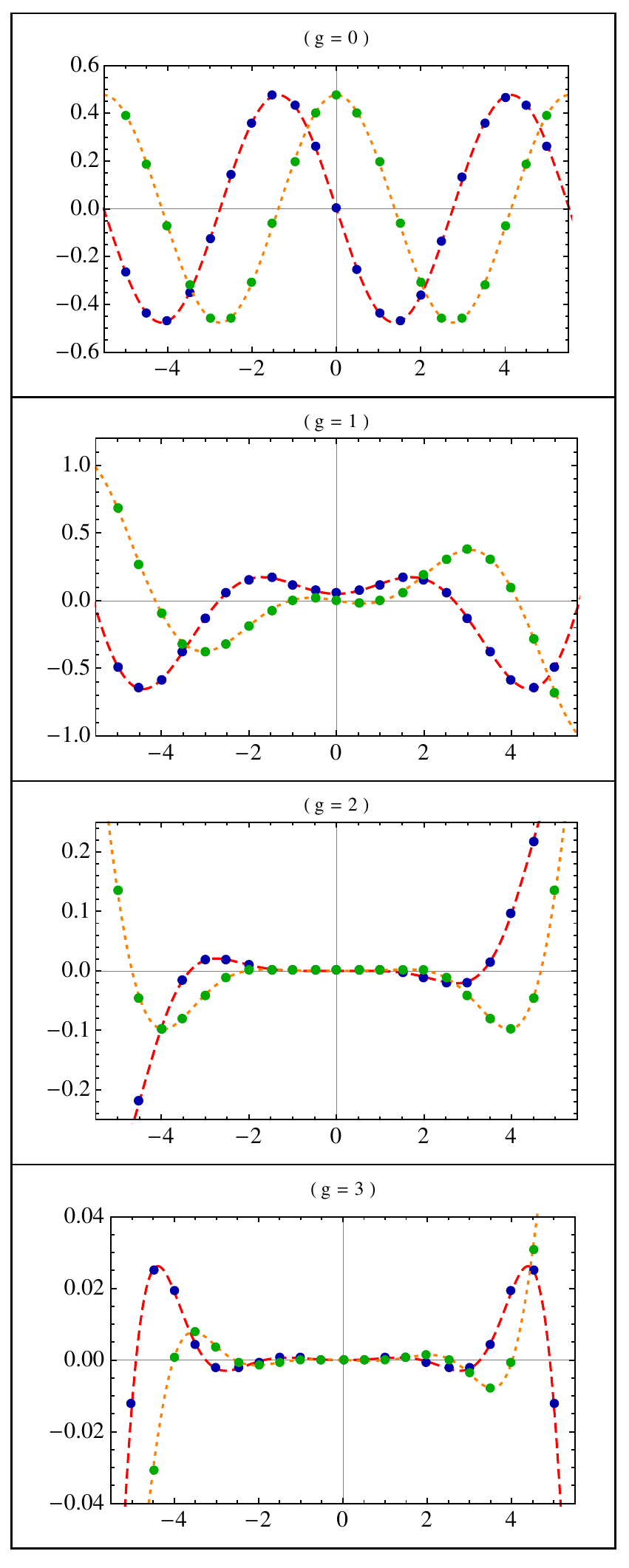}
\end{center}
\vspace{-1\baselineskip}
\caption{Large-order check of $\frac{S_{1,1}}{\rmi\pi}\, F^{(\bfe_1)}_g$ for $g=0,1,2,3$, at three different points in moduli space, $\psi = \frac{1}{2} \rme^{- \rmi \pi /5}, \frac{3}{2} \rme^{+ \rmi\pi/4}, 2$. Moving along the horizontal axis, $x$, is equivalent to changing the value of the propagator around its holomorphic value as $S^{zz} = S^{zz}_{1,\hol} \cdot \left( 1 + \rmi x \right)$. The numerical data is represented by the blue and green dots, for the real and imaginary parts, respectively; and three Richardson transforms were used for each point---but see figure \ref{fig:oneinstlargeorderRTs} for more on these Richardson transforms. The predictions are shown in dashed red and dotted orange lines.}
\label{fig:oneinstlargeorder}
\end{figure}

As we described in section \ref{sec:largeorderanalysisoftheperturbativeexpansion}, for smaller values of the modulus the large-order growth of the perturbative sector behaves rather differently depending on whether $-\pi/3 < \arg(\psi) < +\pi/3$, or $\arg(\psi) = \pm \pi/3$. In the former case, the growth at leading order is determined by $A_1$ alone, while in the latter it is a combination of both $A_1$ and $A_2$, or $A_1$ and $A_3$, respectively. We shall treat both situations separately, and will begin in this subsection by considering (generic) values of $\psi$ whose argument satisfies
\be
-\pi/3 < \arg(\psi) < +\pi/3.
\ee
\noindent
This regime ensures that $| A_1 | < | A_2 |, | A_3 |$, so we can be sure that any contributions from the second and third conifold points will be \textit{exponentially} suppressed (we shall discuss these corrections later on, for the moment one may see figure \ref{fig:instantonactionscomparisonplots}). The numerical analysis we have carried out shows to great accuracy that the growth of the perturbative-sector coefficients is found to have the form
\be
F^{(\bf0)}_{g} \simeq \sum_{h=0}^{+\infty} \frac{\G\left( 2g-1-h \right)}{A_1^{2g-1-h}}\, \frac{S_{1,1}}{\rmi \pi}\, F^{(\bfe_1)}_h + \cdots,
\label{eq:perturbativelargeorder1sector}
\ee
\noindent
where $(\bfe_1) = (1|0 \| 0|0 \| 0|0 \|Ê\cdots)$ is the ``pure'' one-instanton sector of $A_1$ and we have organized the instanton actions in the transseries as in \eqref{eq:organizationoftheinstantonactions}. The subleading terms represented by the ellipses in \eqref{eq:perturbativelargeorder1sector} will be studied in section \ref{sec:resummation}. This particular dependence on the one-instanton sector is expected when we have a standard bridge equation, as explained in subsection \ref{sec:largeorderfromresurgence}, and we found no deviations from this situation at leading order. We can obtain numerical predictions out of large-order, for the different free energies (or, actually, their product with the Stokes constant), by taking the appropriate limit with $g$ going to infinity. For instance,
\be
\frac{S_{1,1}}{\rmi\pi}\, F^{(\bfe_1)}_0 = \lim_{g\to\infty} \frac{A_1^{2g-1}}{\G(2g-1)}\, F^{(\bf0)}_g.
\label{eq:glimitgivingF01}
\ee
\noindent
Recall that both $F^{(\bf0)}_g$ and $F^{(\bfe_1)}_0$ have a complex modulus and a propagator dependence, while the instanton action is just holomorphic. In practice, we pick values for the modulus and the propagator on the right-hand-side of \eqref{eq:glimitgivingF01} and then take the numerical limit, accelerated with the help of Richardson transforms since we only have a finite number of perturbative free energies at our disposal (up to genus $g=114$). For illustration of the results, we have selected three points in moduli space and computed the right-hand-side of \eqref{eq:glimitgivingF01} for several values of the propagator. These results appear as the blue and green dots in the top three plots of figure \ref{fig:oneinstlargeorder}. The numerical value at the origin ($x = 0$), in each plot, corresponds to the holomorphic limit of the free energies.

%%%%%%%%%%%%%%%%%%%%%%%%%%%%%%%%%%%%%%%%%%%%%%%%%%%%%%%%%%%%%%%%%
%%%%%%%%%%%%%%%%%%%%%%%%%%%%%%%%%%%%%%%%%%%%%%%%%%%%%%%%%%%%%%%%%

We can next proceed recursively and compute numerical predictions for \textit{higher loop} one-instanton free energies, $F^{(\bfe_1)}_h$, by simply considering instead the limit
\be
\frac{S_{1,1}}{\rmi\pi}\, F^{(\bfe_1)}_h = \lim_{g\to\infty} \frac{A_1^{2g-1-h}}{\G(2g-1-h)} \left( F^{(\bf0)}_g - \sum_{h'=0}^{h-1} \frac{\G\left( 2g-1-h' \right)}{A_1^{2g-1-h'}}\, \frac{S_{1,1}}{\rmi\pi}\, F^{(\bfe_1)}_{h'} \right).
\label{eq:glimitgivingF0h}
\ee
\noindent
The procedure is exactly the same as outlined above for the $h=0$ case, and the results are displayed in the rest of the plots appearing in figure \ref{fig:oneinstlargeorder}.

Of course our main point now is to show how one can obtain analytical results for the left-hand-side of the above equations, \eqref{eq:glimitgivingF01} and \eqref{eq:glimitgivingF0h}, out of the nonperturbative holomorphic anomaly equations \eqref{eq:NPeq}, which will \textit{very precisely} match against the (numerical) large-order data we just described. Let us thus make these equations concrete for the case of local $\BC\BP^2$, and first focus on the one-instanton sector $\bfn = \bfe_1$. The total instanton action is just $A_1$. Because of the prime, the sum over $\bfm$ in the second line of \eqref{eq:NPeq} is empty, and we are left with just the first one. In this case, the first equation we find is for $F^{(\bfe_1)}_0$. But for $g=0$ the sum of $h$ in \eqref{eq:NPeq} is also empty and so we can directly integrate to obtain
\be
F^{(\bfe_1)}_0(z,S^{zz}) = \rme^{\frac{1}{2}(\p_z A_1)^2 S^{zz}}\, f^{(\bfe_1)}_0(z),
\ee
\noindent
where $f^{(\bfe_1)}_0(z)$ is the holomorphic ambiguity. We fix it by imposing that, in the holomorphic limit, the nonperturbative free energy matches against the one-instanton one-loop result at the conifold \cite{ps09, cesv13},
\be
\frac{S_{1,1}}{\rmi\pi}\, \CF^{(\bfe_1)}_0 = \frac{A_1}{2\pi^2},
\label{eq:oneinsthololimit}
\ee
\noindent
where $S_{1,1}$ is the Stokes constant. Numerically, against large-order data, we find that \eqref{eq:oneinsthololimit} is indeed satisfied; but this can also be shown from (analytical) first principles. Near the conifold point, the perturbative free energies in the holomorphic limit are given by the gap condition \eqref{eq:gapcondition}. Using the holomorphic limit of \eqref{eq:glimitgivingF01} alongside \eqref{eq:gapcondition}, one can find the relation \eqref{eq:oneinsthololimit} as, in the large $g$ limit, the first term of \eqref{eq:gapcondition} (which blows up exactly at the conifold point) dominates---anything else coming after the gap is washed away in the limit (but see \cite{cesv13} for a longer explanation and some calculations). Thus, one obtains
\be
F^{(\bfe_1)}_0(z,S^{zz}) = \frac{\rmi\pi}{S_{1,1}}\, \frac{A_1}{2\pi^2}\, \rme^{\frac{1}{2} \left(\p_z A_1\right)^2 \left(S^{zz}-S^{zz}_{1,\hol}\right)}.
\label{eq:fixedF01analytic}
\ee

The next equation we have to deal with is the one for $F^{(\bfe_1)}_1$, where its right-hand-side now involves $F^{(\bfe_1)}_0$ directly. The concrete expression for the free energy is quite long and not very illuminating, but the result of the integration has the simpler form
\be
F^{(\bfe_1)}_1 (z,S^{zz}) = \rme^{\frac{1}{2} \left(\p_z A_1\right)^2 \left(S^{zz}-S^{zz}_{1,\hol}\right)} \left( f^{(\bfe_1)}_1 (z) + R_1\, S^{zz} + R_2 \left(S^{zz}\right)^2 + R_3 \left(S^{zz}\right)^3 \right),
\ee
\noindent
where $R_i \equiv R_i \left(z,A_1,\p_z A_1\right)$ involves rational functions of $z$, and polynomials in $A_1$ and $\p_z A_1$. The ambiguity, $f^{(\bfe_1)}_1(z)$ is again fixed with the holomorphic constraint of matching the conifold \cite{ps09, cesv13},
\be
\frac{S_{1,1}}{\rmi\pi}\, \CF^{(\bfe_1)}_1 = \frac{1}{2\pi^2},
\ee
\noindent
which again can both be checked numerically against large-order data, and analytically computed from first principles. The exact same procedure applies to the next free energies, where the conifold constraint is now (see \cite{cesv13} for further details on the fixing of the ambiguities)
\be
\frac{S_{1,1}}{\rmi\pi}\, \CF^{(\bfe_1)}_{g\geq 2} = 0.
\label{eq:allzeroesafterg2}
\ee
\noindent
The final structure of the solutions is then\footnote{One might worry that, in the holomorphic limit, the nonperturbative structure of local $\BC\BP^2$ collapses to that of the conifold. This cannot be true of course. It is simple to see that in spite of \eqref{eq:oneinsthololimit} having a somewhat universal structure, the precise function of the modulus appearing in its right-hand-side is an action associated to a \textit{specific} conifold point of our \textit{specific} model---and thus the result is in fact model \textit{dependent} as it should be. One might also worry that, given \eqref{eq:allzeroesafterg2}, the model has no asymptotics of instantons (which was the case for the conifold \cite{ps09}). Again, this is not the case, as one should recall that there are \textit{other} special points in moduli space and these have \textit{not} truncated and are still in fact asymptotic---we shall discuss this issue carefully in the next section.}
\be
F^{(\bfe_1)}_g (z,S^{zz}) = \frac{\rmi\pi}{S_{1,1}}\, \rme^{\frac{1}{2} \left(\p_z A_1\right)^2 \left(S^{zz}-S^{zz}_{1,\hol}\right)}\, \textrm{Pol} \left(S^{zz};3g\right),
\label{eq:fixed1sector}
\ee
\noindent
where $\textrm{Pol} \left(S^{zz};3g\right)$ represents a polynomial of degree $3g$ in the propagator,  whose coefficients involve rational functions of $z$, and powers of $A_1$, $\p_zA_1$ and\footnote{Recall that, since $A_1$ is a period satisfying the Picard--Fuchs equation, we can always trade any third-order derivatives by lower ones, alongside rational functions of $z$.} $\p_z^2A_1$. Recall that $S^{zz}_{1,\hol}$ is a frame dependent quantity, and in this case we are focusing on the first conifold point. Of course the analogous procedure can be applied to the other sectors $(\bfe_2)$ or $(\bfe_3)$, where different holomorphic limits are relevant. Finally, \eqref{eq:fixed1sector} tells us that the undetermined Stokes constant $S_{1,1}$ is part of the free energies, in such a way that it always gets cancelled when we plug \eqref{eq:fixed1sector} into the resurgence relation \eqref{eq:generalperturbativelargeorder}. This means its value cannot be computed using these relations; however, the Stokes constants will also multiply the transseries parameters, $\sigma_i$, in \eqref{eq:Ftransseriesansatz}, and may thus be absorbed in these---which in any case need to be determined by physical conditions. 

%%%%%%%%%%%%%%%%%%%%%%%%%%%%%%%%%%%%%%%%%%%%%%%%%%%%%%%%%%%%%%%%%
\begin{figure}[t!]
\begin{center}
\includegraphics[width=\textwidth]{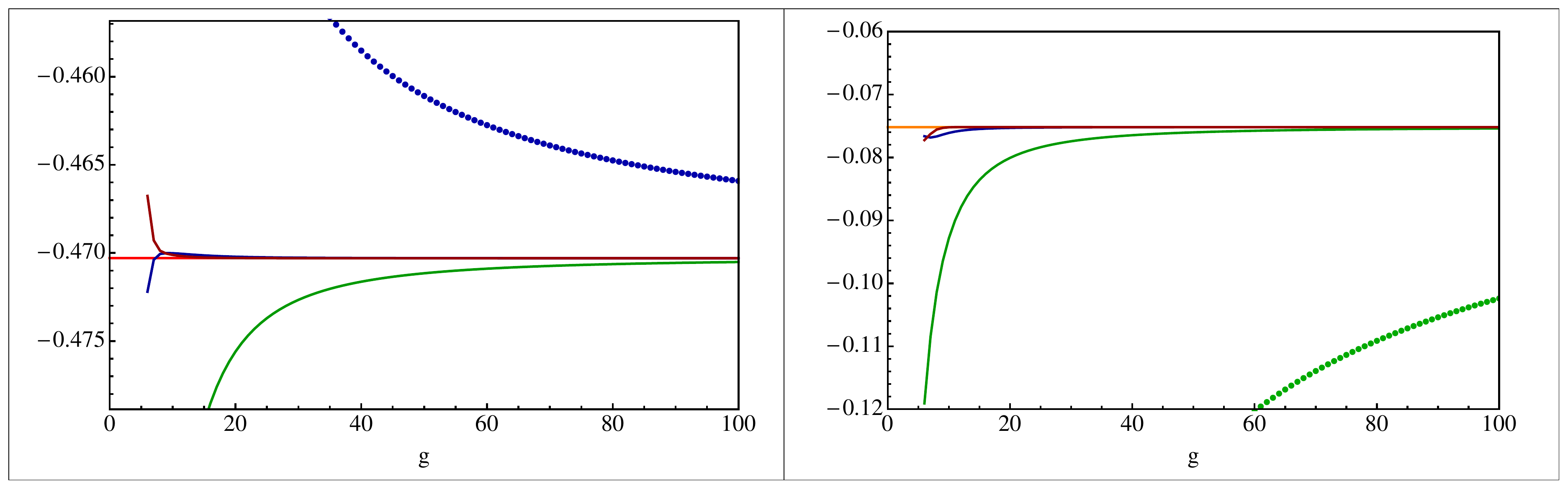}\\
\vspace{-1\baselineskip}
{\scriptsize
\bea
2\,\frac{S_{1,1}}{2\pi \rmi}\,F^{(\bfe_1)}_0 &=& -0.470\,302\,487-0.075\,185\,636\,\rmi  \nonumber\\
\text{3 Richardson Transforms} &=& -0.470\,302\,475-0.075\,185\,625\,\rmi \nonumber
\eea
}
\includegraphics[width=\textwidth]{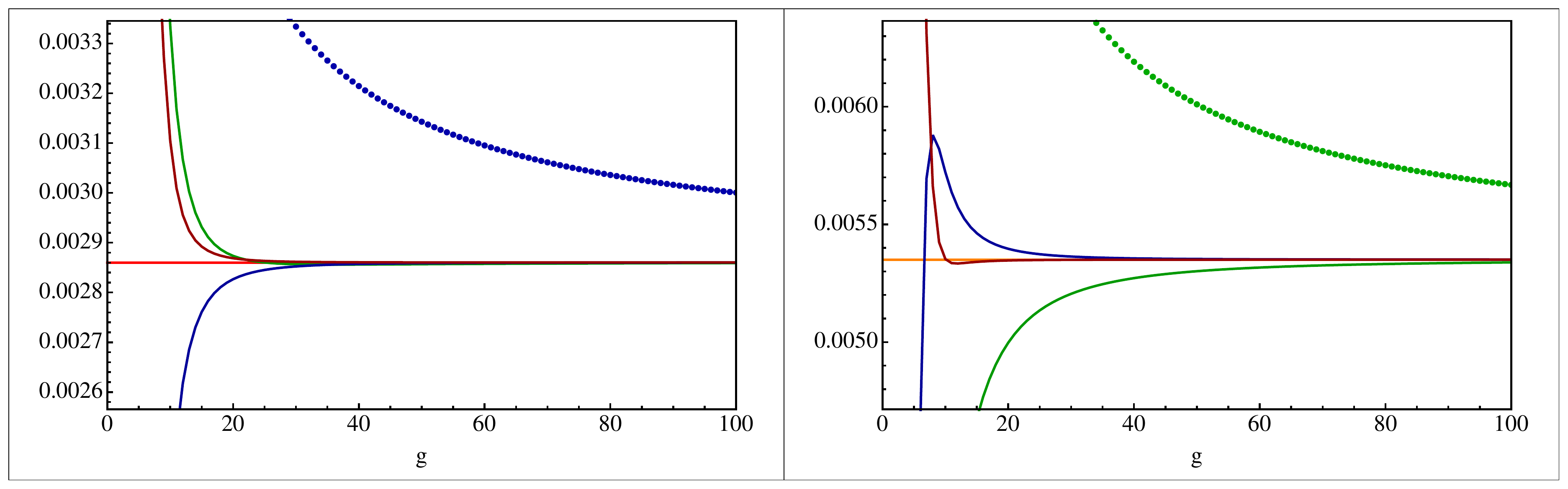}\\
\vspace{-1\baselineskip}
{\scriptsize
\bea
2\,\frac{S_{1,1}}{2\pi \rmi}\,F^{(\bfe_1)}_2 &=& 0.002\,860\,250\,8+0.005\,350\,260\,\rmi  \nonumber\\
\text{3 Richardson Transforms} &=& 0.002\,860\,257\,4+0.005\,350\,255\,\rmi \nonumber
\eea
}
\includegraphics[width=\textwidth]{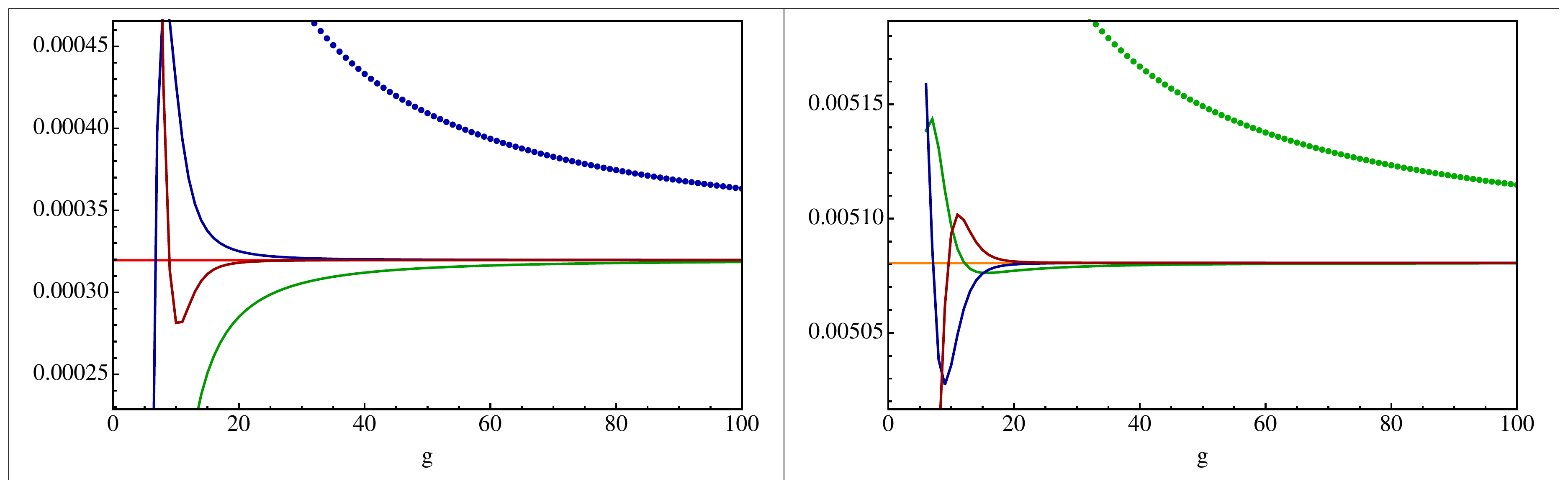}
\vspace{-1\baselineskip}
{\scriptsize
\bea
2\,\frac{S_{1,1}}{2\pi \rmi}\,F^{(\bfe_1)}_4 &=& 0.000\,319\,688\,7+0.005\,080\,627\,80\,\rmi  \nonumber\\
\text{3 Richardson Transforms} &=& 0.000\,319\,687\,9+0.005\,080\,628\,13\,\rmi \nonumber
\eea
}
\end{center}
\vspace{-0.8\baselineskip}
\caption{Real (left) and imaginary (right) parts of sector $(\bfe_1)$ free energies for different values of complex modulus, propagator and loop order. The comparison between analytic and numerical results shows an excellent agreement, which can be improved taking more Richardson transforms.}
\label{fig:oneinstlargeorderRTs}
\end{figure}
%%%%%%%%%%%%%%%%%%%%%%%%%%%%%%%%%%%%%%%%%%%%%%%%%%%%%%%%%%%%%%%%%

We may now go back to figure \ref{fig:oneinstlargeorder}, where, on top of the numerical data, we have plotted the analytical closed form of the very same free energies \eqref{eq:fixed1sector}, as just computed above out of the nonperturbative holomorphic anomaly equations \eqref{eq:NPeq}. The agreement is already visually extremely good, and we can be more quantitative by focusing on some of the points (\textit{i.e.}, fixed $\psi$ and $S^{zz}$). In figure \ref{fig:oneinstlargeorderRTs} we show how the right-hand-side of \eqref{eq:glimitgivingF0h} approaches the expect (analytical) value for large $g$. With the help of a few Richardson transforms one quickly finds agreement to several digits, serving as a very strong check on our proposals.

Let us finish with some remarks. First, note that taking the holomorphic limit of \eqref{eq:glimitgivingF0h} corresponds to the $x = 0$ points in the plots of figure \ref{fig:oneinstlargeorder}. As discussed above, in this case only for $h = 0$ and $h = 1$ are the one-instanton free energies nonzero; this was the key piece of information we used in order to fix the holomorphic ambiguities for the coefficients $F^{(\bfe_1)}_h$. Another thing to notice is that the agreement between large-order numerics and the above analytical results from the holomorphic anomaly equations relies on the Stokes constants being both holomorphic and modulus independent, \textit{i.e.}, actual \textit{constants}. This is not a trivial fact because the modulus and the propagator act, from the point-of-view of resurgence, as external parameters. Therefore, any $g_s$-independent quantity may still be a function of these parameters. However, this is not what we found. If the Stokes constant had any dependence on $\psi$ and $S^{zz}$, \textit{i.e.}, $S_{1,1} \equiv S_{1,1}(\psi,S^{zz})$, \eqref{eq:perturbativelargeorder1sector} would read instead
\bea
F^{(\bf0)}_g &\simeq& \frac{\G(2g-1)}{A_1^{2g-1}}\, \frac{S_{1,1}(\psi,S^{zz})}{\rmi\pi}\, F^{(\bfe_1)}_0 + \cdots \\
&=& \frac{\G(2g-1)}{A_1^{2g-1}} \left(\frac{S_{1,1}(\psi,S^{zz})}{S_{1,1}(\psi,S^{zz}_{1,\hol})}\right) \frac{A_1}{2\pi^2}\, \rme^{\frac{1}{2} \left(\p_zA_1\right)^2 \left(S^{zz}-S^{zz}_\hol\right)} + \cdots.
\eea
\noindent
Recall that the holomorphic ambiguity of $F^{(\bfe_1)}_0$ is fixed precisely in the holomorphic limit, which implies that the Stokes constant in the denominator is the holomorphic limit of the one in the numerator. But the first row of plots in figure \ref{fig:oneinstlargeorder} shows that the above fraction of Stokes constants is equal to one, \textit{i.e.}, that $S_{1,1}$ is at the very least holomorphic. To see that it is also independent of $\psi$, or equivalently, of $z$, we notice that if it were not then the solution of the holomorphic anomaly equation for $F^{(\bfe_1)}_1$ would include the factor $\p_z S_{1,1}$. This is because the holomorphic anomaly equation for this free energy depends on the $z$-derivative of $F^{(\bfe_1)}_0$. The final expression for the free energy, after fixing of the ambiguity, would then depend of the ratio $\frac{\p_z S_{1,1}}{S_{1,1}}$. But the second row of plots in figure \ref{fig:oneinstlargeorder} gives numerical evidence that such a term is not present, hence showing the constancy of the Stokes constant.

%%%%%%%%%%%%%%%%%%%%%%%%%%%%%%%%%%%%%%%%%%%%%%%%%%%%%%%%%%%%%%%%%
\subsection{Perturbative Large-Order: Boundary Analysis}
%%%%%%%%%%%%%%%%%%%%%%%%%%%%%%%%%%%%%%%%%%%%%%%%%%%%%%%%%%%%%%%%%

When the argument of $\psi$ reaches either $+\pi/3$ or $-\pi/3$, $A_1$ ceases to be the only leading instanton action, and has to share dominance with either $A_2$ or $A_3$, respectively (where one has $| A_1 | = | A_2 |$ or $| A_1 | = | A_3 |$). This means that two \textit{different} one-instanton sectors, $(\bfe_1)$ and $(\bfe_2)$, or $(\bfe_1)$ and $(\bfe_3)$, will now provide the leading contribution to the large-order growth of the perturbative coefficients\footnote{Strictly speaking we actually have four sectors, associated to the actions $+A_1$, $-A_1$, $+A_2$ and $-A_2$ controlling the growth at $\arg(\psi) = +\pi/3$ (and similarly for $\pm A_3$ at $\arg(\psi) = -\pi/3$). However, the symmetric sectors $\bfe_1=(1|0 \| 0|0 \| 0|0 \| \cdots)$ and $\bfe_2 = (0|1 \| 0|0 \| 0|0 \| \cdots)$, and so on, turn out to give the same contribution---see the discussion leading up to equation \eqref{eq:generalperturbativelargeorder}.}. In the following we shall focus upon the case $\arg(\psi) = +\pi/3$, for definiteness, but always bear in mind that there is a ``symmetric'' situation for $\arg(\psi) = -\pi/3$.

\begin{figure}[t!]
\begin{center}
\includegraphics[scale=0.4]{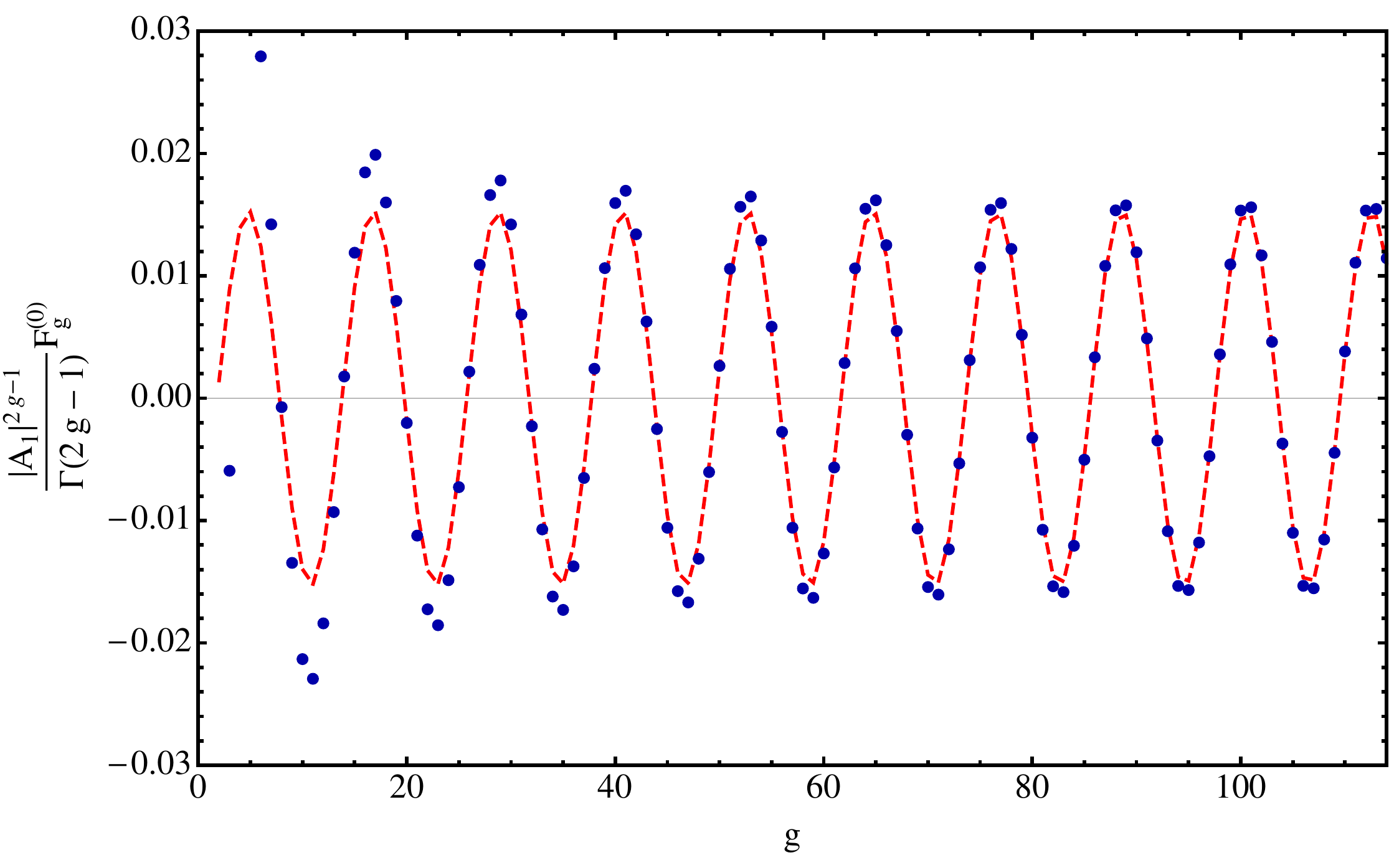} \\
\includegraphics[scale=0.36]{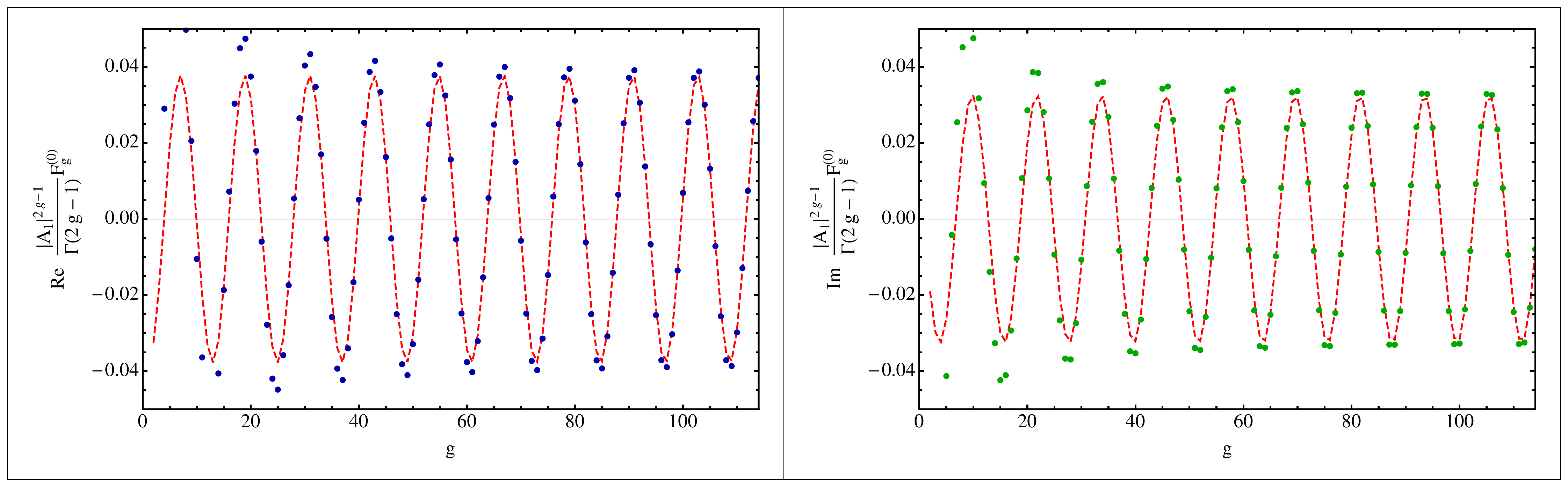}
\end{center}
\vspace{-1\baselineskip}
\caption{Large-order oscillatory behavior of the perturbative sector, due to simultaneous contributions of $A_1$ and $A_2$, for a value of $\psi = 1.1\, \rme^{\rmi\pi/3}$ and different values of the propagator. The plot on top is purely real, as it combines two complex conjugate quantities. The plots below include both real and imaginary parts, as the propagator is no longer chosen to be real. The data-points approach \eqref{eq:oscillatory1} (top) and \eqref{eq:oscillatory2} (bottom), for large $g$, precisely as expected.}
\label{fig:oscillatoryplots}
\end{figure}

The key point of the different large-order behavior at the boundary is that the two instanton actions, $A_1$ and $A_2$, have become complex conjugates of each other, up to a sign. Explicitly
\bea
A_1 &=& - A_2^*, \\
\p_z A_1 &=& - \left(\p_z A_2\right)^*, \\
\p^2_z A_1 &=& - \left(\p^2_z A_2\right)^*, \\
S^{zz}_{1,\hol} &=& + \left(S^{zz}_{2,\hol}\right)^*.
\eea
\noindent
The subindex, $i$, in $S^{zz}_{i,\hol}$ makes reference to the preferred frame associated to the $i$-th conifold point\footnote{Note that one can write the holomorphic limit of the propagator as
\be
S^{zz}_{i,\hol} = -\frac{1}{C_{zzz}(z)} \left( \frac{A_i''(z)}{A_i'(z)} - \tilde{f}^{z}_{zz}(z) \right),
\ee
\noindent
from which we see the relation involving complex conjugation. In here, $C_{zzz}(z)$ and $\tilde{f}^{z}_{zz}(z)$ are given by \eqref{eq:Czzz} and \eqref{eq:ftilde}, respectively.}, $i=1,2,3$. Since \mbox{$|A_1| = |A_2|$}, the large-order growth of the perturbative free energies \eqref{eq:generalperturbativelargeorder} is now given, at leading order, by
\bea
F^{(\bf0)}_g &\simeq& \frac{\G(2g-1)}{A_1^{2g-1}}\, \frac{S_{1,1}}{\rmi\pi}\, F^{(\bfe_1)}_0 + \frac{\G(2g-1)}{A_2^{2g-1}}\, \frac{S_{1,2}}{\rmi\pi}\, F^{(\bfe_2)}_0 + \cdots \\
&=& \frac{\G(2g-1)}{A_1^{2g-1}}\, \frac{A_1}{2\pi^2}\, \rme^{\frac{1}{2} \left(\p_zA_1\right)^2 \left(S^{zz}-S^{zz}_{1,\hol}\right)} + \frac{\G(2g-1)}{A_2^{2g-1}}\, \frac{A_2}{2\pi^2}\, \rme^{\frac{1}{2} \left(\p_zA_2\right)^2 \left(S^{zz}-S^{zz}_{2,\hol}\right)} + \cdots \\
&=& \frac{\G(2g-1)}{A_1^{2g-1}}\, \frac{A_1}{2\pi^2}\, \rme^{\frac{1}{2} \left(\p_zA_1\right)^2 \left(S^{zz}-S^{zz}_{1,\hol}\right)} + \frac{\G(2g-1)}{(A_1^*)^{2g-1}}\, \frac{A_1^*}{2\pi^2}\, \rme^{\frac{1}{2} \left( (\p_zA_1)^* \right)^2 \left( S^{zz}-(S^{zz}_{1,\hol})^* \right)} + \cdots.
\label{eq:almostcxconjugatetoeachother}
\eea
\noindent
Here we have used the explicit form of the free energy $F^{(\bfe_2)}_0$, where $(\bfe_2) = (0|0\|1|0\|0\cdots)$. This nonperturbative sector is computed in the same way as we described the computation of $F^{(\bfe_1)}_0$ in the previous section \ref{sec:largeorderoftheperturbativesector}, but using $A_2$ instead of $A_1$, $S^{zz}_{2,\hol}$ instead of $S^{zz}_{1,\hol}$, and $S_{1,2}$ instead of $S_{1,1}$. Note that the two terms in \eqref{eq:almostcxconjugatetoeachother} are \textit{almost} complex conjugate to each other, and that the sole obstruction is the propagator not being real (in fact, it does not need to be). If for a moment we do choose the propagator to be real, then we can write
\be
F^{(\bf0)}_g \simeq \frac{\G(2g-1)}{A_1^{2g-1}}\, \frac{A_1}{2\pi^2}\, \rme^{\frac{1}{2}\left(\p_zA_1\right)^2 \left(S^{zz}-S^{zz}_{1,\hol}\right)} + (\text{complex conjugate}) + \cdots.
\ee
\noindent
Since the perturbative free energies are real for real $z$ (note that $\arg(\psi) = \pm \pi/3$ implies $z \in \BR^+$) and real $S^{zz}$, this result is of course fully consistent. Writing the instanton action and the one-instanton one-loop free energies in polar form
\be
A_1 = |A_1 |\, \rme^{\rmi \th_{A_1}}, \qquad \mu = | \mu |\, \rme^{\rmi \th_{\mu}} := \frac{A_1}{2\pi^2}\, \rme^{\frac{1}{2} \left(\p_zA_1\right)^2 \left( S^{zz}-S^{zz}_{1,\hol} \right)},
\ee
\noindent
we immediately find the familiar oscillatory behavior in genus, $g$, which we anticipated earlier in section \ref{sec:instantonactions} (see as well, \textit{e.g.}, \cite{msw07}),
\be
F^{(\bf0)}_g \simeq \frac{\G(2g-1)}{|A_1|^{2g-1}}\, 2 |\mu|\, \cos \left( \th_{A_1} \left(2g-1\right) - \th_\mu \right) + \cdots.
\label{eq:oscillatory1}
\ee
\noindent
In the general case, where $S^{zz}$ is complex, we first define
\be
\tmu = |\tmu |\, \rme^{\rmi\th_\tmu} := \frac{A_1^*}{2\pi^2}\, \rme^{\frac{1}{2} \left( (\p_zA_1)^* \right)^2 \left( S^{zz}-(S^{zz}_{1,\hol})^* \right)},
\ee
\noindent
and then obtain a slightly more complicated oscillatory formula,
\bea
F^{(\bf0)}_g &\simeq& \frac{\G(2g-1)}{A_1^{2g-1}}\, \mu + \frac{\G(2g-1)}{(A_1^*)^{2g-1}}\, \tmu + \cdots \\
&=& \frac{\G(2g-1)}{|A_1|^{2g-1}}\, \Big\{ + |\mu|\, \cos\left( \th_{A_1} \left(2g-1\right) - \th_\mu \right) + |\tmu|\, \cos\left( \th_{A_1} \left(2g-1\right) + \th_\tmu \right) \Big\} + \nonumber\\
&+& \rmi\, \frac{\G(2g-1)}{|A_1|^{2g-1}}\, \Big\{ -|\mu|\, \sin\left( \th_{A_1} \left(2g-1\right) - \th_\mu \right) + |\tmu|\, \sin\left( \th_{A_1} \left(2g-1\right) + \th_\tmu \right) \Big\} + \cdots.
\label{eq:oscillatory2}
\eea
\noindent
In order to check both large-order resurgence relations \eqref{eq:oscillatory1} and \eqref{eq:oscillatory2}, we first move the gamma function and the power of $| A_1 |$ over to the left-hand-side, and then plot the numerical data (the numerical large-order values of the ``new'' left-hand-side) against the analytical expressions in the right-hand-side, computed out of the nonperturbative holomorphic anomaly equations. The result, for two different values of the propagator, real and complex, is shown in figure \ref{fig:oscillatoryplots}. Richardson transforms do not work for oscillatory behaviors and thus we cannot use them in here to show quantitative agreement, but it should already be visually evident from the plots that there is a very good match at large-order between numerical data and analytical predictions.

As described, this is essentially the first check on the presence and relevance of the instanton actions $A_2$ and $A_3$, and their respective instanton sectors (to one loop). Later on, in subsection \ref{sec:nonholomorphiccase}, we shall do more quantitative tests on the higher-loop free-energies.

%%%%%%%%%%%%%%%%%%%%%%%%%%%%%%%%%%%%%%%%%%%%%%%%%%%%%%%%%%%%%%%%%
%%%%%%%%%%%%%%%%%%%%%%%%%%%%%%%%%%%%%%%%%%%%%%%%%%%%%%%%%%%%%%%%%

%%%%%%%%%%%%%%%%%%%%%%%%%%%%%%%%%%%%%%%%%%%%%%%%%%%%%%%%%%%%%%%%%
\subsection{Large-Order Behavior of the One-Instanton Sector}
\label{largeorderoftheoneinstantonsector}
%%%%%%%%%%%%%%%%%%%%%%%%%%%%%%%%%%%%%%%%%%%%%%%%%%%%%%%%%%%%%%%%%

In an earlier subsection we addressed nonperturbative free energies associated to the first few loop corrections around the $(\bfe_1)$ one-instanton sector, by analyzing the large-order growth of the perturbative free energies. In particular, we have seen how the numerical results very precisely agree with our analytical calculations for this one-instanton sector, as integrated out of the nonperturbative holomorphic anomaly equations. Now, we want to address these one-instanton free energies themselves, in particular uncover what controls their asymptotic growth at large $g$. Note, however, that while the integration of the nonperturbative coefficients is relatively straightforward, the ``size'' of the resulting free energies grows faster than what was previously found for the perturbative sector. This ``functional growth'' is essentially associated to the coefficients of the polynomials, which involve not only rational functions of $z$ but also the instanton actions, $A_i$, and their first and second derivatives. The outcome of this fact is that it is only reasonable to perform the analytical calculations up to around $g=20$ for the one-instanton sectors. Nonetheless, as we are mainly interested in numerical values of the free energies (this is what is needed in order to check resurgence relations), it is possible to turn to an integration of the higher instanton sectors which is numerical from scratch. More precisely, we shall fix a particular point in moduli space while leaving the propagator variable free, and implement the integration numerically at this point. This allows us to speed up the computation and reach around $g=80$, for a given fixed value of $z$, but still with full analytic dependence on the propagator.

%%%%%%%%%%%%%%%%%%%%%%%%%%%%%%%%%%%%%%%%%%%%%%%%%%%%%%%%%%%%%%%%%
\begin{figure}
\begin{center}
\includegraphics[scale=0.36]{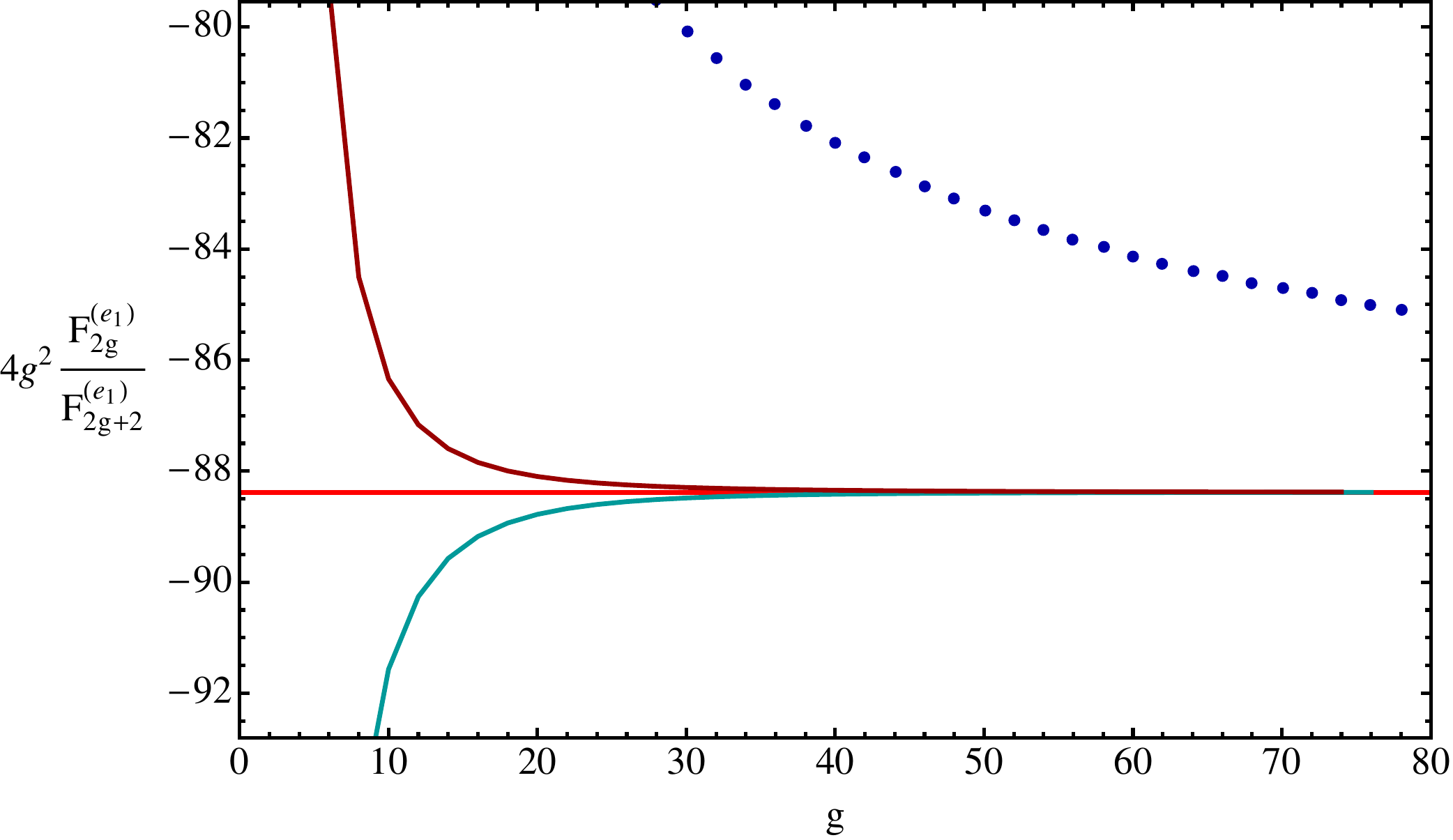}
\hspace{0.3cm}
\includegraphics[scale=0.36]{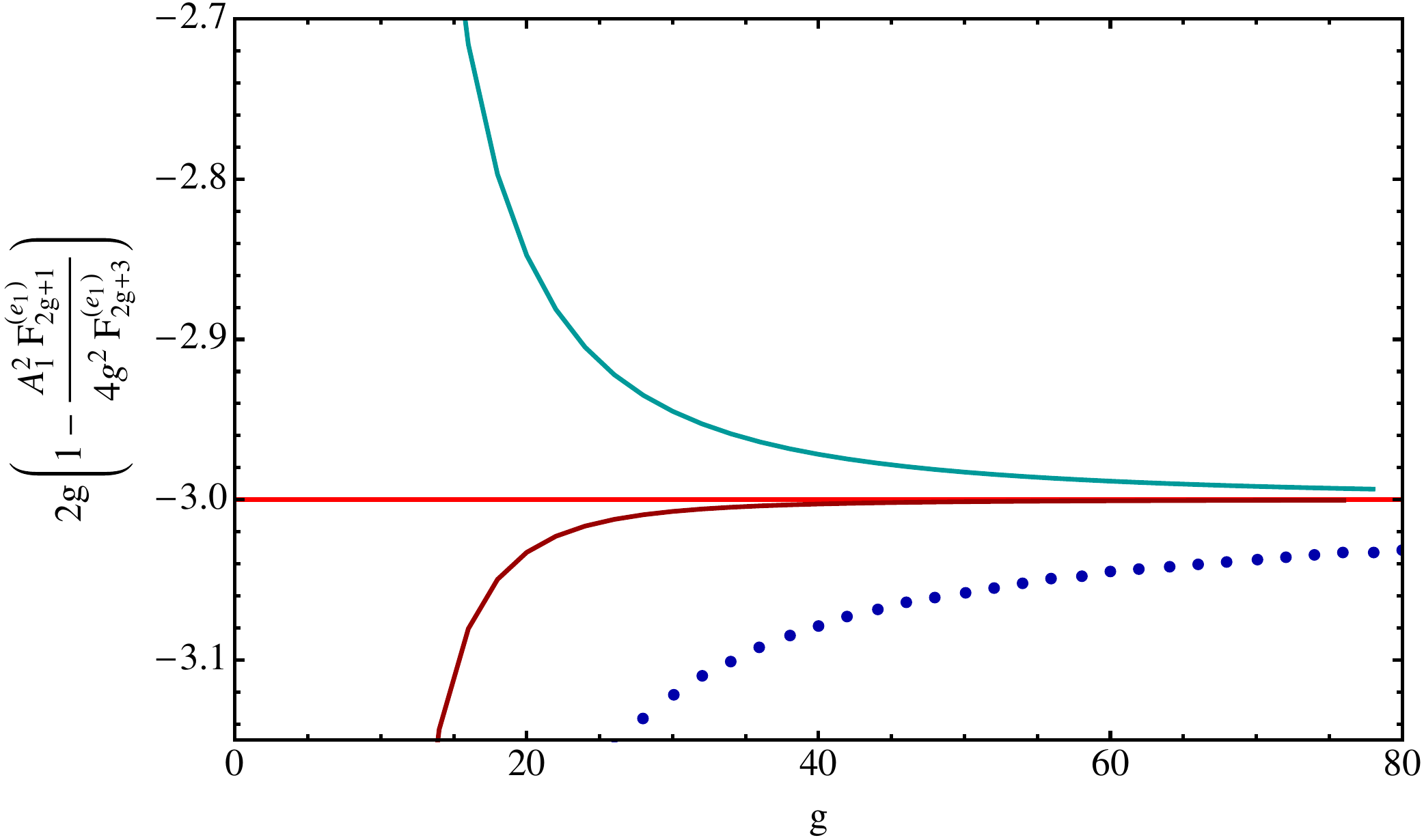}
\end{center}
\vspace{-0.5\baselineskip}
\caption{On the left, we plot a test of \eqref{eq:Asquarefrom1sector} for $\psi = 2$ and $S^{zz} = -2\, S^{zz}_{1,\hol}$. The value of the propagator can take any value except for the holomorphic one, since in that case the free energies are zero and the ratio under consideration would become indeterminate. On the right, we plot a test for \eqref{eq:cfromoneinstsector}, with the same values of $\psi$ and $S^{zz}$. It implies that $c=1$ in \eqref{eq:largeorder1sector}. On both plots, original data and two Richardson transforms are shown.}
\label{fig:instantonactionfrom1sector}
\end{figure}
%%%%%%%%%%%%%%%%%%%%%%%%%%%%%%%%%%%%%%%%%%%%%%%%%%%%%%%%%%%%%%%%%

One now expects two novelties in the large-order relations. On the one hand, addressing asymptotics of instantons, we expect to find a factorial growth in $g$, for $F^{(\bfe_1)}_g$, in comparison with $2g$ for the perturbative sector. On the other hand, as discussed earlier, we further expect to find instanton actions of both signs, $+A_1$ and $-A_1$, appearing in the large-order relation (up until now, this explicit sign had a rather limited appearance due to the merging of the symmetric sectors, $(1|0 \| \cdots )$ and $(0|1 \| \cdots )$, discussed in subsection \ref{sec:largeorderfromresurgence}). The presence of these sectors is now manifest, in the form (see also, \textit{e.g.}, \cite{asv11})
\be
F^{(\bfe_1)}_g \simeq \frac{ \G(g+c) }{ (+A_1)^{g+c} }\, \mu_0(2\bfe_1) + \frac{ \G(g+c) }{ (-A_1)^{g+c} }\,\mu_0(\bfe_{1,1}) + \cdots.
\label{eq:largeorder1sector}
\ee
\noindent
Standard resurgence relations, based on the existence of a bridge equation, would associate the term $\mu_0(2\bfe_1)$ with the free energy of the two-instanton sector $(2\bfe_1) = (2|0 \| \cdots)$, and $\mu_0(\bfe_{1,1})$ with the free energy from the mixed sector $(\bfe_{1,1}) = (1 | 1 \| 0 \cdots)$. We shall see in the remainder of this subsection that this naive expectation is seemingly fulfilled. However, there are also some subtleties which will arise upon consideration of subleading contributions to the perturbative large-order growth, and we shall comment upon these at the end of the next section. The factor $(-1)^g$ in \eqref{eq:largeorder1sector} gives rise to oscillations, but it is simple to get rid of them by looking at either even or odd values of $g$. In figure \ref{fig:instantonactionfrom1sector}, left plot, we check that the large-order growth is indeed factorial in $g$ and controlled by $A_1$ by finding a finite limit for the ratio of coefficients
\be
A_1^2 = \lim_{g\to\infty} 4g^2\, \frac{F^{(\bfe_1)}_{2g}}{F^{(\bfe_1)}_{2g+2}} = \lim_{g\to\infty} 4g^2\, \frac{F^{(\bfe_1)}_{2g+1}}{F^{(\bfe_1)}_{2g+3}}.
\label{eq:Asquarefrom1sector}
\ee
\noindent
As to the value of $c$ in \eqref{eq:largeorder1sector}, it turns out to be equal to $1$, as shown in figure \ref{fig:instantonactionfrom1sector}, right plot, using the relation
\be
-2c-1 = \lim_{g\to\infty} 2g \left( 1- \frac{A_1^2}{4g^2} \frac{F^{(\bfe_1)}_{2g+2}}{F^{(\bfe_1)}_{2g}} \right).
\label{eq:cfromoneinstsector}
\ee

%%%%%%%%%%%%%%%%%%%%%%%%%%%%%%%%%%%%%%%%%%%%%%%%%%%%%%%%%%%%%%%%%
\begin{figure}
\begin{center}
\includegraphics[scale=0.52]{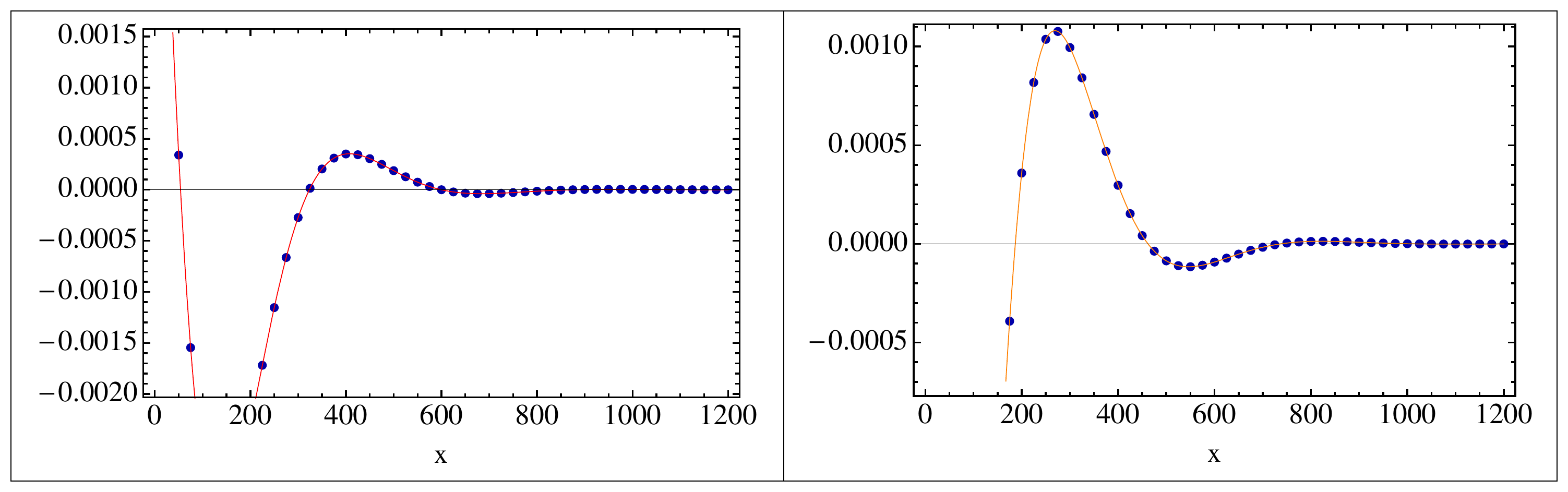}\\
\includegraphics[scale=0.52]{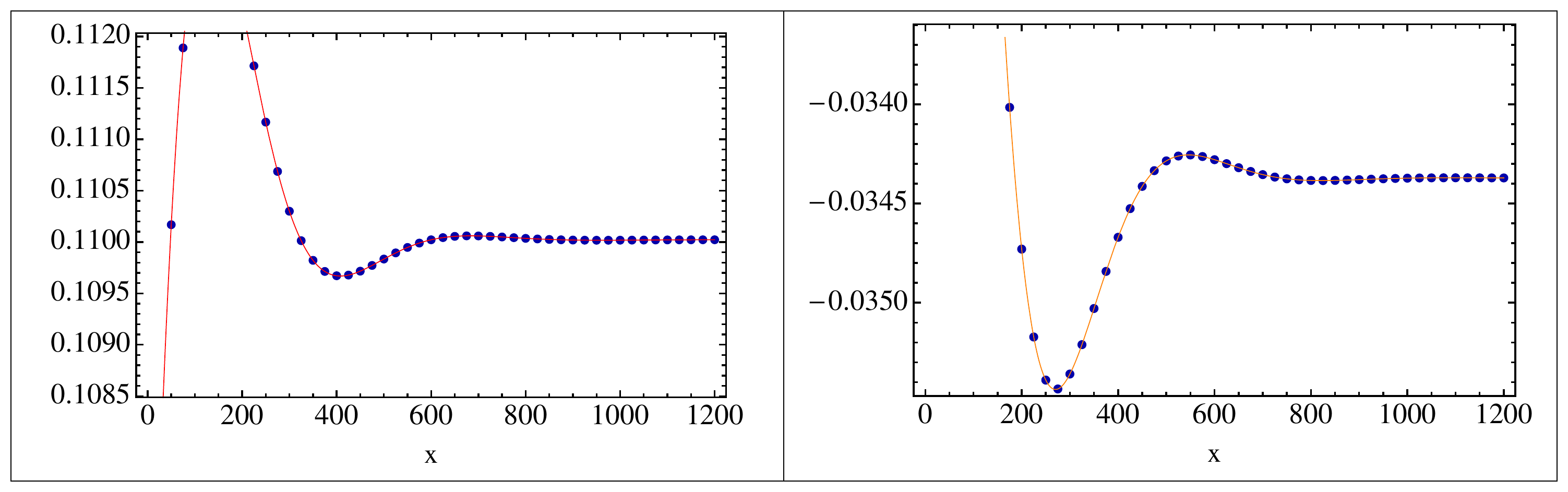}
\end{center}
\vspace{-0.5\baselineskip}
\caption{Real and imaginary parts of $\frac{S_{1,1}^2}{\left(\rmi\pi\right)^2}\, \widehat{F}^{(2\bfe_1)}_0$, following equation \eqref{eq:alt2instg0freeenergy} (top plots), and $\frac{S_{1,1}}{\rmi\pi}\, \frac{\widetilde{S}_{-1,1}}{2\pi\rmi}\, \widehat{F}^{(\bfe_{1,1})}_0$, following equation \eqref{eq:alt11instg0freeenergy} (bottom plots), for $\psi = 2\,\rme^{-\rmi\pi/36}$ and varying propagator $S^{zz} = 10^{-8} \left(1+\rmi x\right)$, compared against numerical data from the large-order growth in \eqref{eq:largeorder1sector}. As usual, the agreement between numerical data and analytical predictions is excellent.}
\label{fig:alphabetaplots}
\end{figure}
%%%%%%%%%%%%%%%%%%%%%%%%%%%%%%%%%%%%%%%%%%%%%%%%%%%%%%%%%%%%%%%%%

Doing similar types of limits we can calculate numerical data concerning the nonperturbative sectors $(2\bfe_1)$ and $(\bfe_{1,1})$ appearing in \eqref{eq:largeorder1sector}. To illustrate this, let us pick one point in moduli space and several values of $S^{zz}$---for which we show the results in figure \ref{fig:alphabetaplots}. The numerical data is very precisely reproduced by the closed form expressions\footnote{Note that since we only have access to the combination $\frac{S_{1,1}}{\rmi\pi}\, F^{(\bfe_1)}_g$, we first had to multiply \eqref{eq:largeorder1sector} by this prefactor involving the Stokes constant, $S_{1,1}$.} (the ``hatted'' notation will be clear shortly)
\bea
\frac{S_{1,1}}{\rmi\pi}\, \mu_0(2\bfe_1) &=& \frac{S_{1,1}^2}{\left(\rmi\pi\right)^2}\, \widehat{F}^{(2\bfe_1)}_0 =\\
&=& - \frac{1}{2} \left( \frac{A_1}{2\pi^2} \right)^2 \rme^{2\cdot\frac{1}{2} \left(\p_z A_1\right)^2 \left( S^{zz}-S^{zz}_{1,\hol} \right)} + \frac{1}{2} \left( \frac{A_1}{2\pi^2} \right)^2 \rme^{4\cdot\frac{1}{2} \left(\p_z A_1\right)^2 \left( S^{zz}-S^{zz}_{1,\hol} \right)}, \label{eq:alt2instg0freeenergy} \\
\frac{S_{1,1}}{\rmi\pi}\, \mu_0(\bfe_{1,1}) &=& \frac{S_{1,1}}{\rmi\pi}\, \frac{\widetilde{S}_{-1,1}}{2\pi\rmi}\, \widehat{F}^{(\bfe_{1,1})}_0 =\\
&=& + \frac{1}{2} \left( \frac{A_1}{2\pi^2} \right)^2 \rme^{2\cdot\frac{1}{2} \left(\p_z A_1\right)^2 \left( S^{zz}-S^{zz}_{1,\hol} \right)} - \frac{1}{2} \left( \frac{A_1}{2\pi^2} \right)^2. \label{eq:alt11instg0freeenergy}
\eea
\noindent
Both analytical expressions for the above nonperturbative free energies, \eqref{eq:alt2instg0freeenergy} and \eqref{eq:alt11instg0freeenergy}, are obtained from the nonperturbative holomorphic anomaly equations \eqref{eq:NPeq}, with $B(2\bfe_1,\bfe_1) = 0 = B(\bfe_{1,1},\bfe_1)$. Just as for the one-instanton sector in subsection \ref{sec:largeorderoftheperturbativesector}, their integration is straightforward, and the only remaining question concerns the fixing of the ambiguity. Now we already know that the left-hand-side of \eqref{eq:largeorder1sector} vanishes in the holomorphic limit, for $g \geq 2$, from comparison against the conifold \eqref{eq:allzeroesafterg2}. This immediately implies that the holomorphic ambiguity of the nonperturbative sectors above, $(2\bfe_1)$ and $(\bfe_{1,1})$, must be fixed by requiring a \textit{vanishing} holomorphic limit, \textit{already} at first order. It is important to note that this fixing departs from the one in subsection \ref{sec:largeorderoftheperturbativesector}, where we were fixing the ambiguity by matching against nonperturbative free energies at the conifold, and we have introduced the ``hatted'' notation above precisely to distinguish nonperturbative free energies computed out of the same nonperturbative holomorphic anomaly equations, but having their ambiguities fixed differently (either against conifold, no hat, or against zero, hatted). Indeed, the two-instanton sector of the conifold is only vanishing for $g\ge2$ \cite{ps09}, implying that the $(2\bfe_1)$ sector of local $\BC\BP^2$ does \textit{not} have its ambiguity fixed against the conifold two-instanton sector. This is not surprising: while nonholomorphically the $(\bfe_1)$ one-instanton sector of local $\BC\BP^2$ is described by an asymptotic series, this series truncates in the $S^{zz}_{1,\hol}$ holomorphic limit and the (original) large-order of its coefficients becomes meaningless. As such, whatever two-instanton sector controlled the large-order growth away from holomorphicity, $\mu_0(2\bfe_1)$, it must vanish in the holomorphic limit. This could imply that the transseries is likely including more sectors than naively expected, including sectors associated to different (allowed) possibilities of ambiguity fixing, and we shall return to this point later on.

The numerical analysis further supports that the ambiguity should be fixed by requiring a vanishing limit at \textit{all} loop orders as we shall see next (leading up to \eqref{eq:extendedlargorder1sector}). Being more specific about these other orders, note that the free energies $\widehat{F}^{(2\bfe_1)}_h$, for general $h$, depend on two exponentials (with exponents 2 and 4 as in \eqref{eq:alt2instg0freeenergy}) which are multiplied by polynomials in the propagator of the same degree \cite{cesv13},
\be
\widehat{F}^{(2\bfe_1)}_h = \rme^{2\,\frac{1}{2} \left(\p_z A_1\right)^2 \left( S^{zz}-S^{zz}_{1,\hol} \right)}\, \text{Pol} \left(S^{zz};3h\right) + \rme^{4\,\frac{1}{2} \left(\p_z A_1\right)^2 \left( S^{zz}-S^{zz}_{1,\hol} \right)}\, \text{Pol} \left(S^{zz};3h\right).
\ee
\noindent
On the other hand, the mixed sector $\widehat{F}^{(\bfe_{1,1})}_h$ is zero for odd $h$. This is a consequence of resonance ($A_1+(-A_1) = 0$), and has been seen in other examples such as Painlev\'e equations or the quartic matrix model \cite{asv11,sv13} (but see also \cite{cesv13} for a general discussion and results in the context of the holomorphic anomaly equation). The nonzero free energies have the general form, for $h$ even,
\be
\widehat{F}^{(\bfe_{1,1})}_h = \rme^{2\,\frac{1}{2} \left(\p_z A_1\right)^2 \left( S^{zz}-S^{zz}_{1,\hol} \right)}\,\text{Pol}\left(S^{zz};5\frac{h}{2}\right) + \text{Pol}\left(S^{zz};3\frac{h}{2}-1\right).
\ee
\noindent
A generic description of the propagator structure of the free energies is explained in more detail in \cite{cesv13} (and also see appendix \ref{ap:structure}). We can now check the validity of these higher-loop free energies, as they appear when considering \eqref{eq:largeorder1sector} fully nonholomorphic
\be
F^{(\bfe_1)}_g \simeq \sum_{h=0}^{+\infty} \left\{ \frac{ \G(g+1-h) }{ (+A_1)^{g+1-h} }\, \frac{S_{1,1}}{\rmi\pi}\, \widehat{F}^{(2\bfe_1)}_h + \frac{ \G(g+1-h) }{ (-A_1)^{g+1-h} }\, \frac{\widetilde{S}_{-1,1}}{2\pi\rmi}\, \widehat{F}^{(\bfe_{1,1})}_h \right\} + \cdots.
\label{eq:extendedlargorder1sector}
\ee
\noindent
For instance, one can do the exercise of computing the partial sums of a slight rearrangement of \eqref{eq:extendedlargorder1sector},
\be
\frac{ A_1^{g+1} }{ \G(g+1) }\, \frac{S_{1,1}}{\rmi\pi}\, F^{(\bfe_1)}_g \simeq \sum_{h=0}^{h^*} \frac{\G(g+1-h)}{\G(g+1)}A_1^h \left\{ \frac{S_{1,1}}{\rmi\pi}\, \frac{S_{1,1}}{\rmi\pi}\,  \widehat{F}^{(2\bfe_1)}_h + (-1)^{g+1-h}\,\frac{S_{1,1}}{\rmi\pi}\, \frac{\widetilde{S}_{-1,1}}{2\pi\rmi}\, \widehat{F}^{(\bfe_{1,1})}_h \right\}.
\label{eq:1sectorpartialsums}
\ee
\noindent
For large values of $h^*$ the series is expected to start diverging asymptotically but, before that happens, and having in mind the mechanism of optimal truncation, the partial sums will still give a good approximation to the left-hand-side. For example, fixing genus $g=75$, $\psi = 2$ and $S^{zz}=10^{-8}$, we find that the difference between the left and right-hand sides of \eqref{eq:1sectorpartialsums} becomes smaller and smaller as $h^*$ increases (recall that $\widehat{F}^{(\bfe_{1,1})}_h$ is zero when $h$ is odd, and that is why the difference between the left and right-hand sides of \eqref{eq:1sectorpartialsums} reduces drastically for $h^*$ even):
\begin{center}
\begin{tabular}{lll}
$h^*$ & RHS & LHS $-$ RHS \\
\hline\hline
0     & $0.112\,257\,517\,800$ & $+8\cdot 10^{-4}$ \\
1     & $0.113\,083\,826\,046$ & $-3\cdot 10^{-5}$ \\
2     & $0.113\,054\,511\,927$ & $+3\cdot 10^{-9}$ \\
3     & $0.113\,054\,512\,813$ & $+2\cdot 10^{-9}$ \\
4     & $0.113\,054\,514\,589$ & $-7\cdot 10^{-11}$ \\
5     & $0.113\,054\,514\,517$ & $+1\cdot 10^{-12}$ \\
\hline\hline
LHS & $0.113\,054\,514\,518$ & 
\end{tabular}
\end{center}

%%%%%%%%%%%%%%%%%%%%%%%%%%%%%%%%%%%%%%%%%%%%%%%%%%%%%%%%%%%%%%%%%
%%%%%%%%%%%%%%%%%%%%%%%%%%%%%%%%%%%%%%%%%%%%%%%%%%%%%%%%%%%%%%%%%
\section{Exponentially Subleading Contributions and Resummation}\label{sec:resummation}
%%%%%%%%%%%%%%%%%%%%%%%%%%%%%%%%%%%%%%%%%%%%%%%%%%%%%%%%%%%%%%%%%
%%%%%%%%%%%%%%%%%%%%%%%%%%%%%%%%%%%%%%%%%%%%%%%%%%%%%%%%%%%%%%%%%

\begin{figure}[t!]
\begin{center}
\includegraphics[scale=0.40]{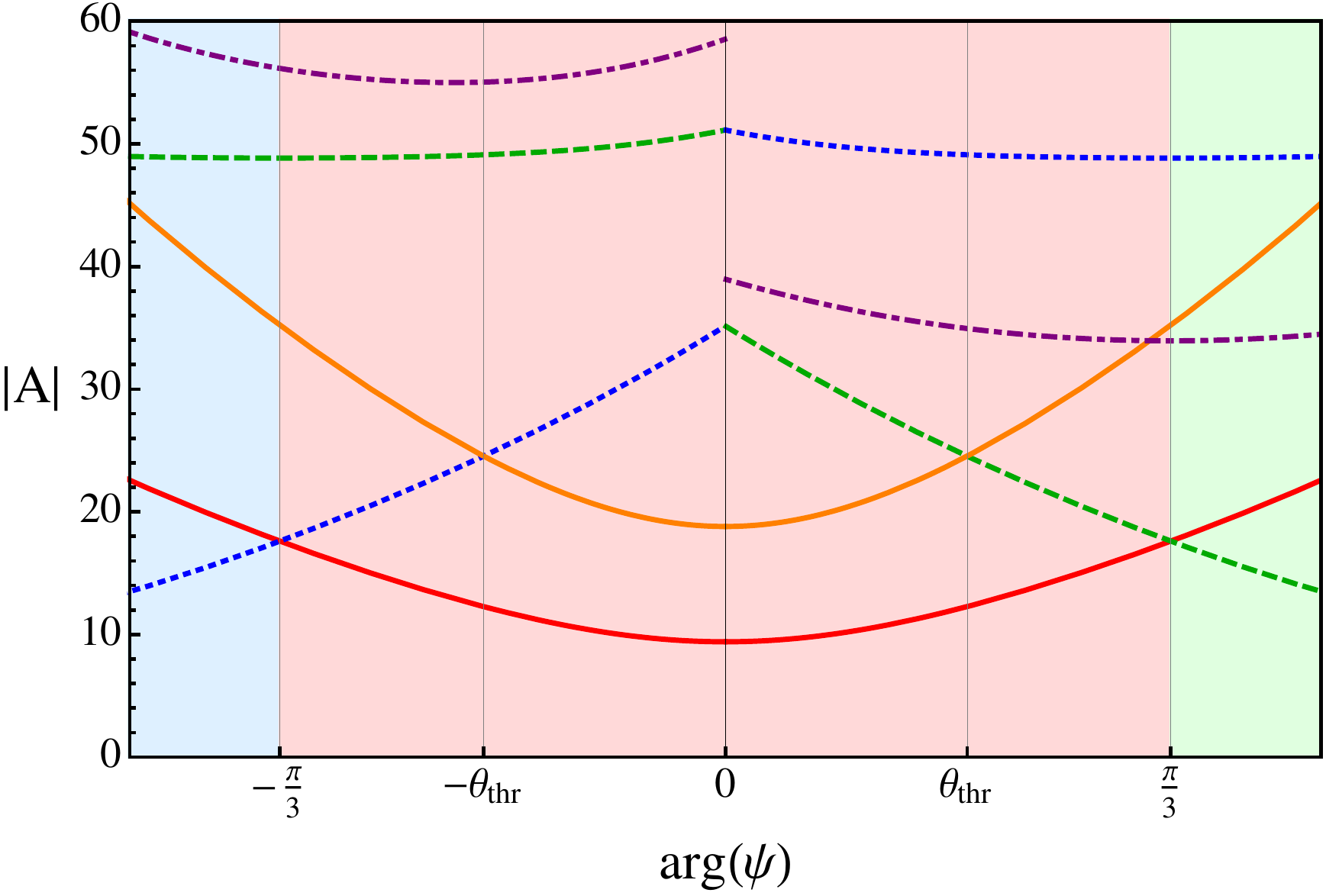}
\hspace{0.3cm}
\includegraphics[scale=0.40]{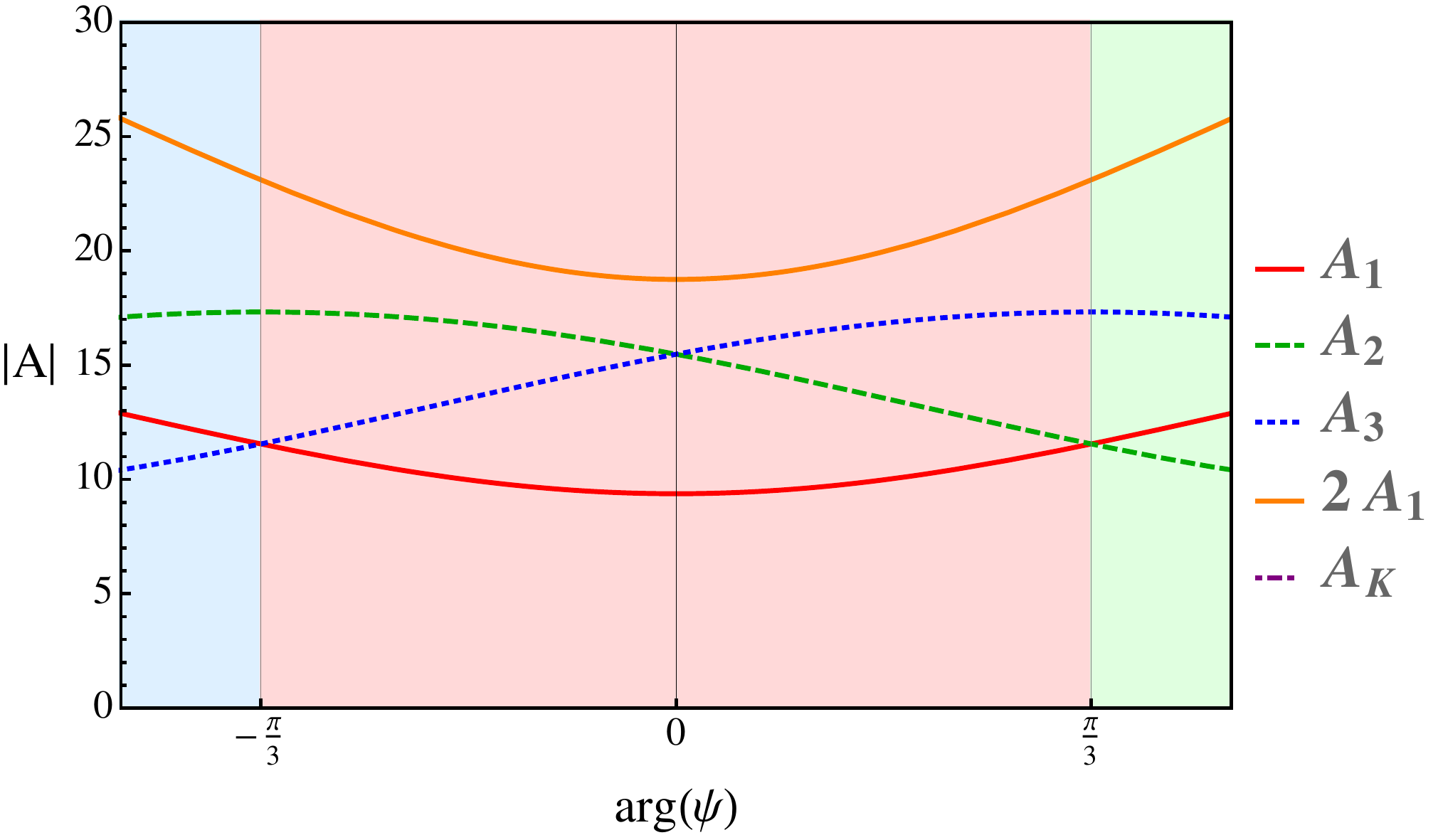} 
\end{center}
\vspace{-1\baselineskip}
\caption{The absolute values of the three conifold actions, and also of $2A_1$, for fixed $| \psi |$ and varying argument. On the left, $| \psi | = 2$, while on the right, $| \psi | = \frac{1}{4}$. If $| \psi |$ is big enough, there exists a value $\theta_{\text{thr}}$ such that for $\arg(\psi)> \th_{\text{thr}}$, $| A_2| < | 2 A_1 |$, and for $\arg(\psi)< -\th_{\text{thr}}$, $| A_3| < | 2 A_1 |$. Otherwise, $|2A_1|$ is always greater than the two other instanton actions. For $| \psi | = 2$ we have also depicted the large-radius instanton action, $A_{\text{K}}$, which is always subleading from the standpoint of the present subsection (we have not included this large-radius action for $| \psi | = \frac{1}{4}$ as, for this value of the modulus, it has already moved into another Riemann sheet; see subsection \ref{sec:largeradiusactions}).}
\label{fig:instantonactionscomparisonplots}
\end{figure}

In subsection \ref{sec:largeorderoftheperturbativesector}, we studied the main contribution to the large-order growth of the perturbative free energies, and we saw that at small values of the modulus it is controlled by the conifold $A_1$ and its associated instanton-sector coefficients, $F^{(\bfe_1)}_h$. There are, in principle, infinitely many subleading contributions in \eqref{eq:perturbativelargeorder1sector} coming from other sectors of the transseries. The most relevant ones will be those with the smallest instanton action, in absolute value, because they will be less suppressed as $g$ goes to infinity (the large-order limit). Focusing on the other conifold instanton actions, $A_2$ and $A_3$, note how they appear in \eqref{eq:generalperturbativelargeorder} with (multi-instanton) dependence $( k A_i )^{-2g}$. This implies that, depending on the point in moduli space\footnote{For the moment we will have in mind small values of the modulus, such that the large-radius instanton action associated to the K\"ahler parameter never plays a role in our subsequent analysis within subsections \ref{sec:holomorphiccase} and \ref{sec:nonholomorphiccase}.}, we can have different situations:
\bea
&&| A_1 | < | A_2 | < | 2A_1 | < | A_3 | < \cdots, \\
&&| A_1 | < | A_3 | < | 2A_1 | < | A_2 | < \cdots, \\
&&| A_1 | < | 2A_1 | < | A_{2,3} | <  \cdots, \\
&&| A_1 | < | A_2 | < | A_3 | < \cdots. 
\eea
\noindent
One example where we can see the first three situations is when we fix the absolute value of $\psi$ to, say, $| \psi | = 2$, and vary its argument. We illustrate in figure \ref{fig:instantonactionscomparisonplots} how in different regions one finds different possibilities. We shall study these cases, first in the holomorphic limit (with respect to the first conifold point), and later allowing the propagator to take any values. We will distinguish between these two situations as the latter involves a resummation of a divergent series---the complete leading one-instanton contribution to the large-order growth---, while in the former the series truncates to just two terms and no resummation is required.

%%%%%%%%%%%%%%%%%%%%%%%%%%%%%%%%%%%%%%%%%%%%%%%%%%%%%%%%%%%%%%%%%
%%%%%%%%%%%%%%%%%%%%%%%%%%%%%%%%%%%%%%%%%%%%%%%%%%%%%%%%%%%%%%%%%

%%%%%%%%%%%%%%%%%%%%%%%%%%%%%%%%%%%%%%%%%%%%%%%%%%%%%%%%%%%%%%%%%
\subsection{Analysis in the Holomorphic Limit}
\label{sec:holomorphiccase}
%%%%%%%%%%%%%%%%%%%%%%%%%%%%%%%%%%%%%%%%%%%%%%%%%%%%%%%%%%%%%%%%%

Throughout, whenever we mention the holomorphic limit we shall implicitly be referring to the frame of the \textit{first} conifold point; which means $S^{zz} \to S^{zz}_{1,\hol}$. As discussed earlier, in this limit the $(\bfe_1)$ free energies are nonzero only for $h=0$ and $h=1$, that is,
\be
\frac{S_{1,1}}{\rmi\pi}\, \CF^{(\bfe_1)}_0 = \frac{A_1}{2\pi^2}, \qquad \frac{S_{1,1}}{\rmi\pi}\, \CF^{(\bfe_1)}_1 = \frac{1}{2\pi^2},
\ee
\noindent
with all following ones vanishing. This means that \eqref{eq:perturbativelargeorder1sector} reduces to the much simpler
\be
\CF^{(\bf0)}_g \simeq \sum_{h=0}^{+1} \frac{ \G(2g-1-h) }{ A_1^{2g-1-h} }\, \frac{S_{1,1}}{\rmi\pi}\, \CF^{(\bfe_1)}_h + \cdots,
\label{eq:1insttruncated}
\ee
\noindent
where the dots indicate the subleading terms we wish to study in the following. For the moment, notice the finite upper bound in the sum.

\begin{figure}[t]
\begin{center}
\includegraphics[scale=0.35]{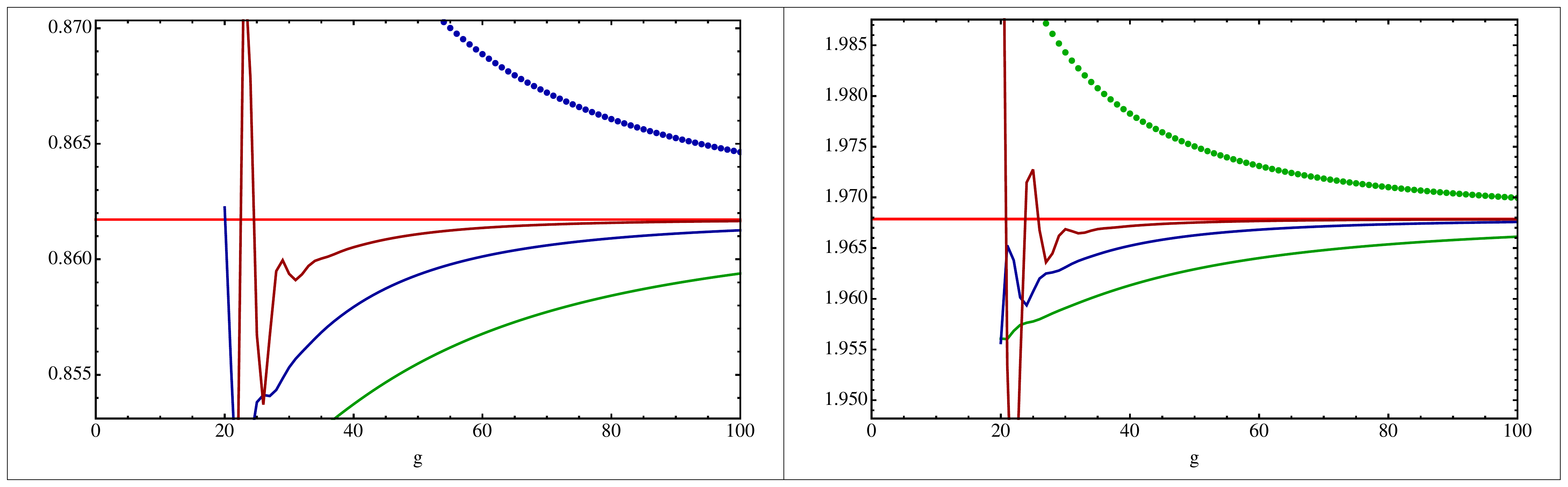}\\
\vspace{-1\baselineskip}
{\scriptsize
\bea
\left(\frac{A_2}{A_1}\right)^2 &=& 0.861\,721\,3+1.967\,864\,55\,\rmi  \nonumber\\
\text{5 Richardson Transforms} &=& 0.861\,720\,5+1.967\,864\,37\,\rmi \nonumber
\eea
}
\end{center}
\vspace{-1\baselineskip}
\caption{Numerical value for the limit \eqref{eq:limitratiocurlyX}, with $\psi = 2\,\rme^{\rmi\pi/4}$. We plot both real and imaginary parts, along with the first Richardson transforms. The agreement is better than one part in $10^6$.}
\label{fig:Xratiocase2}
\end{figure}

\begin{figure}[ht!]
\begin{center}
\includegraphics[scale=0.35]{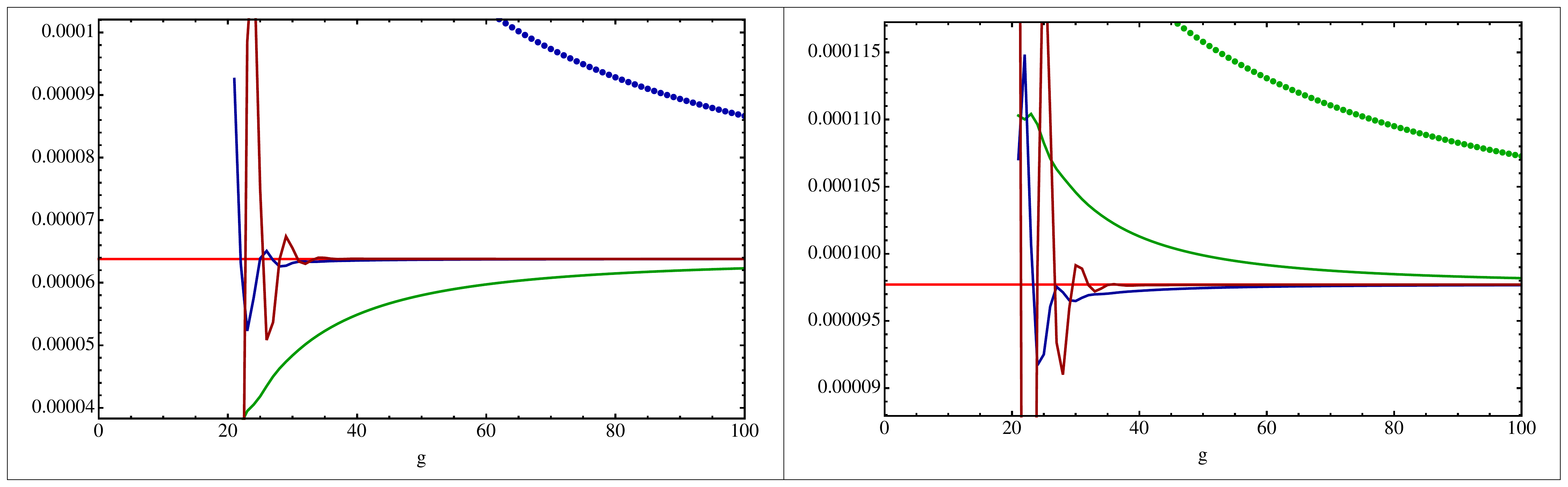}
\vspace{-0.1cm}
{\scriptsize
\bea
\frac{S_{1,2}}{\rmi\pi}\,F^{(\bfe_2)}_0 &=& 0.000\,063\,797\,45+0.000\,097\,703\,4\,\rmi  \nonumber\\
\text{3 Richardson Transforms} &=& 0.000\,063\,797\,90+0.000\,097\,702\,4\,\rmi \nonumber
\eea
}
\includegraphics[scale=0.35]{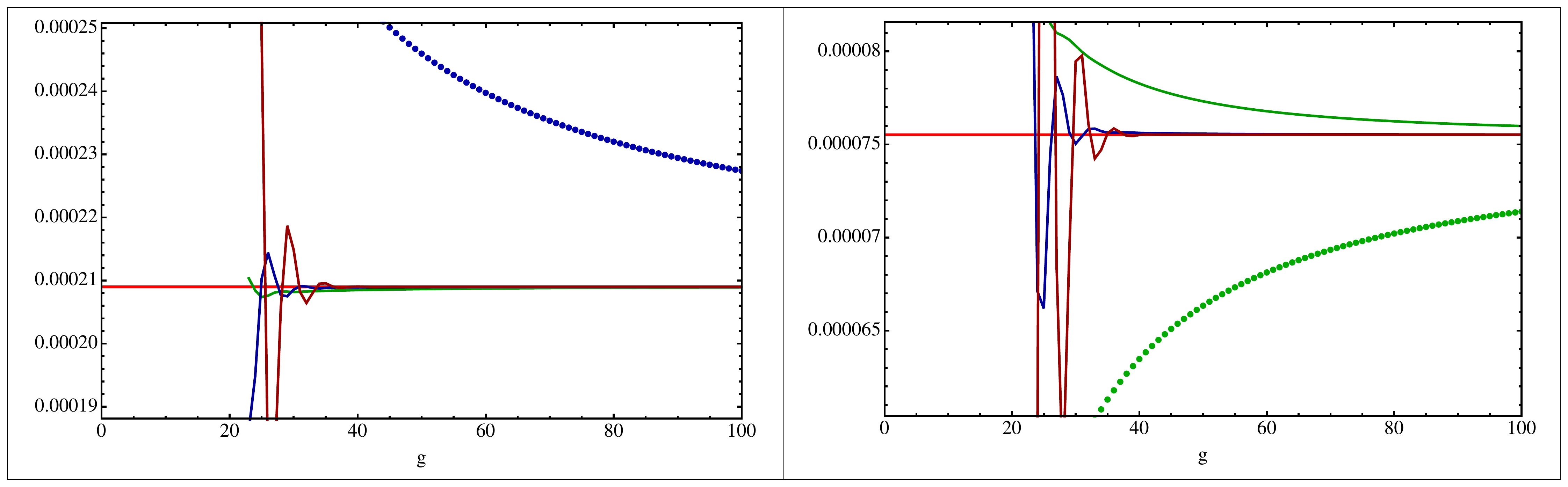}
\vspace{-0.2cm}
{\scriptsize
\bea
\frac{S_{1,2}}{\rmi\pi}\,F^{(\bfe_2)}_1 &=& 0.000\,209\,011\,81+0.000\,075\,524\,080\,\rmi  \nonumber\\
\text{3 Richardson Transforms} &=& 0.000\,209\,011\,64+0.000\,075\,524\,095\,\rmi \nonumber
\eea
}
\includegraphics[scale=0.35]{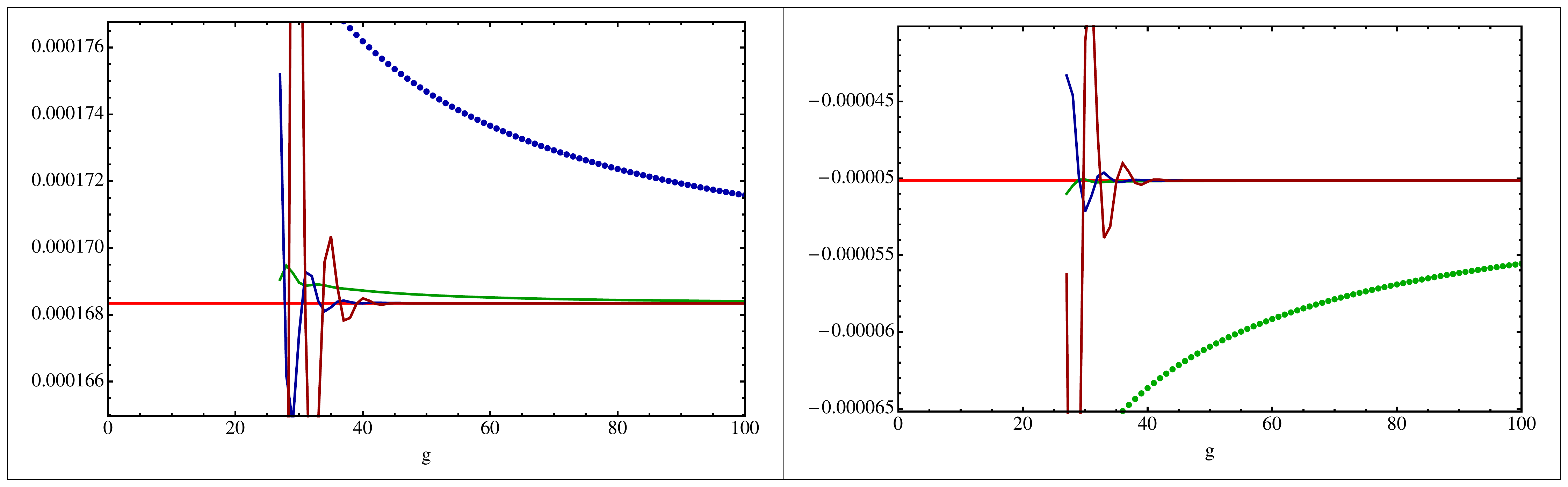}
\vspace{-0.2cm}
{\scriptsize
\bea
\frac{S_{1,2}}{\rmi\pi}\,F^{(\bfe_2)}_2 &=& 0.000\,168\,340\,320-0.000\,050\,145\,41\,\rmi  \nonumber\\
\text{3 Richardson Transforms} &=& 0.000\,168\,340\,333-0.000\,050\,145\,39\,\rmi \nonumber
\eea
}
\end{center}
\vspace{-1\baselineskip}
\caption{$(\bfe_2)$ free energy coefficients, at $\psi = 2\,\rme^{\rmi\pi/4}$, up to three loops. Numerical results after several Richardson transforms are compared to analytic expressions such as \eqref{eq:holoother1sectorh0} and \eqref{eq:holoother1sectorhH}.}
\label{fig:otheroneinsthololargeorder}
\end{figure}

Let us first focus on a region where strictly
\be
| A_1 | < | A_2 | < \cdots.
\ee
\noindent
As can be seen in figure \ref{fig:instantonactionscomparisonplots}, left plot, this happens if $\arg(\psi) > \theta_{\text{thr}}$. Let us define
\be
\CX^{(\bfe_1)}_g := \frac{A_1^{2g-1}}{\G(2g-1)} \left\{ \CF^{(\bf0)}_g - \sum_{h=0}^{+1} \frac{ \G(2g-1-h) }{ A_1^{2g-1-h} }\, \frac{S_{1,1}}{\rmi\pi}\, \CF^{(\bfe_1)}_h \right\},
\label{eq:holoXdef}
\ee
\noindent
where the $\CX^{(\bfe_1)}_g$ coefficients precisely probe the exponentially subleading corrections to the perturbative large-order behavior we are interested in analyzing. One can first check numerically that these quantities decrease exponentially with $g$, as $(A_1/A_2)^{2g-1}$. This is simply verified by comparing against the numerical limit
\be
\lim_{g\to\infty} \frac{\CX^{(\bfe_1)}_g}{\CX^{(\bfe_1)}_{g+1}} = \left( \frac{A_2}{A_1} \right)^2, 
\label{eq:limitratiocurlyX}
\ee
\noindent
as shown in figure \ref{fig:Xratiocase2}. This is also the first direct check on the value of $A_2$, as computed via \eqref{eq:Aitci}. Next, we can go further in the numerical analysis and find the \textit{asymptotic} behavior for the coefficients \eqref{eq:holoXdef}
\be
\CX^{(\bfe_1)}_g \simeq \left( \frac{A_1}{A_2} \right)^{2g-1} \sum_{h=0}^{+\infty} \frac{\G(2g-1-h)}{\G(2g-1)}\, \frac{S_{1,2}}{\rmi\pi}\, A_2^h\, F^{(\bfe_2)}_h (z,S^{zz}_{1,\hol}) + \cdots,
\label{eq:holoXother1sector}
\ee
\noindent
which is controlled by $F^{(\bfe_2)}_h (z,S^{zz}_{1,\hol})$. These are the free energies for the $(\bfe_2)$ sector, the one-instanton of the \textit{second} conifold point, computed out of the holomorphic anomaly equations similarly to what we did earlier for the $(\bfe_1)$ sector, and evaluated in the holomorphic limit of the \textit{first} conifold-point frame. Recall that $(\bfe_2) = (0 | 0 \| 1 | 0 \| 0 \ldots)$ and, for example,
\be
\frac{S_{1,2}}{\rmi\pi}\, F^{(\bfe_2)}_0 (z,S^{zz}_{1,\hol}) = \frac{A_2}{2\pi^2}\, \rme^{\frac{1}{2} \left(\p_z A_2\right)^2 \left(S^{zz}_{1,\hol}-S^{zz}_{2,\hol}\right)}.
\label{eq:holoother1sectorh0}
\ee
\noindent
In fact, \textit{all} these free energies are nonzero
\be
F^{(\bfe_2)}_h (z,S^{zz}_{1,\hol}) = \frac{\rmi\pi}{S_{1,2}}\, \rme^{\frac{1}{2} \left(\p_z A_2\right)^2 \left(S^{zz}_{1,\hol}-S^{zz}_{2,\hol}\right)}\, \text{Pol} \left(S^{zz}_{1,\hol}; 3h\right),
\label{eq:holoother1sectorhH}
\ee
\noindent
and figure \ref{fig:otheroneinsthololargeorder} shows very precise numerical checks, out of \eqref{eq:holoXother1sector}, of these free energies up to three loops. One important point to realize is that these coefficients grow \textit{factorially} fast. This explicitly shows that even in a holomorphic limit (in this case, the specific limit $S^{zz} \to S^{zz}_{1,\hol}$) there still are nontrivial asymptotics of instantons, albeit, of course, of instanton sectors \textit{other} than the ones associated to the chosen holomorphic frame. It is simple to see how this occurs. First, note that \eqref{eq:holoother1sectorhH} is formally equivalent to $F^{(\bfe_1)}_h$ in \eqref{eq:fixed1sector}, once we exchange first and second conifold points. These coefficients $F^{(\bfe_1)}_h$ were shown to be asymptotic according to \eqref{eq:largeorder1sector}, for every value of the propagator except $S^{zz}_{1,\hol}$, implying that the coefficients $F^{(\bfe_2)}_h$ will be asymptotic for all values of the propagator except $S^{zz}_{2,\hol}$. But this is exactly the case: the implicit holomorphic limit in \eqref{eq:holoother1sectorhH} is with respect to the first and \textit{not} the second conifold point. 

Needless to mention, a similar calculation may be done with respect to $A_3$ by going to a suitable point in moduli space where strictly $| A_1 | < | A_3 | < \cdots$.

%%%%%%%%%%%%%%%%%%%%%%%%%%%%%%%%%%%%%%%%%%%%%%%%%%%%%%%%%%%%%%%%%
%%%%%%%%%%%%%%%%%%%%%%%%%%%%%%%%%%%%%%%%%%%%%%%%%%%%%%%%%%%%%%%%%

%%%%%%%%%%%%%%%%%%%%%%%%%%%%%%%%%%%%%%%%%%%%%%%%%%%%%%%%%%%%%%%%%
%%%%%%%%%%%%%%%%%%%%%%%%%%%%%%%%%%%%%%%%%%%%%%%%%%%%%%%%%%%%%%%%%

\begin{figure}[t]
\begin{minipage}[c]{0.48\textwidth}
\centering
\includegraphics[scale=0.40]{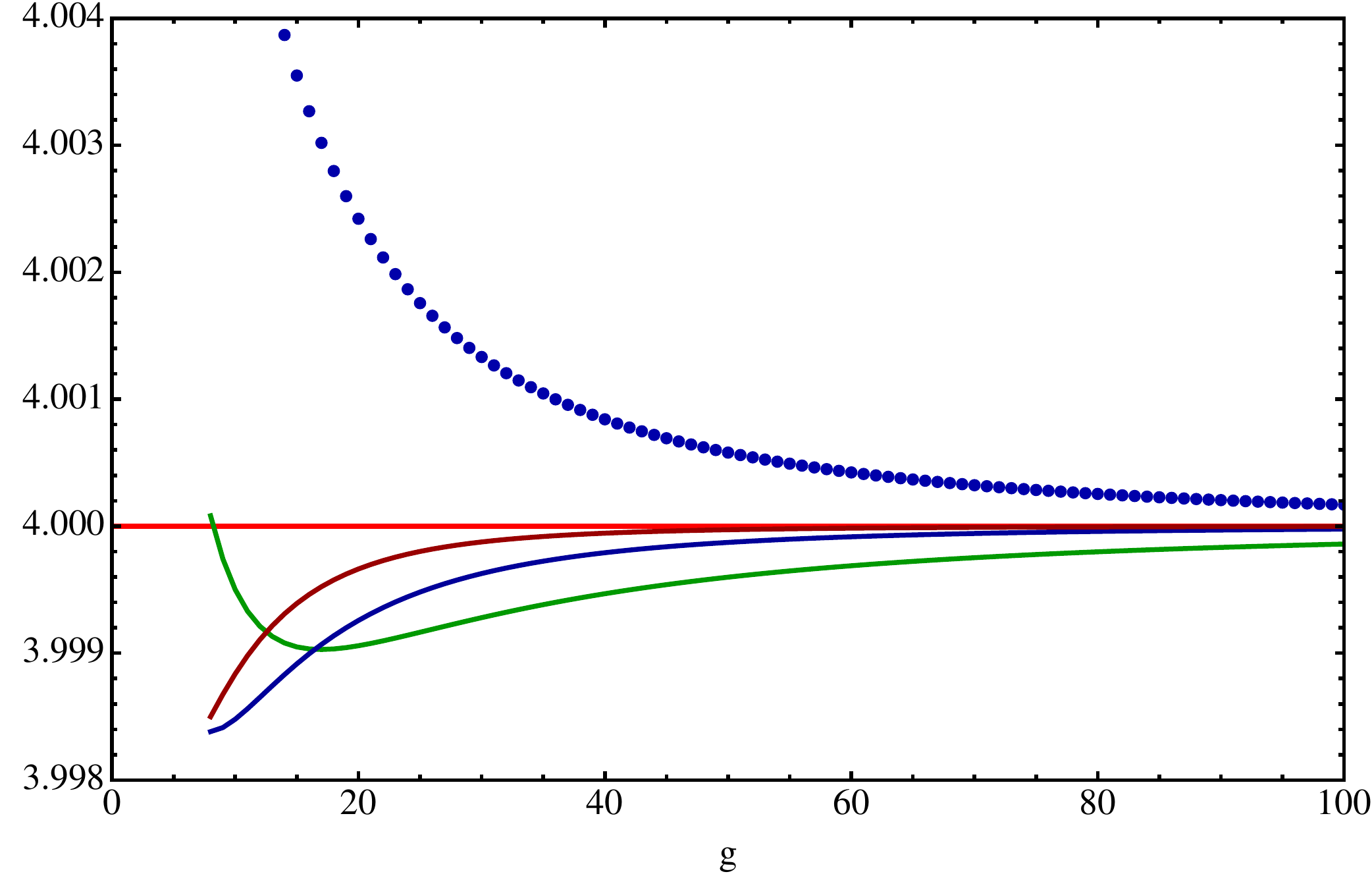}
\end{minipage}  
\begin{minipage}[c]{0.48\textwidth}
{\scriptsize
\bea
\left(\frac{2A_1}{A_1}\right)^2 &=& 4.000\,000\,0  \nonumber\\
\text{3 Richardson Transforms} &=& 3.999\,997\,3 \nonumber
\eea
}
\end{minipage}
\vspace{-1\baselineskip}
\caption{Numerical limit of the ratio of consecutive $\CX^{(\bfe_1)}_g$, for $\psi = 2$, testing the two-instanton action. The theoretical value is precisely $2^2 = 4$, and the large-order agreement is excellent.}
\label{fig:Xratiocase1}
\end{figure}

\begin{figure}[t]
\begin{center}
\includegraphics[scale=0.36]{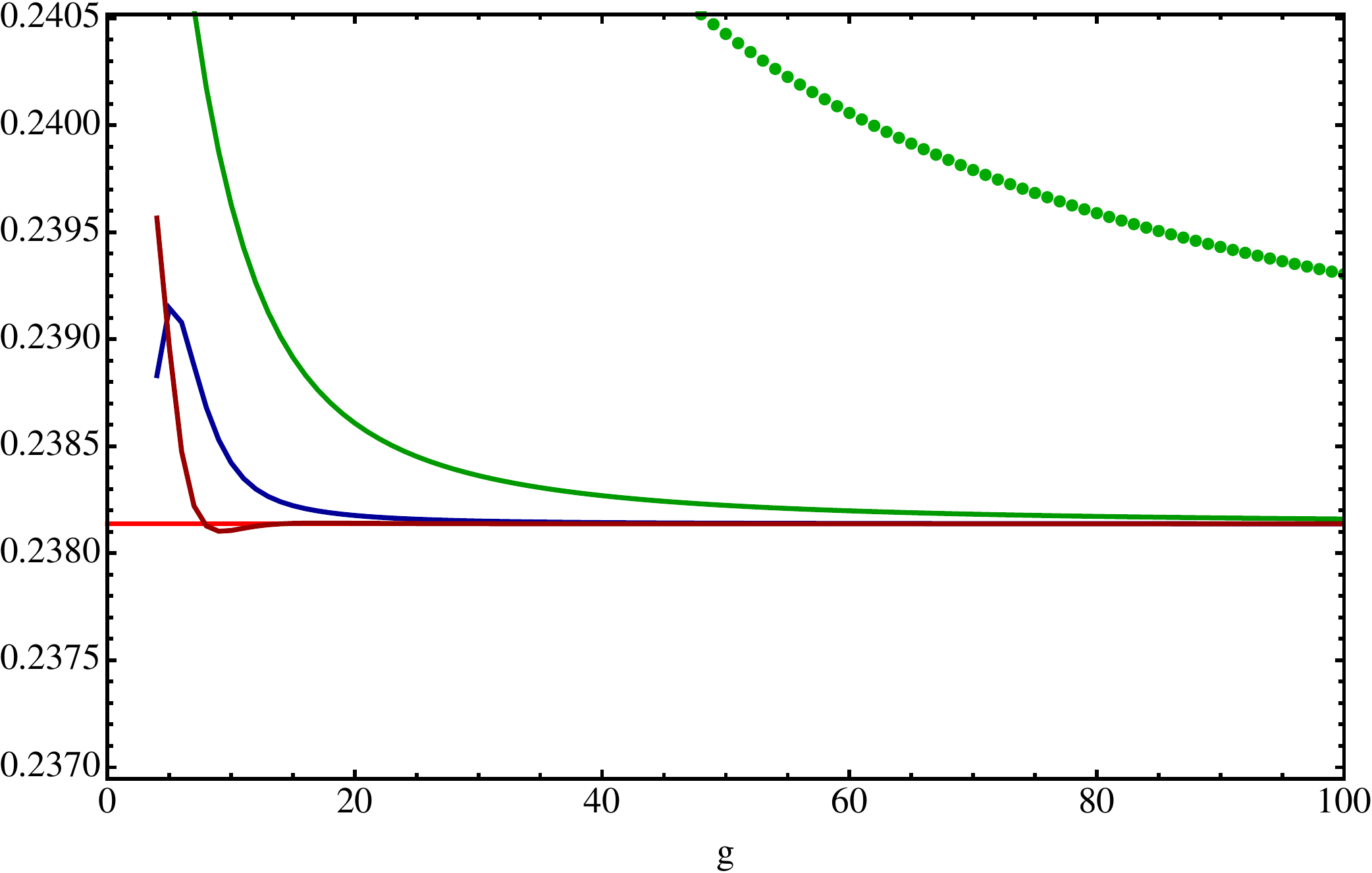} 
\hspace{0.3cm} 
\includegraphics[scale=0.37]{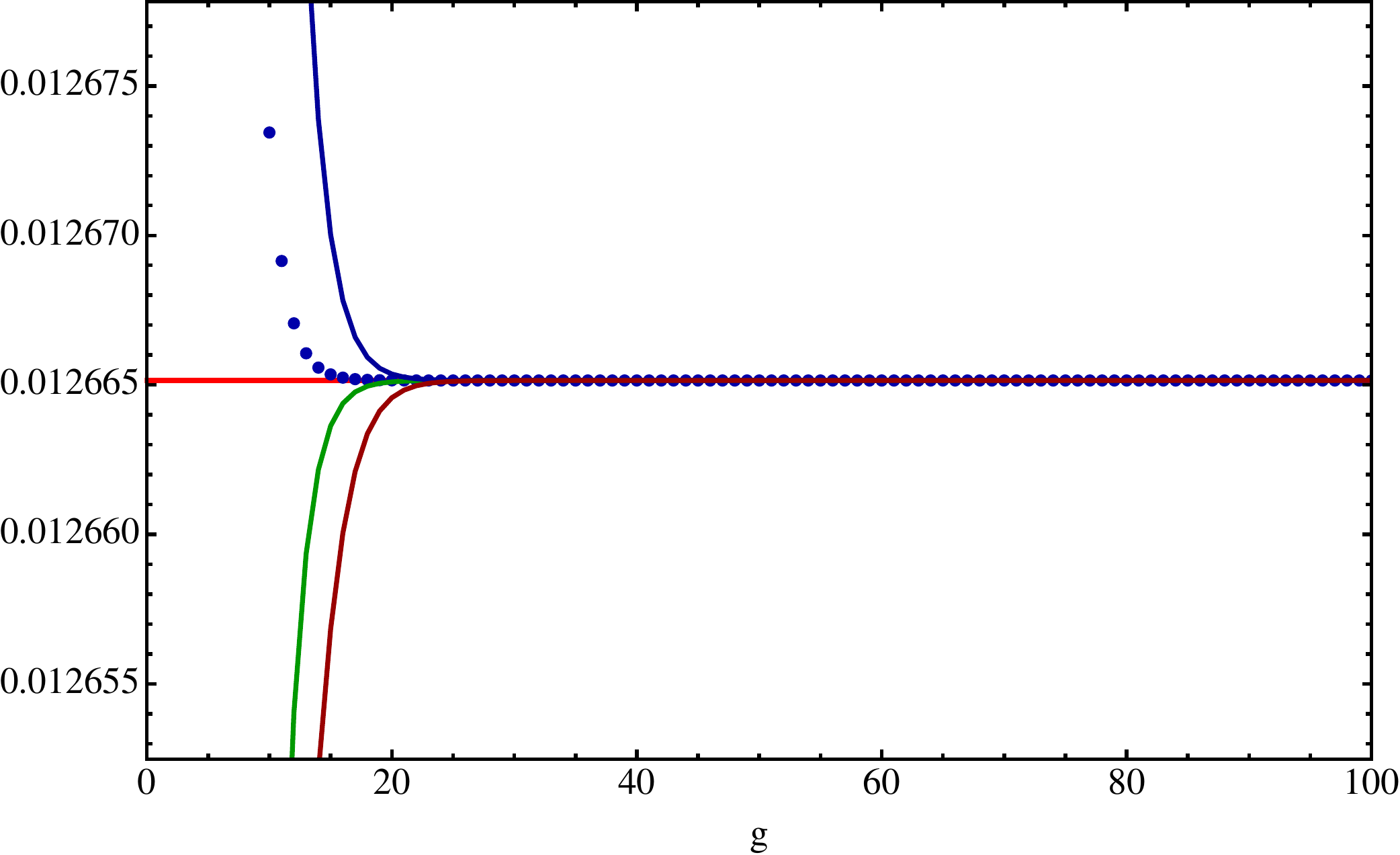}  
{\scriptsize
\begin{align}
\frac{S_{1,1}^2}{\rmi\pi}\,\widetilde{\CF}^{(2\bfe_1)}_0 &= 0.238\,137\,202\,745\,\rmi  & \frac{S_{1,1}^2}{\rmi\pi}\,\widetilde{\CF}^{(2\bfe_1)}_0 &=  0.012\,665\,147\,955   \nonumber\\
\text{3 Richardson Transforms} &= 0.238\,137\,208\,909\,\rmi & \text{No Richardson Transforms} &= 0.012\,665\,147\,955 \nonumber
\end{align}
}
\end{center}
\vspace{-1\baselineskip}
\caption{Numerical tests of the holomorphic free-energy coefficients \eqref{eq:holo2sector}, for the $(2\bfe_1)$ sector, with $\psi = 2$. Numerical results after several or no Richardson transforms are compared against the analytic expressions, with impressive agreement. All higher-loop coefficients are zero.}
\label{fig:twoinstantonholo}
\end{figure}

Let us next go to a point in moduli space where $-\theta_{\text{thr}} < \arg(\psi) < + \theta_{\text{thr}}$, so that it is instead $2A_1$ which controls the exponentially subleading contribution to the perturbative sector, see figure \ref{fig:instantonactionscomparisonplots}. The free energies that will now appear in the large-order growth of $\CX^{(\bfe_1)}_g$ are expected to arise from a two-instanton sector associated to $A_1$, but one finds that they actually \textit{truncate} in this case; \textit{i.e.}, there are only two of them and they coincide with the nonperturbative free energies of the conifold model \cite{ps09}. 
Explicitly, the large-order is found to be
\be
\CX^{(\bfe_1)}_g \simeq \frac{1}{2^{2g-1}} \sum_{h=0}^{+1} \frac{\G(2g-1-h)}{\G(2g-1)}\, \frac{S_{1,1}^2}{\rmi\pi} \left(2A_1\right)^h \widetilde{\CF}^{(2\bfe_1)}_h + \cdots,
\ee
\noindent
where
\be
\frac{S_{1,1}^2}{\rmi\pi}\, \widetilde{\CF}^{(2\bfe_1)}_0 = \frac{1}{2}\, \frac{A_1}{2\pi^2}, \qquad \frac{S_{1,1}^2}{\rmi\pi}\, \widetilde{\CF}^{(2\bfe_1)}_1 = \frac{1}{2^2}\, \frac{1}{2\pi^2}.
\label{eq:holo2sector}
\ee
\noindent
Note that we have now represented these free energies with a tilde, as they do \textit{not} correspond to the holomorphic limit of the two-instanton sector $\widehat{F}^{(2\bfe_1)}_h$ in \eqref{eq:alt2instg0freeenergy} (recall that that holomorphic limit vanished at all orders). One thus finds more than one two-instanton sector, and we shall comment more on this point in the following subsections. For the moment let us just mention that the nonholomorphic extension of $\widetilde{\CF}^{(2\bfe_1)}_h$ turns out to be a \textit{composite} object which can be written as the sum of two free energies. The numerical checks for the dominant instanton action, and for these free energies, are displayed in figures \ref{fig:Xratiocase1} and \ref{fig:twoinstantonholo}.

%%%%%%%%%%%%%%%%%%%%%%%%%%%%%%%%%%%%%%%%%%%%%%%%%%%%%%%%%%%%%%%%%
%%%%%%%%%%%%%%%%%%%%%%%%%%%%%%%%%%%%%%%%%%%%%%%%%%%%%%%%%%%%%%%%%

%%%%%%%%%%%%%%%%%%%%%%%%%%%%%%%%%%%%%%%%%%%%%%%%%%%%%%%%%%%%%%%%%
%%%%%%%%%%%%%%%%%%%%%%%%%%%%%%%%%%%%%%%%%%%%%%%%%%%%%%%%%%%%%%%%%

\begin{figure}[t]
\begin{center}  
\includegraphics[scale=0.43]{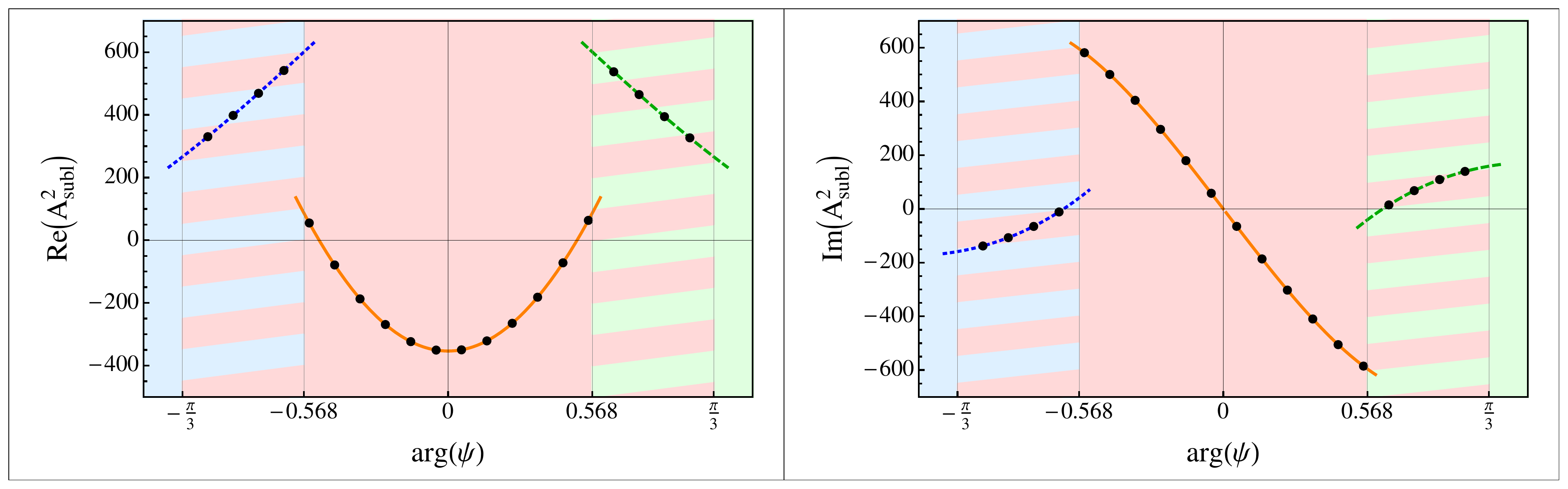}\\
\vspace{0.3cm}
\includegraphics[scale=0.43]{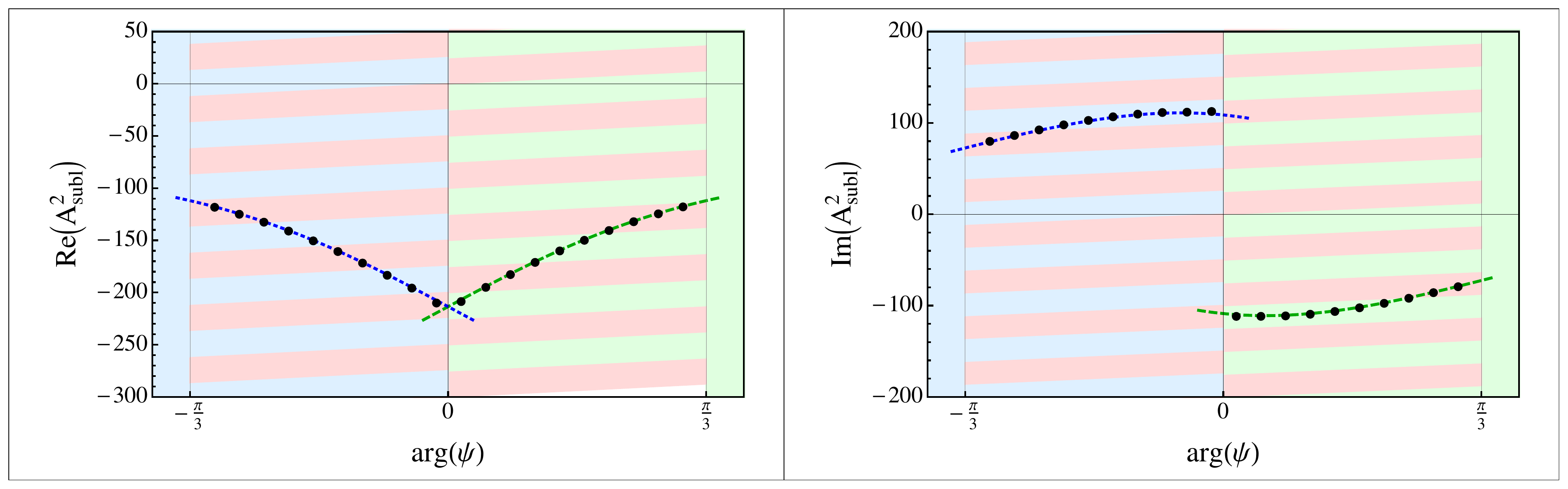}\\
\includegraphics[scale=0.45]{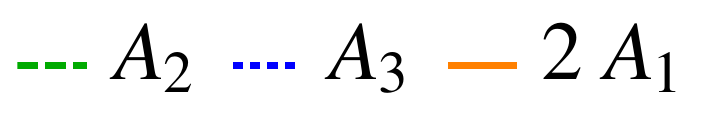}
\end{center}
\vspace{-1\baselineskip}
\caption{The real and imaginary parts of the subleading instanton action squared, $A_{\text{subl}}^2$, where $A_{\text{subl}}$ can be $2A_1$ (red), $A_2$ (green) or $A_3$ (blue) depending on the point in moduli space. The numerical values are obtained from ratios of coefficients $\CX^{(\bfe_1)}_g$, for $| \psi | = 2$ (top) and for $| \psi | = \frac{1}{4}$ (bottom). The results are in correspondence with figure \ref{fig:instantonactionscomparisonplots}. Note that the jumps are not due to branch cuts---not even the bottom one; the cuts start at $|\psi|\geq 1$---but to a change in dominance of the subleading instanton action.}
\label{fig:transition}
\end{figure}

We conclude this subsection by performing a systematic large-order computation of the instanton actions which control the subleading contribution to the perturbative sector. This information is contained in the sequence $\CX^{(\bfe_1)}_g$, and may be extracted numerically using a limit such as \eqref{eq:limitratiocurlyX} at different points in moduli space. In figure \ref{fig:transition} we show examples for the two situations we described earlier in figure \ref{fig:instantonactionscomparisonplots}: the first in which we see three different subleading instanton actions $2A_1$, $A_2$ and $A_3$; and the second in which $2A_1$ never controls the subleading contributions to the perturbative free energies. The transition between dominating instanton actions occurs whenever there is a change in the function $\min\{|2A_1|,|A_2|,|A_3|\}$. 

%%%%%%%%%%%%%%%%%%%%%%%%%%%%%%%%%%%%%%%%%%%%%%%%%%%%%%%%%%%%%%%%%
%%%%%%%%%%%%%%%%%%%%%%%%%%%%%%%%%%%%%%%%%%%%%%%%%%%%%%%%%%%%%%%%%%

%%%%%%%%%%%%%%%%%%%%%%%%%%%%%%%%%%%%%%%%%%%%%%%%%%%%%%%%%%%%%%%%%
\subsection{The General Nonholomorphic Case}
\label{sec:nonholomorphiccase}
%%%%%%%%%%%%%%%%%%%%%%%%%%%%%%%%%%%%%%%%%%%%%%%%%%%%%%%%%%%%%%%%%

Having understood the simpler holomorphic case, let us next turn on the full nonholomorphic dependence included in the propagator. The first implication this has is that the leading contribution to the perturbative sector, \eqref{eq:perturbativelargeorder1sector}, no longer truncates as in \eqref{eq:1insttruncated}. One thus has to deal with the whole series as in \eqref{eq:perturbativelargeorder1sector}, which, not surprisingly, is asymptotically divergent. Let us then first define the nonholomorphic version of \eqref{eq:holoXdef},
\be
X^{(\bfe_1)}_g := \frac{A_1^{2g-1}}{\G(2g-1)} \left\{ F^{(\bf0)}_g - \sum_{h=0}^{+\infty} \frac{ \G(2g-1-h) }{ A_1^{2g-1-h} }\, \frac{S_{1,1}}{\rmi\pi}\, F^{(\bfe_1)}_h \right\},
\label{eq:nonholoXdef}
\ee
\noindent
which probes the exponentially subleading corrections to the perturbative large-order. Consequentially, the divergent series we now have to analyze is
\be
I(g) := \sum_{h=0}^{+\infty} \frac{\G(2g-1-h)}{\G(2g-1)}\, \frac{S_{1,1}}{\rmi\pi}\, A_1^h\, F^{(\bfe_1)}_h.
\label{eq:toberesummed}
\ee
\noindent
The functional dependence of this function is in $g$, and it proves useful to rewrite it as a series in $1/g$. By expanding the ratio of gamma functions for large $g$,
\be
\frac{\G\left( 2g-1-h \right)}{\G(2g-1)} = \left(2g\right)^{-h} \left( 1+ \frac{h\left(h+3\right)}{4g} + \cdots \right),
\ee
\noindent
it is simple to obtain
\be
I(g) = \sum_{m=0}^{+\infty} \frac{a_m}{g^m},
\label{eq:I1series}
\ee
\noindent
where the coefficients $a_m$ depend on various free energies and powers of the instanton action. They grow factorially fast with $m$ because the one-instanton free energies do so, and the series \eqref{eq:I1series} is thus asymptotic. The first approach to make sense of \eqref{eq:I1series} is to use optimal truncation, which keeps only the partial sum that gives the best approximation to the actual result, for a given value of $g$, before the factorial growth takes over \cite{bo78}. The error associated to the truncation is of the order of the last term considered in the partial sum. 

A more powerful approach to resummation involves Pad\'e approximants and Borel resummation. The Borel transform removes the factorial growth of the $a_m$'s by dividing by $m!$. But since we only have finitely many coefficients ($m \leq m_{\max}$) we cannot exactly sum the Borel transform, which would now have a finite radius of convergence, and thus the need for the Pad\'e approximant. In particular, the Pad\'e approximant will incorporate (as poles of its rational expression) approximations to some of the Borel singularities, and that will improve the precision of the resummation. In summary, we have to consider
\be
\textrm{BP}[I](g) := g \int_0^{+\infty} \rmd\xi\, \rme^{-\xi /g}\, \textrm{Pad\'e}\left( \sum_{m=0}^{m_{\max}} \frac{a_m}{m!}\xi^m \right).
\label{eq:BP}
\ee
\noindent
In practice, we shall use the ``diagonal'' Pad\'e approximant, because this empirically gives the best results (see, \textit{e.g.}, \cite{bo78}). A somewhat surprising fact is that we will not always need to use the Borel--Pad\'e approach in order to be able to access the subleading contributions. The reason lies in the comparison between the optimal truncation error and the order of magnitude of the subleading terms \cite{bpv78a}. To illustrate this point with actual numbers, let us consider a definite value of the modulus, $\psi = 2\,\rme^{2\pi\rmi/9}$, for which $A_2$ controls the subleading contributions. Assigning a specific value to the propagator, such as $S^{zz} = 10^{-8}$, we can compute the optimal truncation (OT) of \eqref{eq:I1series} and its intrinsic error, and also the order of magnitude of the subleading contribution:
\be
\text{Exact}-\text{OT} \sim 10^{-24},Ê\qquad \text{OT error} \sim 10^{-33}, \qquad \left(A_2/A_1\right)^{-(2g-1)} \sim 10^{-24},
\label{eq:OTcomparison}
\ee
\noindent
where the exact value is given by $\frac{A_1^{2g-1}}{\G(2g-1)}\, F^{(\bf0)}_g$ (we fixed a value of $g=50$). We can see that the difference between exact and optimal truncation values is substantially larger than its error. That difference is what the subleading terms will provide. So, in this situation optimal truncation is precise enough to ensure that $X^{(\bfe_1)}_g$ will have reliable accuracy to study subleading terms. On the other hand, if we go to a point where $2A_1$ controls those subleading terms, such as $\psi = 2$, we find that all numbers above are now of the same order (the last quantity in \eqref{eq:OTcomparison} changes to $(2A_1/A_1)^{-(2g-1)}$), around $\sim 10^{-32}$. Then, it is Borel--Pad\'e which improves the precision sufficiently in order to provide an accurate enough $X^{(\bfe_1)}_g$.

Before showing the results, let us further comment on the integration appearing in \eqref{eq:BP}. Some of the poles of the Pad\'e approximant may lie on the real axis, along the naive contour of integration. This is certainly the case when the $a_m$ coefficients are real. A solution to such presence of poles is to deform the contour slightly above or below the real axis, but this induces a nonperturbative ambiguity. In order to avoid this extra piece, we will use the Cauchy principal value prescription \cite{r91, p99}. This is equivalent to a combination of both upper and lower contour deformations, that is, to the average of both lateral resummations. Indeed, the following formula is satisfied
\be
\mathcal{S}_{\pm} I (g) = \dashint_0^{+\infty} f(\xi,g)\, \rmd\xi \mp \frac{1}{2}\, 2\pi \rmi\, \sum_{\substack{\text{poles:} \\ p \in \BR^+ }} \mathop{\text{Res}}\limits_{\xi\to p} f(\xi,g),
\ee
\noindent
where $\mathcal{S}_{\pm}$ denote lateral resummations (above or below), $\dashint$ indicates principal value, and $f$ is the integrand in \eqref{eq:BP}. In more general contexts, in which one is faced with the task of resumming a whole transseries, nonperturbative ambiguities appear at each sector, but with the prescription of the median resummation (\cite{b80, z81b}, see \cite{as13} for a very general discussion), those ambiguities end up canceling each other. Taking the principal value prescription is a shortcut on this general result. We must also take into account what happens when the coefficients $a_m$  are no longer real. This is the generic situation that appears as soon as we change $\psi$ and $S^{zz}$ from their real values. In that case, the poles that lay on the real axis move slightly into the complex plane. It turns out that we have to deform the principal value contour accordingly, in order to capture the nonperturbative information of the poles but not the ambiguity. If we were to take a real contour, we would effectively be performing one of the lateral resummations. 

\begin{figure}[ht!]
\begin{center}
\includegraphics[width=\textwidth]{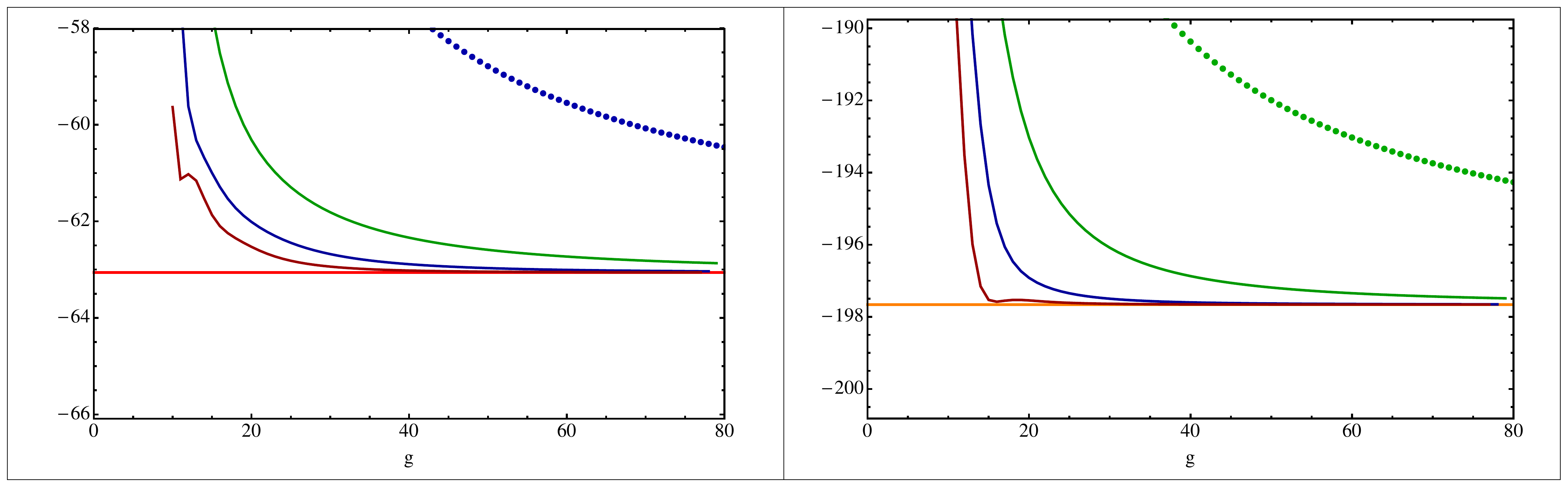} 
{\scriptsize
\bea
\frac{S_{1,2}}{\rmi\pi}\,F^{(\bfe_2)}_0 &=& -63.059\,684-197.660\,388\,9\,\rmi \nonumber\\
\text{5 Richardson Transforms} &=& -63.059\,646-197.660\,384\,1\,\rmi \nonumber
\eea
}
\includegraphics[width=\textwidth]{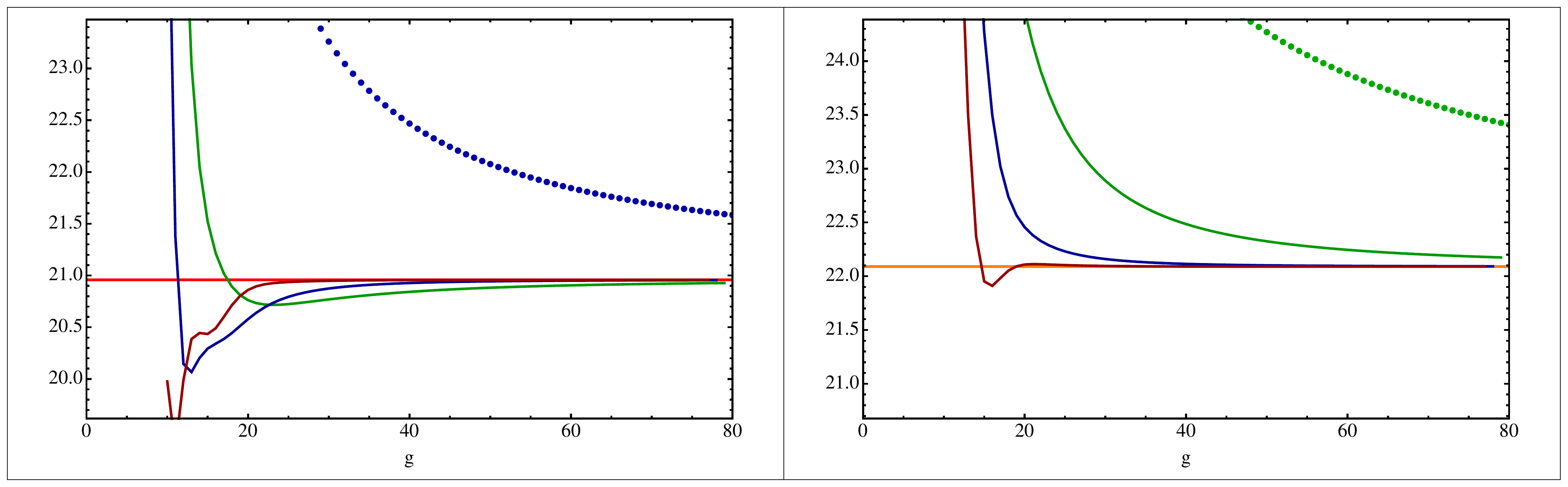}  
{\scriptsize
\bea
\frac{S_{1,2}}{\rmi\pi}\,F^{(\bfe_2)}_1 &=& 20.958\,451\,4+22.089\,526\,167\,\rmi \nonumber\\
\text{5 Richardson Transforms} &=& 20.958\,453\,5+22.089\,526\,107\,\rmi \nonumber
\eea
}
\includegraphics[width=\textwidth]{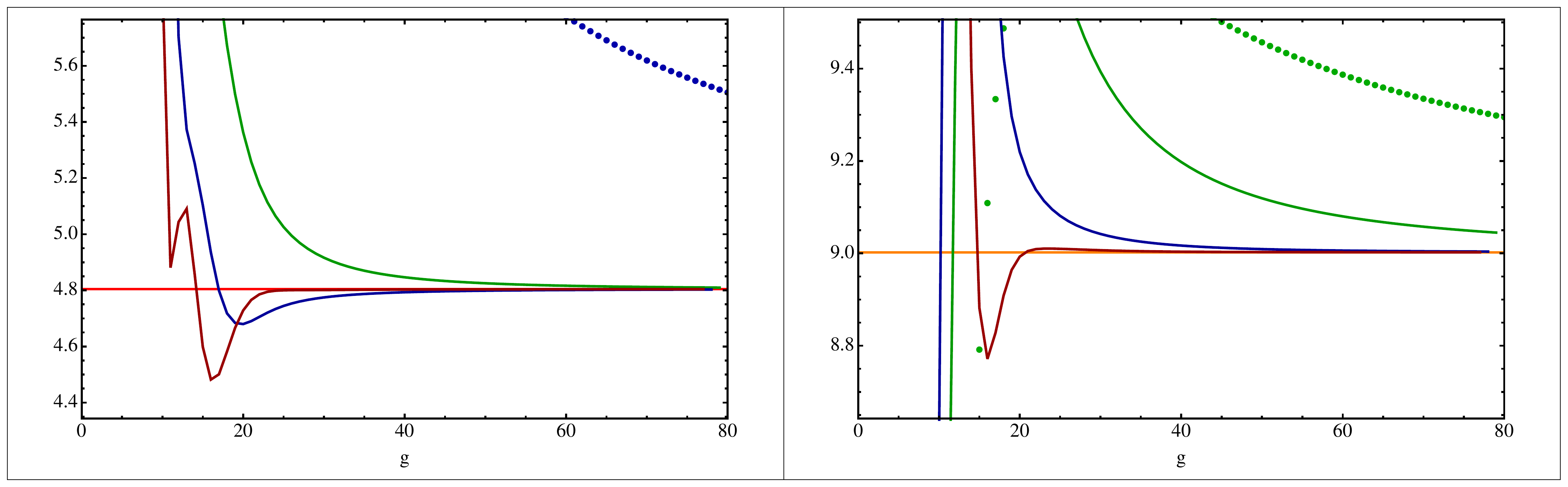}  
{\scriptsize
\bea
\frac{S_{1,2}}{\rmi\pi}\,F^{(\bfe_2)}_2 &=& 4.804\,186\,3+9.002\,264\,0\,\rmi \nonumber\\
\text{5 Richardson Transforms} &=& 4.804\,185\,4+9.002\,263\,2\,\rmi \nonumber
\eea
}
\end{center}
\vspace{-0.5\baselineskip}
\caption{Nonholomorphic counterpart of figure \ref{fig:otheroneinsthololargeorder}, for $\psi = 2\,\rme^{\rmi\pi/4}$ and $S^{zz}=10^{-5}$.}
\label{fig:otheroneinstlargeorder}
\end{figure}

Let us now go back to the same regions in moduli space that we studied earlier in subsection \ref{sec:holomorphiccase}. In the first region, $A_2$ controls the asymptotic growth of $X^{(\bfe_1)}_g$. In this case, the resummation of the one-instanton contribution may be computed by optimal truncation. A numerical analysis then shows that $X^{(\bfe_1)}_g$ behaves as
\be
X^{(\bfe_1)}_g \simeq \left( \frac{A_1}{A_2} \right)^{2g-1}\, \sum_{h=0}^{+\infty} \frac{\G(2g-1-h)}{\G(2g-1)}\, \frac{S_{1,2}}{\rmi\pi}\, A_2^h\, F^{(\bfe_2)}_h + \cdots,
\ee
\noindent
which is the analogue of \eqref{eq:holoXother1sector} when the propagator is no longer evaluated at $S^{zz} = S^{zz}_{1,\hol}$, but allowed to roam free. We show the excellent agreement between numerical and analytical values, for specific values of $\psi$ and $S^{zz}$, in figure \ref{fig:otheroneinstlargeorder}.

%%%%%%%%%%%%%%%%%%%%%%%%%%%%%%%%%%%%%%%%%%%%%%%%%%%%%%%%%%%%%%%%%
%%%%%%%%%%%%%%%%%%%%%%%%%%%%%%%%%%%%%%%%%%%%%%%%%%%%%%%%%%%%%%%%%

\begin{figure}[ht!]
\begin{center}
\includegraphics[scale=0.41]{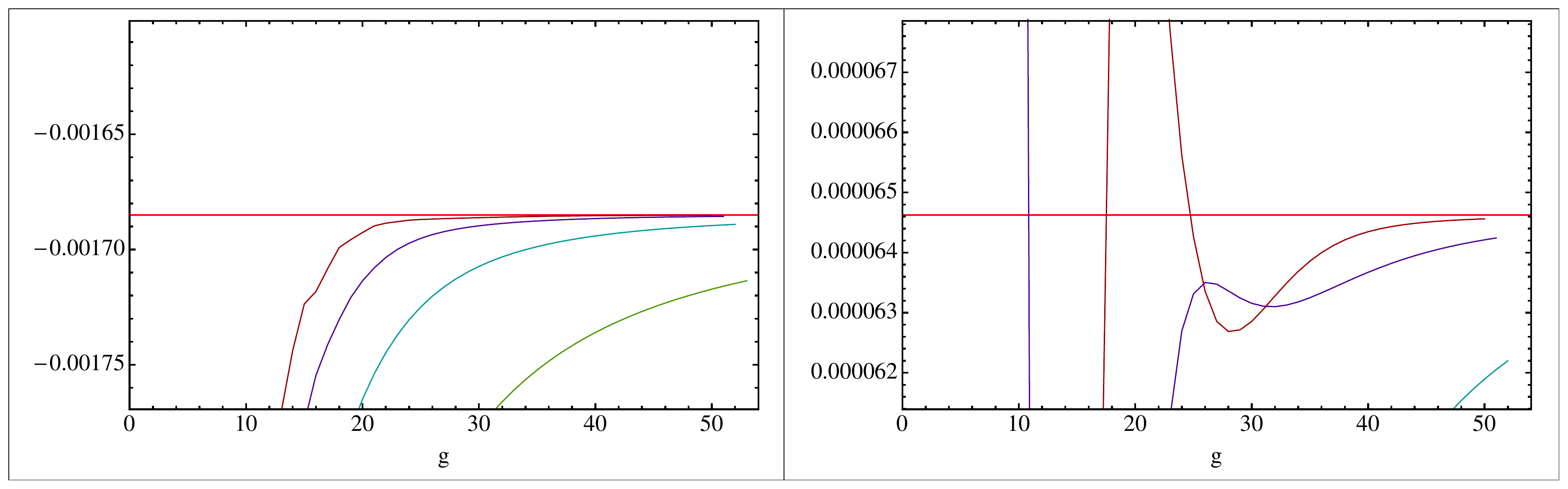} 
{\scriptsize
\bea
\frac{(S_{1,1})^2}{\rmi\pi}\, \widetilde{F}^{(2\bfe_1)}_0 &=& -0.001\,685\,034+0.000\,064\,624\,8\,\rmi \nonumber\\
\text{5 Richardson Transforms} &=& -0.001\,685\,043+0.000\,064\,622\,0\,\rmi \nonumber
\eea
}
\includegraphics[scale=0.41]{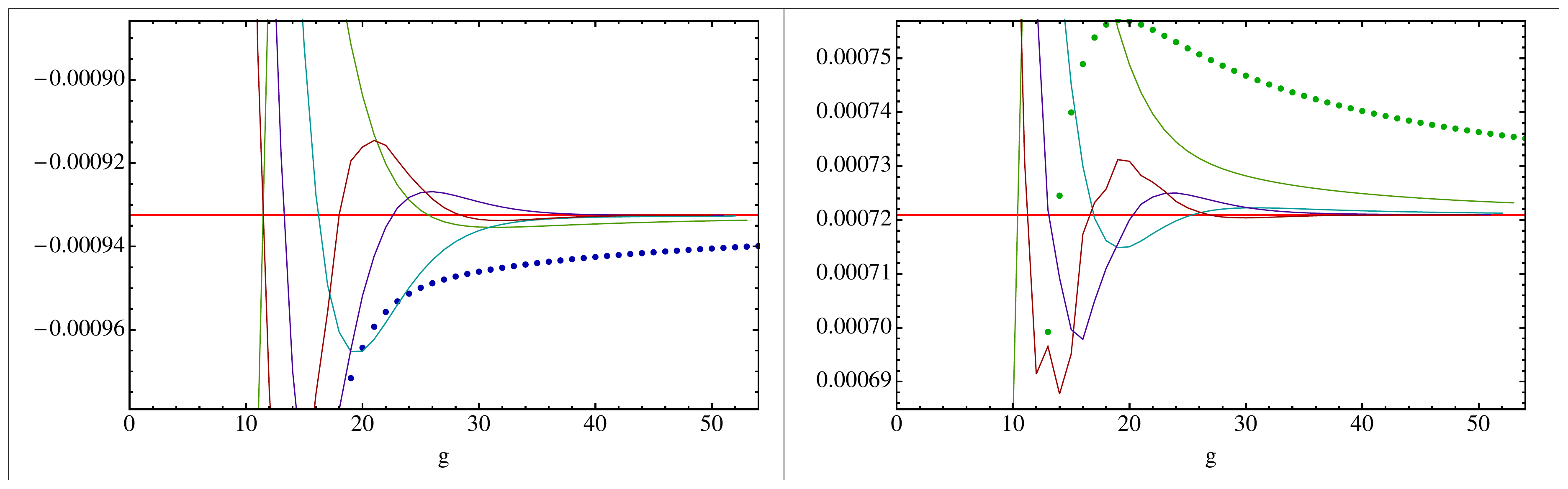}  
{\scriptsize
\bea
\frac{(S_{1,1})^2}{\rmi\pi}\, \widetilde{F}^{(2\bfe_1)}_1 &=& -0.000\,932\,452+0.000\,720\,907\,\rmi \nonumber\\
\text{5 Richardson Transforms} &=& -0.000\,932\,426+0.000\,720\,911\,\rmi \nonumber
\eea
}
\includegraphics[scale=0.41]{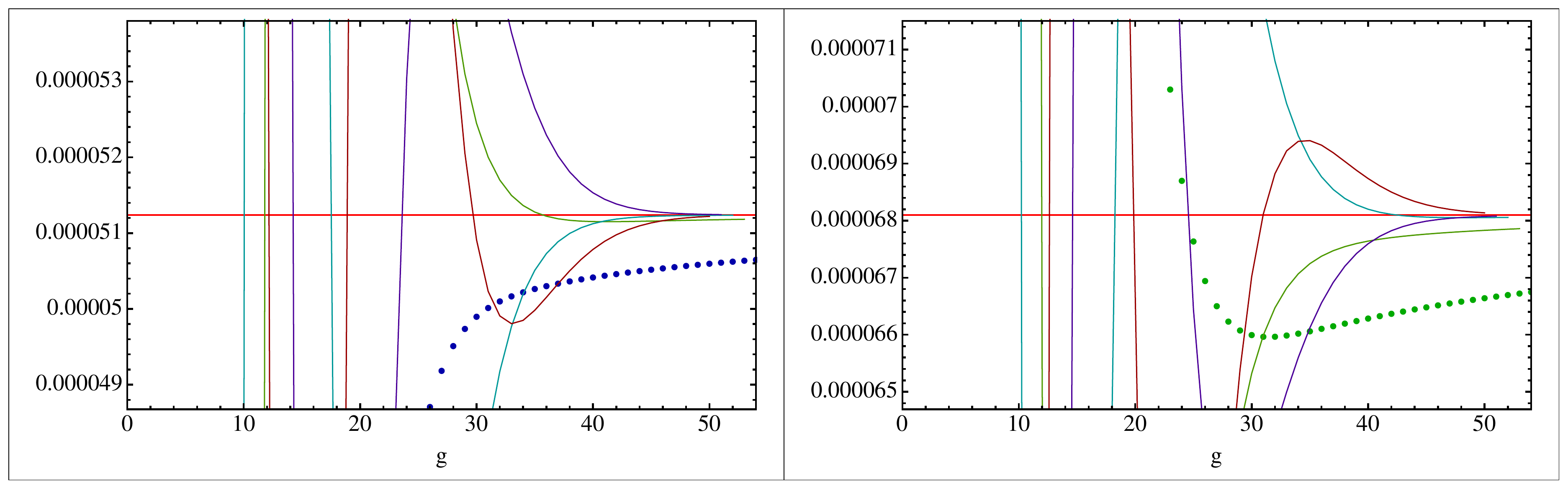}  
{\scriptsize
\bea
\frac{(S_{1,1})^2}{\rmi\pi}\, \widetilde{F}^{(2\bfe_1)}_2 &=& 0.000\,051\,238+0.000\,068\,099\,\rmi \nonumber\\
\text{5 Richardson Transforms} &=& 0.000\,051\,285+0.000\,068\,018\,\rmi\nonumber
\eea
}
\end{center}
\vspace{-0.5\baselineskip}
\caption{The usual tests of $\frac{(S_{1,1})^2}{\rmi\pi}\, \widetilde{F}^{(2 \bfe_1)}_h$, for $h=0,1,2$. We have used values $\psi = 2\,\rme^{-\rmi\pi/12}$, $S^{zz}= 10^{-8}$. The left plots show the real part and the right ones show the imaginary part.}
\label{fig:twoinstantonex1}
\end{figure}

The second region of interest involves free energies associated to a two-instanton sector $(2\bfe_1)$,
\bea
X^{(\bfe_1)}_g &=& \frac{A_1^{2g-1}}{\G(2g-1)} \left\{ F^{(\bf0)}_g - \text{BP}\left( \sum_{h=0}^{+\infty} \frac{\G(2g-1-h)}{A_1^{2g-1-h}}\, \frac{S_{1,1}}{\rmi\pi}\, F^{(\bfe_1)}_h \right) \right\} \\
&\simeq& \frac{1}{2^{2g-1}}\, \sum_{h=0}^{+\infty} \frac{\G(2g-1-h)}{\G(2g-1)}\, \frac{S_{1,1}^2}{\rmi\pi} \left( 2A_1 \right)^h \widetilde{F}^{(2\bfe_1)}_h + \cdots.
\label{eq:Xlargeordertwoinst}
\eea
\noindent
Now the resummation is performed with the technique of Borel--Pad\'e explained above. As we have already seen in the holomorphic limit, the two-instanton sector $\widetilde{F}^{(2\bfe_1)}_h$ which we find in the large-order behavior of \eqref{eq:Xlargeordertwoinst} is \textit{not} quite the same which appeared in the large-order behavior of the one-instanton free energies, \eqref{eq:alt2instg0freeenergy}. Off holomorphicity, we may now learn that this sector is in fact a \textit{composite} object, in the sense that it can be expressed in terms of two different two-instanton free energies computed from the holomorphic anomaly equation. One of these free energies is exactly the $\widehat{F}^{(2\bfe_1)}_h$ sector which we encountered in subsection \ref{largeorderoftheoneinstantonsector}. Its holomorphic ambiguity is fixed by requiring a vanishing holomorphic limit at all orders. The second free energy that builds up $\widetilde{F}^{(2\bfe_1)}_h$ has instead its holomorphic ambiguity fixed in a manner akin to what we saw for the one-instanton sector, \textit{i.e.}, it has its holomorphic ambiguity fixed against the conifold result \cite{ps09} (see, \textit{e.g.}, subsection \ref{sec:largeorderoftheperturbativesector}, or simply consider the fixing in \eqref{eq:holo2sector} for $h=0,1$ and zero for the rest). In parallel to what we did for the one-instanton sector, we denote these free energies by the usual notation $F^{(2\bfe_1)}_h$. In this case, we find out of large-order that 
\be
\widetilde{F}^{(2\bfe_1)}_h = F^{(2\bfe_1)}_h - \widehat{F}^{(2\bfe_1)}_h.
\ee
\noindent
For instance, for $h=0$ the second term is \eqref{eq:alt2instg0freeenergy} and 
\be
F^{(2\bfe_1)}_0 = \widehat{F}^{(2\bfe_1)}_0 + \frac{\rmi\pi}{\left(S_{1,1}\right)^2}\, \frac{1}{2}\, \frac{A_1}{2\pi^2}\, \rme^{4\,\frac{1}{2} \left(\p_z A_1\right)^2 \left(S^{zz}-S^{zz}_{1,\hol}\right)}.
\ee
\noindent
We can see the numerical results in figure \ref{fig:twoinstantonex1}, showing how our analytical result for $\widetilde{F}^{(2\bfe_1)}_h$ matches the numerics to very good precision.

%%%%%%%%%%%%%%%%%%%%%%%%%%%%%%%%%%%%%%%%%%%%%%%%%%%%%%%%%%%%%%%%%
%%%%%%%%%%%%%%%%%%%%%%%%%%%%%%%%%%%%%%%%%%%%%%%%%%%%%%%%%%%%%%%%%

%%%%%%%%%%%%%%%%%%%%%%%%%%%%%%%%%%%%%%%%%%%%%%%%%%%%%%%%%%%%%%%%%
\subsection{On the Construction of the Transseries Solution}
\label{sec:comments}
%%%%%%%%%%%%%%%%%%%%%%%%%%%%%%%%%%%%%%%%%%%%%%%%%%%%%%%%%%%%%%%%%

Having gathered a large amount of high-precision tests on the resurgent structure of the topological string in the local $\BC\BP^2$ Calabi--Yau background, one may now ask how far will this allow us to go in uncovering and eventually fully constructing the transseries structure of its free energy. Although this is at present a goal still out of our reach, we can nonetheless already make clear that the transseries structure of the full topological-string free energy will be more complicated than in previous examples concerning matrix models and their double-scaling limits, \textit{e.g.}, \cite{gikm10, asv11, sv13}.

Already in subsection \ref{sec:largeradiusactions} we have seen that the string free energy must have an intricate multi-sheeted Borel structure, with Borel singularities moving between sheets as one tours moduli space. Then, starting in subsection \ref{largeorderoftheoneinstantonsector} but mainly throughout the present section, we have found that the two-instanton contributions, appearing in the resurgence relations we have studied, do not correspond to a single sector, \textit{i.e.}, a single $(2\bfe_1)$ free energy. Even momentarily ignoring the fact that one finds free energies with their holomorphic ambiguities fixed according to different prescriptions, the minimalistic assumption based on the existence of a bridge equation for the topological-string free energy, and of a transseries built upon standard multi-instanton sectors, would have imposed that $\widehat{F}^{(2\bfe_1)}_h$ (from the large-order of the one-instanton sector \eqref{eq:alt2instg0freeenergy}) and $\widetilde{F}^{(2\bfe_1)}_h$ (from the subleading large-order of the perturbative sector \eqref{eq:Xlargeordertwoinst}) would have been the same. They are not; the numerical analysis tells us very definitely that this is not the case, implying the resurgence relations have to be generalized to accommodate this fact. Furthermore, the numerical analysis also tells us that, starting at the two-instanton level, there are (at least) two ``classes'' of nonperturbative free energies entering the game: those whose holomorphic ambiguity is fixed against nonperturbative results at the pure conifold \cite{ps09}, and those whose holomorphic ambiguity is fixed to vanish. Again, the topological string transseries must be constructed in order to accommodate this fact. Classes of transseries solutions to the nonperturbative holomorphic anomaly equations were already constructed in \cite{cesv13}, but one cannot exclude the existence of further, more complicated, sets of transseries solutions one might still have to address. At this point, we do not yet have a complete framework which can predict the resurgence properties we have uncovered but we can certainly have a look at what possible ingredients such a framework might include. Before that, let us just stress once again the very important point that every sector appearing in the resurgence relations we have studied was computed \textit{directly} from the nonperturbative holomorphic anomaly equations set forth in \cite{cesv13}.
 
The best way to see which options are available for generalizations is to go back to basics and study which possible singularities the Borel transforms of different transseries sectors can have. Resurgence tells us that the singularity structure of the Borel transform determines the large-order growth of a given sector (see, \textit{e.g.}, \cite{m10, asv11} for reviews and examples). This can be studied systematically by making use of alien derivatives (alien differential operators signaling the Borel singularities) and of their associated Stokes automorphism (roughly speaking, the exponential of the alien derivatives, gluing together left and right Borel resummations and, as such, dictating the large-order relations from first principles). We are not going to be very rigorous in the following, as our purpose is just to give an idea of what possible generalizations are available within the standard framework of resurgence (but we refer the reader to, \textit{e.g.}, \cite{asv11} for a light technical introduction to the previous concepts of alien calculus). All we need to have in mind is that the knowledge of the Stokes automorphism acting on a given sector of the transseries completely determines its large-order growth, and that in order to know this automorphism we first need to find the alien derivatives; see \eqref{eq:stokesautomorphism} below.

Let us first focus on the large-order behavior of the one-instanton sector, $F^{(\bfe_1)}_g$, which we explicitly addressed in subsection \ref{largeorderoftheoneinstantonsector}. Its growth at leading order, \eqref{eq:largeorder1sector}, is dictated by the pole of its Borel transform at $A_1$ (and also at $-A_1$ but let us just deal with the former), and it involves a two-instanton sector as we saw in \eqref{eq:alt2instg0freeenergy}. The alien derivative\footnote{Note that the alien derivative, $\Delta_\omega$, that we use here, and the (pointed) alien derivative which was mentioned in section \ref{sec:largeorderfromresurgence}, are related by an exponential factor, $\sim \rme^{-\omega/g_s}$.} captures this information as
\be
\Delta_{A_1} F^{(\bfe_1)} = \widehat{a}\, \widehat{F}^{(2\bfe_1)},
\label{eq:aliendderatA}
\ee
\noindent
where $\widehat{a}$ is a proportionality factor related to the Stokes constant. Equation \eqref{eq:aliendderatA} codifies the fact that the large-order of $F^{(\bfe_1)}$ includes the sector $\widehat{F}^{(2\bfe_1)}$ at its leading order. The Stokes automorphism, ${\underline{\mathfrak{S}}}$, is given by 
\be
{\underline{\mathfrak{S}}} - \1 = \rme^{-A_1/g_s}\, \Delta_{A_1} + \rme^{-2A_1/g_s} \left( \Delta_{2A_1} + \frac{1}{2}\Delta_{A_1}^2 \right) + \cdots,
\label{eq:stokesautomorphism}
\ee
\noindent
up to second order. The leading contribution in the large-order relation of $F^{(\bfe_1)}$ is determined by the first term in \eqref{eq:stokesautomorphism}, and exactly reproduces the relation \eqref{eq:largeorder1sector} (its second term, depending on $-A_1$, is obtained from the alien derivatives at opposite instanton actions); see \cite{asv11}.

Let us next compare the above discussion with the large-order behavior of the perturbative sector, up to second order contributions (\textit{i.e.}, including contributions arising from the second term in \eqref{eq:stokesautomorphism}). Inspired by the actual relations we found numerically, we should have something of the form
\bea
\Delta_{A_1} F^{(\bf0)} &=& \alpha\, F^{(\bfe_1)}, \\
\Delta_{2A_1} F^{(\bf0)} &=& \beta\, F^{(2\bfe_1)}.
\label{eq:alienderat2A}
\eea
\noindent
Note that here equation \eqref{eq:alienderat2A} is somewhat unconventional comparing to nontrivial examples, \textit{e.g.}, \cite{asv11}, although in the simpler example of topological strings in the resolved conifold, structures of this type do appear \cite{ps09, asv11}. In fact, with a standard bridge equation this right-hand-side would be zero; we are thus facing the first required generalization. As to the term $\Delta_{A_1}^2$ in \eqref{eq:stokesautomorphism}, it is of course determined by the previous alien derivative, as
\be
\Delta_{A_1}^2 F^{(\bf0)} = \alpha\,\widehat{a}\, \widehat{F}^{(2\bfe_1)}.
\ee
\noindent
In this case we see that the large-order behavior of the perturbative free energies will incorporate $F^{(\bfe_1)}$ at leading order, and both $F^{(2\bfe_1)}$ and $\widehat{F}^{(2\bfe_1)}$ at subleading order. This setting is the minimal modification we can do in order to explain the numerics we found in this and in the previous section. However, other details still need to be made to work: the precise coefficients of the combinations of $F^{(2\bfe_1)}$ and $\widehat{F}^{(2\bfe_1)}$ appearing at large-order, and the precise factorial growths, $\G(g+1)$ and $\G(2g-1)$ for the one-instanton and perturbative sectors, respectively (\textit{i.e.}, compare the factorial growths in \eqref{eq:extendedlargorder1sector} and \eqref{eq:Xlargeordertwoinst}). On this last point, it may be necessary to drop the (possibly naive) assumption that the resurgent functions we are working with are simple resurgent functions, in a technical sense (also see \cite{absu14}). For this class of resurgent functions, the Borel transform around \textit{any} singularity has only simple poles and logarithmic branch-cuts. But for resurgent functions with higher-order poles as Borel singularities one has to generalize \eqref{eq:aliendderatA}, and others, to
\be
\Delta_{A_1} F^{(\bfe_1)} = \widehat{a}\, g_s^\gamma\, \widehat{F}^{(2\bfe_1)},
\ee
\noindent
for some $\gamma$. This power $\gamma$ will be present in the gamma function of the large-order as an additive constant. However, a naive try with the data we have does not succeed, so some other ingredients have to included. More complicated generalizations may have to be considered in order to fully reproduce the exact factorial growth. 

It is also important to notice that the fact that the resulting transseries for the topological string free-energy seems to be so intricate (and so much more than in the matrix model examples studied in \cite{asv11, sv13}) cannot be dissociated from its nonholomorphic nature. In fact, if one considers the holomorphic limit of the above discussion, things simplify considerably. Taking the holomorphic limit associated to, say, the first conifold point, implies that there is no asymptotics of instantons related to $A_1$, and thus the above ``hatted'' quantities all vanish. In particular, the composite two-instanton sector we considered earlier in this section is now just a ``regular'' two-instanton sector with its holomorphic ambiguity fixed against conifold data. While understanding the full transseries structure of the topological string solution to the holomorphic anomaly equations, and thus including nonholomorphic dependence, is our main goal in this line of work, we must also stress that one first natural step may be to start off addressing some holomorphic limit instead. One natural ground for this approach may be considering backgrounds with matrix model duals, where a natural holomorphic limit steps in.

One aspect of the transseries that we have not fully explored in this paper deals with resonance, the phenomenon where different nonperturbative sectors may appear in the transseries with the same exponential instanton weight. In subsection \ref{sec:instantonactions} we already discussed that resonance will be part of the topological string free energy, not only due to the appearance of the different instanton actions in symmetric pairs, but also due to dependence relations such as \eqref{eq:leadstoresonance}. The fact that this relation involves the three conifold actions, and the constant-map action, may also point to the fact that our present example may have a rather intricate set of resonant sectors. At the same time, we have found that the resurgence relations see nonperturbative sectors which at first look functionally identical, but which in fact have their holomorphic ambiguities fixed differently (either against the conifold or against zero). Because a transseries is a sum over all possible semiclassical sectors, one may wonder if these different fixings of the ambiguity actually lead to distinct semiclassical sectors, in which case they would certainly need to be included as distinct sectors in the transseries. Should this be the case one could even envisage the appearance of further resonance due to these ``same weight'' contributions, of sectors with differently-fixed holomorphic-ambiguity. Although the final picture is certainly unclear at this stage, what is clear is that the full topological string transseries is very likely highly resonant, and one will probably not be able to construct this complete transseries without properly addressing this phenomenon. Note that in known string theoretic examples \cite{gikm10, asv11, sv13}, resonance leads to the appearance of nonperturbative logarithmic sectors, accompanying powers of $\log g_s$ (which is in fact nonanalytic at $g_s = 0$, just like the familiar $\rme^{-1/g_s}$). We described in \cite{cesv13} how these sectors are compatible with the nonperturbative holomorphic anomaly equations. However, one problem to be solved is fixing their holomorphic ambiguities. It is certainly not excluded that such sectors may actually include some of the free energies we are now labeling with $(2\bfe_1)$, and thus that they might be relevant for the large-order relations studied in this paper. Note however that the specific use of logarithms is not actually mandatory from the equations (in contrast to the case of Painlev\'e and matrix models, where the ``string equation'' fixes the logarithmic structure), and in fact any other nonanalytic function may take over its role. Integrating the holomorphic anomaly equations for these classes of more exotic sectors we do not find anything that could rule them out, as their functional form is compatible with the propagator dependence we find from the numerics. 

Finally, once the transseries ingredients are made clear, as just described above, one may turn to the Borel structure. As we have seen in subsection \ref{sec:largeradiusactions}, we are now dealing with a branched Borel structure, with several Riemann sheets, and where singularities may move between sheets as one moves along moduli space. While this was made very clear for the large-radius singularity---moving away from the principal sheet of the perturbative sector---, a similar behavior may occur for the conifold actions once one probes different points of moduli space, and even around either perturbative or multi-instanton sheets. To clarify this situation, one would have to move to the principal sheet of some chosen instanton sector and then perform a similar Pad\'e analysis of the Borel structure around that sector. It is also important to notice that this multi-branched Borel structure may be linked to the previous issue of finding new nonperturbative sectors (beyond those related to ``generalized'' instantons as in \cite{gikm10, asv11, sv13}), associated to different ambiguity fixings. In fact, in the case of the conifold the Borel structure is much simpler: its Riemann structure is just the complex plane with an infinite set of poles \cite{ps09}. In the present scenario, as one approaches conifold points, singularities in the principal sheet of the perturbative sector must approach the singularities (the poles) of the conifold complex Borel plane, while singularities in other sheets should disappear. This could be an explanation for the different ambiguity fixings, but, again, Pad\'e analysis near (multi) instanton sectors will be required in order to further study the singularity structure of the full Borel surface and thus draw more definite conclusions.

In summary, the resurgence properties of the topological-string free energy are not as well understood as those of matrix models, for which the existence of a ``string equation'' and thus a standard bridge equation dictates the form of the resurgence relations precisely. For topological strings the role of the numerical large-order analysis will certainly be of primary importance in order to unveil the resurgent structure of the nonperturbative free energy. At this point we can only present a range of possibilities which probably will participate in the ultimate understanding of this issue. A road towards this goal will involve going further into the large-order of various sectors, but also applying these techniques to other examples for which more about the nonperturbative structure is known. Nonetheless, our present work alongside \cite{cesv13} already goes a long way in uncovering the resurgent transseries structure of topological strings.

%%%%%%%%%%%%%%%%%%%%%%%%%%%%%%%%%%%%%%%%%%%%%%%%%%%%%%%%%%%%%%%%%
%%%%%%%%%%%%%%%%%%%%%%%%%%%%%%%%%%%%%%%%%%%%%%%%%%%%%%%%%%%%%%%%%
\section{Conclusions and Outlook}\label{sec:conclusions}
%%%%%%%%%%%%%%%%%%%%%%%%%%%%%%%%%%%%%%%%%%%%%%%%%%%%%%%%%%%%%%%%%
%%%%%%%%%%%%%%%%%%%%%%%%%%%%%%%%%%%%%%%%%%%%%%%%%%%%%%%%%%%%%%%%%

In the present paper we have further addressed our proposal \cite{cesv13} concerning the nonperturbative extension of the holomorphic anomaly equations. We have constructed in detail a very explicit example, beginning to work out the structure of the resurgent transseries describing the nonperturbative free-energy of closed topological-strings in the mirror of the local $\BC\BP^2$ Calabi--Yau background. The structure of this transseries was further checked numerically, to very high precision, by making use of a variety of large-order resurgence relations. All data present in these resurgence relations is computable from the nonperturbative version of the holomorphic anomaly equations we put forward and, as such, we believe our present results both validate this proposal \cite{cesv13} and further make clear that resurgent transseries methods are very promising techniques in order to address the nonperturbative structure of generic (closed) string theories.

Of course many problems still remain open for future investigations. One natural course of research is to begin working out further examples, in order to see how resurgent transseries apply in more intricate situations or with different geometrical characteristics. Perhaps the next natural step in this ladder would be to address the example of topological strings in the local $\BC\BP^1 \times \BC\BP^1$ geometry \cite{cesv15}, although, of course, many other examples quickly come to mind.

Still on what concerns our local $\BC\BP^2$ example, further study is required in order to have a more detailed picture of the full free-energy transseries, and all nonperturbative sectors it may encompass. On the one hand, we have unveiled resonance effects which are known to lead to logarithmic sectors within matrix model contexts, see \cite{gikm10, asv11, sv13}. It is important to clarify if in the present case of topological string theory resonance also leads to the appearance of logarithmic sectors (which are certainly allowed as our general analysis indicates \cite{cesv13}), or perhaps to the appearance of more complicated nonanalytic structures we have still to unveil. In fact, as we already mentioned earlier in the paper, resonance is only well understood in matrix model examples due to the existence of a ``string equation'', and this is not generically available for the topological string\footnote{Unless, of course, via large $N$ duality, but this is perhaps a harder question within the local $\BC\BP^2$ example \cite{amv02}.}. As such, understanding the full consequences of resonance and any new nonperturbative sectors it may lead to in the present context is clearly a fundamental problem for future research. On the other hand, we have seen that, at perturbative level, resurgence already predicts that the string free energy has an intricate multi-branched Borel structure; and that, at the two-instanton level, resurgence predicts that the string free energy has an intricate structure of Borel singularities---and these two problems may very likely be related. Again, it is also relevant to clarify these structures as they will make definite predictions concerning the full transseries structure we are now facing. Recall that understanding which sectors appear in the transseries, via resurgence, is not decoupled from having a proper understanding of the (would-be) bridge equation in the present context, and this is an open question partially due to the unavailability of a ``string equation''. In this regard, one could envisage starting to understand what types of bridge equations might be adequate within topological string contexts by addressing this question in simpler geometrical examples such as the conifold \cite{ps09}. Then, once the complete structure of the resurgent transseries is clarified, including both resonant sectors and a clean understanding of all starting powers, one may start addressing questions concerning the \textit{resummation} of this transseries---perhaps even yielding an associated partition function which may have some interpretation as a (generalized) theta function (see, \textit{e.g.}, \cite{em08}). 

Within the context of theta functions, it would also be very interesting to address the issues of modular versus holomorphic properties which were put forward in \cite{em08} (and their relation to the question of background independence), and see how they actually materialize within very explicit examples. This also raises the interesting question of what the interplay is between modularity and resurgence. Since the perturbative free energies have specific modular properties, so must the nonperturbative expressions describing their large-order behavior. But can we make stronger claims at the level of individual instanton sectors? Can modular properties further constrain the general form of resurgence relations? These are very intriguing questions for future research.

Finally, yet another course of action arises due to the relation between topological strings and matrix models \cite{dv02, dv02a, emo07}. Via large $N$ duality, one may effectively use the holomorphic anomaly equations (in their holomorphic limit) to compute matrix model data, and, in our particular set-up, one may now use the nonperturbative holomorphic anomaly equations to compute multi-instanton matrix model data. In this context, it would be very interesting to obtain resurgent transseries for different matrix models, possibly even matrix models associated to localizable gauge theories, as they could open the door for finite $N$ calculations via transseries resummations.

%%%%%%%%%%%%%%%%%%%%%%%%%%%%%%%%%%%%%%%%%%%%%%%%%%%%%%%%%%%%%%%%%
\acknowledgments
We would like to thank Murad Alim, In\^es Aniceto, Babak Haghighat, Chris Howls, Albrecht Klemm, Boris Pioline, David Sauzin, Emanuel Scheidegger, Andr\'e Voros, and, especially, Alba Grassi, Marcos Mari\~no, and Szabolcs Zakany, for useful discussions, comments and/or correspondence. RCS and RS would further like to thank CERN TH--Division for hospitality, where a large part of this work was conducted. RCS and JDE are supported in part by MICINN and FEDER (grant FPA2011-22594), Xunta de Galicia (GRC2013-024), and the Spanish Consolider-Ingenio 2010 Programme CPAN (CSD2007-00042). The research of RS was partially supported by the FCT--Portugal grants PTDC/MAT/119689/2010 and EXCL/MAT-GEO/0222/2012. The research of MV was partially supported by the European Research Council Advanced Grant EMERGRAV. The Centro de Estudios Cient\'\i ficos (CECs) is funded by the Chilean Government through the Centers of Excellence Base Financing Program of Conicyt.
%%%%%%%%%%%%%%%%%%%%%%%%%%%%%%%%%%%%%%%%%%%%%%%%%%%%%%%%%%%%%%%%%

\newpage

%%%%%%%%%%%%%%%%%%%%%%%%%%%%%%%%%%%%%%%%%%%%%%%%%%%%%%%%%%%%%%%%%
%%%%%%%%%%%%%%%%%%%%%%%%%%%%%%%%%%%%%%%%%%%%%%%%%%%%%%%%%%%%%%%%%
\appendix
%%%%%%%%%%%%%%%%%%%%%%%%%%%%%%%%%%%%%%%%%%%%%%%%%%%%%%%%%%%%%%%%%
%%%%%%%%%%%%%%%%%%%%%%%%%%%%%%%%%%%%%%%%%%%%%%%%%%%%%%%%%%%%%%%%%

%%%%%%%%%%%%%%%%%%%%%%%%%%%%%%%%%%%%%%%%%%%%%%%%%%%%%%%%%%%%%%%%%
%%%%%%%%%%%%%%%%%%%%%%%%%%%%%%%%%%%%%%%%%%%%%%%%%%%%%%%%%%%%%%%%%
\section{The Local $\mathbb{C}\mathbb{P}^2$ Model: Structural Data}
\label{ap:structure}
%%%%%%%%%%%%%%%%%%%%%%%%%%%%%%%%%%%%%%%%%%%%%%%%%%%%%%%%%%%%%%%%%
%%%%%%%%%%%%%%%%%%%%%%%%%%%%%%%%%%%%%%%%%%%%%%%%%%%%%%%%%%%%%%%%%

In this appendix we describe in more detail some structural aspects of the nonperturbative free energies which appeared in the main part of this paper. Their dependence on the propagator was already analyzed in detail in \cite{cesv13}, where we found a combination of both exponentials and polynomials. The explicit dependence on the complex structure modulus is now more complicated. It involves rational functions of $z$, as well as the instanton actions along with their first and second derivatives (recall that we can trade any higher derivative for lower ones, using the Picard--Fuchs equation). Also, the holomorphic limits of the propagator may be written in terms of the instanton actions and rational functions of $z$, as
\be
S^{zz}_{i,\hol} = -\frac{1}{C_{zzz}(z)}\left( \frac{A_i''(z)}{A_i'(z)} - \tilde{f}^z_{zz}(z) \right).
\label{eq:explicitSzzihol}
\ee
\noindent
As far as we have checked, the inverse power of $A_i'(z)$ in \eqref{eq:explicitSzzihol} always gets cancelled in the final expressions for the nonperturbative free energies, so that the dependence on the instanton actions and their derivatives is also polynomial.

The one-instanton free energies, associated to any of the instanton actions, $\pm A_i$, are given by the product of an exponential in the propagator times a polynomial, 
\be
F^{(1)}_g (z,S^{zz}) = \rme^{\frac{1}{2} \left(\p_z A\right)^2 \left(S^{zz}-S^{zz}_\hol\right)}\, \sum_{k=0}^{3g} p^{(1)}_{g,k}(z)[A,\p_z A,\p_z^2 A]\, \left(S^{zz}\right)^k,
\ee
\noindent
with the ``coefficients''
\be
p^{(1)}_{g,k}(z)[A,\p_z A,\p_z^2 A] = \sum_{\eta \in \CX^{(1)}_{g,k}} p^{(1)}_{g,k,\eta}(z)\, A^{\eta_0} \left(\p_z A\right)^{\eta_1} \left(\p_z^2 A\right)^{\eta_2}.
\ee
\noindent
Here, we have denoted by $A$ any of the conifold instanton actions, or their negatives. The $p^{(1)}_{g,k,\eta}(z)$ are rational functions of $z$, and each $\eta \in \CX^{(1)}_{g,k}$ is of the form $\eta = (k;\eta_0,\eta_1,\eta_2)$. We can give a heuristic formula for the different vectors $\eta \in \CX^{(1)}_{g,k}$, for fixed $g$, starting from the case $k=3g$ where there is only one $\eta = (3g;1,3g,0) \in \CX^{(1)}_{g,k=3g}$, and then applying the recursion
\bea
\left\{\eta \in \CX^{(1)}_{g,k} \right\} &=& \left. \Big\{ \widetilde{\eta}+\lambda\,\, \right|\,\, \widetilde{\eta} \in \CX^{(1)}_{g,k+1},  \\
&&  \lambda \in \left\{ (-1;0,0,0), (-1;0,-2,0), (-1;0,-1,+1), (-1;-1,+1,0) \right\} \Big\}^\star. \nonumber
\eea
\noindent
The star (${}^\star$) indicates that we have discarded $\eta$'s of the form
\begin{gather}
(k;1,0,0), \quad (k;0,1,0), \quad (k;0,0,1), \\
(k;0,\eta_1,\eta_2) \quad \text{if} \quad \eta_1+\eta_2=3g+1, \quad (k;0,0,0) \quad \textrm{if} \quad k\leq 3\left[ \frac{g}{2} \right],
\end{gather}
\noindent
or $\eta$'s with some negative component, should they appear in the set. For $g=1$ we also have to discard $(0;1,1,0)$. For example, the genus $g=1$ free energy, written in a schematic form where we have omitted the rational functions of $z$, $p^{(1)}_{1,k,\eta}(z)$, is
\bea
F^{(1)}_1 (z,S^{zz}) &\sim& \rme^{\frac{1}{2} \left(\p_z A\right)^2 \left(S^{zz}-S^{zz}_\hol\right)}\, \Big\{ \left(S^{zz}\right)^3 A A'^3 + \left(S^{zz}\right)^2 \left(A A'+ A A'^3 + A A'^2 A''\right) + \nonumber \\
&+& \left(S^{zz}\right) \left(A A' + A A'^3 + A A'' + A A'^2 A'' + A A' A''^2\right) + \nonumber \\
&+& 1 + A'^2 + A A'^3 + A' A'' + A A'^2 A'' + A A' A''^2 + A A''^3\, \Big\}.
\eea
\noindent
From the properties of the $\eta$'s one can further see\footnote{For the top case, $k=3g$, $(-1)^{\sum_{i=0}^{2}\eta_i} = (-1)^{3g+1}= (-1)^{g+1}$. Then, by induction on $k$, we find the desired result since the addition of any $\lambda$ does not change parity.} that under a change of sign in $A$ the free energies acquire a sign or not, depending on the parity of $g$;
\be
\left. F^{(1)}_g \right|_{A\to -A} = (-1)^{g+1}\, F^{(1)}_g.
\label{eq:F1AtominusA}
\ee
\noindent
Restoring the full notation from the main body of the text, this means that
\be
\frac{S_{1,i}}{2\pi \rmi}\, F^{(\bfep_{2i-1})}_g = (-1)^{g+1}\, \frac{\widetilde{S}_{-1,i}}{2\pi \rmi}\, F^{(\bfep_{2i})}_g,
\ee
\noindent
which is one of the conditions ensuring a topological genus expansion for the perturbative sector. 

In practice, and using a \textit{Mathematica} symbolic code, we computed one-instanton free energies for the sectors $(1|0 \| 0 \cdots)$ and its symmetric $(0|1 \| 0 \cdots)$ up to genus $g = 21$ in closed analytic form. However, the computation becomes rather impractical past this point (and even before that, the process requires a great amount of computer memory). Switching instead to a high-precision numerical representation for the modulus dependence, allows us to calculate one-instanton free energies up to genus $g = 80$. Each computation is valid for a particular point in moduli space, but the dependence in the propagator remains analytic throughout the calculation.

As discussed in the main text, the large-order of the one-instanton free energies is controlled, to leading order, by the sectors $\widehat{F}^{(2\bfe_1)}_h$ and $\widehat{F}^{(\bfe_{1,1})}_h$, see \eqref{eq:1sectorpartialsums}. For the two-instanton sector $\widehat{F}^{(2\bfe_1)}_g$ the structure of the free energy is 
\bea
\widehat{F}^{(2\bfe_1)}_g &=& \rme^{2\cdot\frac{1}{2} \left(\p_z A\right)^2 \left( S^{zz}-S^{zz}_{\hol} \right)}\, \sum_{k=0}^{3g} p^{(2\bfe_1)}_{2;g,k}(z)[A,\p_z A,\p_z^2 A]\, \left(S^{zz}\right)^k + \\
&+& \rme^{4\cdot\frac{1}{2} \left(\p_z A\right)^2 \left( S^{zz}-S^{zz}_{\hol} \right)}\, \sum_{k=0}^{3g} p^{(2\bfe_1)}_{4;g,k}(z)[A,\p_z A,\p_z^2 A]\, \left(S^{zz}\right)^k,
\eea
\noindent
where
\be
p^{(2\bfe_1)}_{r;g,k}(z)[A,\p_z A,\p_z^2 A] = \sum_{\eta \in \CX^{(2\bfe_1)}_{r;g,k}} p^{(2\bfe_1)}_{r;g,k,\eta}(z)\, A^{\eta_0} \left(\p_z A\right)^{\eta_1} \left(\p_z^2 A\right)^{\eta_2}, \qquad r=2,4.
\ee
\noindent
The sets $\CX^{(2\bfe_1)}_{r;g,k}$ are similar to $\CX^{(1)}_{g,k}$ but we have not found an equally simple way to describe them. They are also equal to each other. In the same schematic form we used before, we find
\bea
\widehat{F}^{(2\bfe_1)}_1 &\sim& \rme^{2\, \frac{1}{2} \left(\p_z A\right)^2 \left(S^{zz}-S^{zz}_\hol\right)}\, \Big\{ \left(S^{zz}\right)^3 A^2 A'^3 + \left(S^{zz}\right)^2 \left(A^2 A'+ A^2 A'^3 + A^2 A'^2 A''\right) + \nonumber\\
&+& \left(S^{zz}\right) \left(A^2 A' + A A'^2 + A^2 A'^3 + A^2 A'' + A^2 A'^2 A'' + A^2 A' A''^2\right) + \nonumber\\
&+& A + A A'^2 + A^2 A'^3 + A A' A'' + A^2 A'^2 A'' + A^2 A' A''^2 + A^2 A''^3\, \Big\} + \nonumber\\
&+& \rme^{4\cdot\frac{1}{2} \left(\p_z A\right)^2 \left(S^{zz}-S^{zz}_\hol\right)}\, \Big\{ \text{ the very same as above } \Big\}.
\eea
\noindent
As for the mixed sector $\widehat{F}^{(\bfe_{1,1})}$, the general structure of the free energy is instead
\bea
\widehat{F}^{(\bfe_{1,1})}_0 &=& \frac{2\pi\rmi}{2 S_{1,1}}\, \frac{2\pi\rmi}{2 \widetilde{S}_{-1,1}} \left(\frac{A}{2\pi^2}\right)^2 \left( \rme^{2\, \frac{1}{2} \left(\p_z A\right)^2 \left( S^{zz}-S^{zz}_{\hol} \right)} - 1 \right), \\
\widehat{F}^{(\bfe_{1,1})}_{2g} &=& \rme^{2\cdot\frac{1}{2} \left(\p_z A\right)^2 \left( S^{zz}-S^{zz}_{\hol} \right)}\, \sum_{k=0}^{5g} p^{(\bfe_{1,1})}_{2;2g,k}(z)[A,\p_z A,\p_z^2 A]\, \left(S^{zz}\right)^k + \\
&+& \sum_{k=0}^{3g-1} p^{(\bfe_{1,1})}_{0;2g,k}(z)[A,\p_z A,\p_z^2 A]\, \left(S^{zz}\right)^k, \qquad g>0.
\eea
\noindent
Due to resonance, $\widehat{F}^{(\bfe_{1,1})}_{\text{odd}} = 0$. The coefficients $p^{(\bfe_{1,1})}_{2;2g,k}(z)$ and $p^{(\bfe_{1,1})}_{0;2g,k}(z)$ are similar to the previous ones for the one and two instanton sectors and it would be interesting to find simple closed formulae for them. As an example, $\widehat{F}^{(\bfe_{1,1})}_2$, again omitting any rational functions of $z$, is
\bea
\widehat{F}^{(\bfe_{1,1})}_2 &\sim& \rme^{2\cdot\frac{1}{2} \left(\p_z A\right)^2 \left( S^{zz}-S^{zz}_{\hol} \right)}\, \Big\{ \left(S^{zz}\right)^5 A^2 A'^4 + \left(S^{zz}\right)^4 \left(A^2 A'^2+A^2 A'^4+A^2 A'^3 A''\right) + \nonumber\\
&+& \left(S^{zz}\right)^3 \left(A^2 A'+A^2 A'^3+A^2 A'^4 + A^2 A' A'' + A^2 A'^3 A'' + A^2 A'^2 A''^2\right) + \nonumber\\
&+&  \left(S^{zz}\right)^2 \left( A A' + A^2 A'^2 + A A'^3 + A'^4 + A^2 A'^4 + A^2 A' A'' \right. + \nonumber\\
&& \hspace{1cm} \left. +\, A A'^2 A'' + A^2 A'^3 A'' + A^2 A''^2 + A^2 A'^2 A''^2 + A^2 A' A''^3 \right) + \nonumber\\
&+& S^{zz} \left(A A' + A'^2 + A^2 A'^2 + A A'^3 + A'^4 + A^2 A'^4 + A A'' + A^2 A' A'' + A A'^2 A'' \right. + \nonumber\\
&& \hspace{1cm} \left. +\, A'^3 A'' + A^2 A'^3 A'' + A^2 A''^2 + A A' A''^2 + A^2 A'^2 A''^2 + A^2 A' A''^3 + A^2 A''^4 \right) + \nonumber\\
&+& 1 + A'^2 + A^2 A'^2 + A A'^3 + A'^4 + A^2 A'^4 + A' A'' + A^2 A' A'' + A A'^2 A'' + A'^3 A'' + \nonumber\\
&+& A^2 A'^3 A'' + A^2 A''^2 + A A' A''^2 + A'^2 A''^2 + A^2 A'^2 A''^2 + A A''^3 + A^2 A' A''^3 + A^2 A''^4\, \Big\} + \nonumber\\
&+& \Big\{ \left(S^{zz}\right)^2 A A' + S^{zz} \left(A A' + A'^2 + A A''\right) + 1 + A'^2 + A' A''\, \Big\}.
\eea
\noindent
In practice, we have computed both these nonperturbative free energies, $\widehat{F}^{(2\bfe_1)}_g$ and $\widehat{F}^{(\bfe_{1,1})}_g$, up to genus $g=8$, but to reach higher orders one must switch to a seminumerical approach.

The two-instanton contribution which appears in the large-order growth of the perturbative coefficients at subleading order, $\widetilde{F}^{(2\bfe_{1,1})}_g$, is a combination of two-instanton free energies computed from the holomorphic anomaly equations. One of them is $\widehat{F}^{(2\bfe_1)}_g$ as above; while the other, $F^{(2\bfe_1)}_g$, is calculated in the same way but with a nonvanishing holomorphic limit (see \eqref{eq:holo2sector}). The structure of $F^{(2\bfe_1)}_g$ is the same as that of $\widehat{F}^{(2\bfe_1)}_g$, in terms of propagator dependence. As to their difference, $\widetilde{F}^{(2\bfe_{1,1})}_g = F^{(2\bfe_1)}_g - \widehat{F}^{(2\bfe_1)}_g$, it is simpler; it only has one exponential term,
\be
\widetilde{F}^{(2\bfe_{1,1})}_g = \rme^{4\cdot\frac{1}{2} \left(\p_z A\right)^2 \left( S^{zz}-S^{zz}_{1,\hol} \right)}\, \sum_{k=0}^{3g} \tilde{p}^{(2\bfe_1)}_{4;g,k}(z)[A,\p_z A,\p_z^2 A]\, \left(S^{zz}\right)^k.
\ee
\noindent
We have computed these free energies up to genus $g = 8$. An analogous relation to \eqref{eq:F1AtominusA} is satisfied for the two-instanton contributions $\widetilde{F}^{(2\bfe_1)}_g$, implying that the topological genus expansion of the perturbative sector is explicitly checked up to second order, for local $\BC\BP^2$.

\newpage

%%%%%%%%%%%%%%%%%%%%%%%%%%%%%%%%%%%%%%%%%%%%%%%%%%%%%%%%%%%%%%%%%
%%%%%%%%%%%%%%%%%%%%%%%%%%%%%%%%%%%%%%%%%%%%%%%%%%%%%%%%%%%%%%%%%

\bibliographystyle{plain}
%\bibliography{papers}

\end{document}